\documentclass[preprint, groupedaddress, showkeys, aps, pre, nofootinbib, longbibliography]{revtex4-2}
\usepackage{graphicx}% Include figure files
\usepackage{bm}% bold math
\usepackage{amssymb, amsmath}
\usepackage{algpseudocode}
\usepackage{booktabs}
\usepackage{xcolor}
\usepackage[acronym]{glossaries}
\usepackage[caption=false]{subfig}

\renewcommand{\deg}[1]{d_{#1}}
\newcommand{\degi}[1]{d_{#1}^{\mathrm{in}}}
\newcommand{\dego}[1]{d_{#1}^{\mathrm{out}}}
\newcommand{\bardegi}[1]{\bar{d}_{#1}^{\mathrm{in}}}
\newcommand{\bardego}[1]{\bar{d}_{#1}^{\mathrm{out}}}
\newcommand{\adjm}[1]{A_{#1}}
\newcommand{\baradjm}[1]{\bar{A}_{#1}}
\newcommand{\qval}[1]{q_{#1}}
\newcommand{\barqval}[1]{\bar{q}_{#1}}
\newcommand{\qVal}{Q}
\newcommand{\adjM}{\mathbf{A}}
\newcommand{\graph}{\mathcal{G}}
\newcommand{\nodes}{\mathcal{V}}
\newcommand{\bnpnodes}[2]{\mathcal{V}_{#1}^{(#2)}}
\newcommand{\edges}{\mathcal{E}}
\newcommand{\communities}{\mathcal{K}}
\newcommand{\community}{\mathcal{C}}
\newcommand{\tcommunity}{\mathcal{T}}
\newcommand{\columns}{\mathcal{U}}
\newcommand{\abnpnodes}{\mathcal{N}}
\newcommand{\partition}{\mathcal{P}}

\newcommand{\decx}[1]{x_{#1}}
\newcommand{\decy}[1]{y_{#1}}

\newcommand{\decz}[1]{z_{#1}}
\newcommand{\decVz}{\mathbf{z}}
\newcommand{\decv}[1]{v_{#1}}
\newcommand{\decw}[1]{w_{#1}}
\newcommand{\dece}[1]{e_{#1}}
\newcommand{\bardece}[1]{\bar{e}_{#1}}

\newcommand{\decVe}{\mathbf{e}}

\newcommand{\qcol}[1]{p_{#1}}

\newcommand{\wedg}[1]{g_{#1}}
\newcommand{\wnod}[1]{h_{#1}}
\newcommand{\relimb}{\tau}
\newcommand{\tol}[1]{\varepsilon_{#1}}
\newcommand{\scon}[1]{\epsilon_{#1}}
\newcommand{\gpc}{K}

\newcommand{\dual}[1]{\pi_{#1}}
\newcommand{\ddual}[1]{\lambda_{#1}}

\newcommand{\gprun}{N_{\mathrm{GP}}}
\newcommand{\gpridx}{\gamma}

\newcommand{\rval}[2]{b_{#1}^{\left(#2\right)}}
\newcommand{\barrval}[2]{\bar{b}_{#1}^{\left(#2\right)}}
\newcommand{\ridx}{o}

\newcommand{\setres}{\mathcal{R}}
\newcommand{\numres}{r}
\newcommand{\maxres}[1]{b_{\mathrm{max}}^{\left(#1\right)}}
\newcommand{\rvec}[1]{\mathbf{b}_{#1}}
\newcommand{\maxVres}{\mathbf{b}_{\max}}
\newcommand{\louexcess}{E}
\newcommand{\objpri}{O}

\newcommand{\mlpairset}{\mathcal{P}_{\mathrm{ML}}}
\newcommand{\clpairset}{\mathcal{P}_{\mathrm{CL}}}
\newcommand{\barclpairset}{\bar{\mathcal{P}}_{\mathrm{CL}}}
\newcommand{\prob}[1]{\mathbf{P}_{#1}}
\newcommand{\nidx}{i}
\newcommand{\niidx}{j}
\newcommand{\niiidx}{l}
\newcommand{\onidx}{\alpha}
\newcommand{\oniidx}{\beta}
\newcommand{\loumidx}{\mu}
\newcommand{\cidx}{k}
\newcommand{\colidx}{u}
\newcommand{\iteridx}{t}
\newcommand{\bnpidx}{\nu}
\newcommand{\numnodes}{n}
\newcommand{\numedges}{m}
\newcommand{\numcom}{k}
\newcommand{\loulidx}{\ell}
\newcommand{\louscore}{F}
\newcommand{\loures}{R}
\newcommand{\lougain}{\Delta}

\newacronym{milp}{MILP}{mixed-integer linear programming}
\newacronym{miqp}{MIQP}{mixed-integer quadratic programming}
\newacronym{lp}{LP}{linear programming}
\newacronym{clouvain}{C--Louvain}{constrained Louvain}
\newacronym{cggp}{CG--GP}{column generation with heuristic graph-partitioning pricing}
\newacronym{cgl}{CG--L}{column generation with exact MILP pricing}
\newacronym{bapgpl}{BaP--GP--L}{branch-and-price with graph-partitioning and exact MILP pricing}
\newacronym{bapgpq}{BaP--GP--Q}{branch-and-price with graph-partitioning and exact quadratic programming pricing}
\newacronym{hmc}{HMC}{Hyundai Motor Company}

\makeatletter
\renewcommand{\fnum@figure}{\textbf{Figure~\thefigure}}
\makeatother

\newcounter{algorithm}
\renewcommand{\thealgorithm}{\arabic{algorithm}}
\newenvironment{ruledalgorithm}[2][hbt!]{%
  \begin{figure}[#1]
  \refstepcounter{algorithm}%
  \par\medskip\noindent
  \begin{minipage}{\linewidth}
  \hrule
  \vspace{0.25em}
  {\raggedright\noindent\textbf{Algorithm~\thealgorithm. #2}\par}
  \vspace{0.20em}
  \hrule
  \vspace{0.20em}
}{%
  \vspace{0.20em}
  \hrule
  \end{minipage}
  \par\smallskip
  \end{figure}
}


\begin{document}
\raggedbottom

% \preprint{APS/123-QED}

\title{Constrained modularity maximization in complex networks via branch-and-price}
% Title candidates
% Constrained modularity maximization in complex networks via graph partitioning and column generation
% A branch-and-price algorithm for constrained modularity maximization in complex networks
% Constrained modularity maximization in complex networks: a branch-and-price approach

\author{Jaewoo Kim}
\author{Seungyeop Lee}
\email[Contact Author: ]{sylee1509@kaist.ac.kr}
\author{Jaemyung Ahn}
\affiliation{Department of Aerospace Engineering, Korea Advanced Institute of Science and Technology, Daehak-Ro 291, Daejeon 34141, Republic of Korea}
\author{Euihyeon Choi}
\affiliation{Daniel Guggenheim School of Aerospace Engineering, Georgia Institute of Technology, Atlanta, Georgia 30332, United States}
% \author{Eun Suk Suh}
%  \email{essuh@snu.ac.kr}
% \affiliation{Graduate School of Engineering Practice, Seoul National University, Gwanak-gu, Seoul 08826, Republic of Korea}
\author{Jeong-Hoon Kim}
\affiliation{Architecture System Planning Team, Hyundai Motor Company, Seongnam-si, Gyeonggi-do 13529, Republic of Korea}

\date{\today}% It is always \today, today,

\begin{abstract}
Modularity maximization is widely used to identify community structure in complex networks. In many applications, however, community detection must also incorporate prior knowledge and domain-specific requirements beyond network topology. We study constrained modularity maximization with restrictions on the number of communities, must-link and cannot-link relations, and general resource constraints. While fast heuristics can provide high-quality solutions, they generally lack optimality guarantees, whereas exact mathematical-programming approaches can become computationally demanding as the network size and constraint complexity increase. We develop a unified optimization framework offering complementary heuristic and exact solution methods. These include a constrained Louvain heuristic for feasible solution generation, a column-generation heuristic with graph-partitioning-based pricing, and exact branch-and-price algorithms using mixed-integer linear programming and sparse mixed-integer quadratic programming formulations for the pricing subproblem. We further develop a cutting-plane procedure that provides independent upper bounds for assessing solution quality when exact optimization cannot be completed. Experiments on 11 unconstrained networks with up to 1,005 nodes and 30 constrained instances with up to 379 nodes demonstrate the complementary roles of these methods. The column-generation heuristic rapidly improves the constrained Louvain solutions across the tested constraint classes, while branch-and-price using the sparse quadratic formulation solves 29 of the 30 constrained instances to optimality within a one-hour time limit. This formulation becomes particularly effective on larger, sparse, undirected networks by substantially reducing the size of the exact pricing model. Overall, the framework enables solution methods to be selected according to the desired balance between computational time, solution quality, and optimality guarantees.
\end{abstract}


\maketitle

\newpage
\section{Introduction}
\label{sec:introduction}
Complex networks provide a compact representation of interacting units in natural, technological, biological, and social systems. Their structural organization has long been studied as a source of collective behavior in complex systems~\cite{strogatz2001exploring,albert2002statistical,newman2003structure,boccaletti2006complex}. Community structure is a widely studied property of this organization, representing a partition of a network into groups of nodes that are densely connected internally and sparsely connected externally~\cite{girvan2002community,fortunato2010community,fortunato2016community}.

Community detection seeks to identify groups of nodes that represent meaningful modules in a network. Although many definitions of communities have been proposed, they differ in the structural principles used to evaluate a partition, such as edge density, separation from the rest of the network, flow persistence, or statistical significance~\cite{fortunato2010community,fortunato2016community}. Modularity maximization is one of the most influential approaches for formulating this ambiguous inference problem as an explicit optimization problem. Newman and Girvan~\cite{newman2004finding} introduced modularity as a measure of the excess density of within-community edges relative to a random-graph null model. The resulting scalar objective has made modularity a standard basis for algorithmic community detection.

Maximizing this objective, however, is computationally challenging: from an algorithmic standpoint, the modularity maximization problem is NP-hard~\cite{brandes2008modularity}. Consequently, much of the literature has focused on heuristic algorithms that obtain high-quality partitions for large networks at manageable computational cost. Early approaches include divisive algorithms based on edge betweenness~\cite{girvan2002community}, greedy agglomerative algorithms~\cite{clauset2004finding}, and spectral methods based on the modularity matrix~\cite{newman2006modularity}. The Louvain method substantially improved scalability by combining local node moves with hierarchical aggregation~\cite{blondel2008fast}, and the Leiden algorithm further improved this framework by addressing the possibility of disconnected or poorly connected communities~\cite{traag2019louvain}. These algorithms have made modularity-based community detection practical in many settings. However, they are designed primarily for unconstrained modularity maximization, where a partition is evaluated solely based on network topology.

In many applications, however, network topology alone is insufficient to determine the desired community structure. From an optimization perspective, modularity assesses partitions according to their structural quality, while side constraints restrict the feasible set according to additional information or requirements. These constraints can serve at least two distinct roles~\cite{ganji2017declarative}. First, they can guide inference by ruling out partitions that are inconsistent with prior information. Known community assignments for a subset of nodes~\cite{allahverdyan2010community}, node annotations or metadata~\cite{newman2016structure,peel2017ground}, and pairwise relationships~\cite{eaton2012spin}, for example, may provide information about community structure beyond that contained in topology alone. Such guidance is particularly relevant given the known limitations of modularity, including the resolution limit~\cite{fortunato2007resolution}, degeneracy in the modularity landscape~\cite{good2010performance}, and difficulties in recovering communities with heterogeneous sizes~\cite{lancichinetti2011limits}. Second, constraints can impose admissibility requirements that need not convey information about latent community structure. Feasible partitions may instead be required to satisfy bounds on community number or size~\cite{ganji2017declarative}, prescribed relationships between nodes~\cite{ganji2017declarative}, attribute balance or composition requirements and capacity limits~\cite{sinha2020design,kim2025mathematical}, or internal-cohesion conditions~\cite{cafieri2015adding}. In summary, constrained modularity maximization preserves modularity as the objective while using side constraints to define the admissible set of partitions.

Such constraints change the nature of the community detection problem. Existing modularity-maximization algorithms may return partitions that violate side constraints because they optimize topology-driven modularity without accounting for additional feasibility requirements. One possible response is to run an unconstrained algorithm and then post-process the resulting partition. For spatial regionalization, for example, Wang et al.~\cite{wang2021network} used a final refinement phase in spatially constrained Louvain and Leiden variants: after an aspatial partition is obtained, the procedure splits noncontiguous communities and merges undersized units to enforce contiguity and minimum region size. Such repair steps can restore feasibility, but because they are applied separately from the original optimization process, they do not optimize modularity over the constrained feasible set. As a result, the quality of the final solution may depend strongly on the problem structure and implementation details. A second response is to incorporate constraint violations into the objective through penalty terms. In engineering design, Sinha et al.~\cite{sinha2020design} modified a community-detection algorithm using a penalty-based approach to design constraints. This approach allows existing heuristic search procedures to account for constraints during optimization. However, because constraint violations are penalized rather than excluded through explicit feasibility constraints, the final solution is not necessarily guaranteed to satisfy all constraints; both feasibility and solution quality may depend on the penalty formulation and the underlying search procedure. For hard constraints, a more direct alternative is to encode the constraints explicitly in the mathematical programming model, making feasibility an integral part of the problem rather than a post hoc correction or penalty term.

Mathematical programming provides a direct way to address constrained modularity maximization by formulating side constraints explicitly and optimizing modularity over the resulting feasible set of partitions. Early formulations showed that modularity maximization can be expressed as a mathematical programming model and solved to global optimality on small instances while accommodating constraints such as community-size and balance constraints~\cite{xu2007finding}. Subsequent studies developed relaxations and solution-quality bounds~\cite{agarwal2008modularity}, incorporated cohesion and other side constraints directly into optimization models~\cite{cafieri2015adding,ganji2018lagrangian}, and introduced decomposition-based approaches such as column generation and branch-and-price~\cite{aloise2010column,sato2019enhanced,aref2024bayan}. These approaches offer advantages that are difficult to obtain from conventional heuristics, including explicit representation of constraints, feasibility guarantees, and, when exact optimization or valid bounds are available, certificates or bounds on solution optimality. These advantages, however, come at a computational cost. Whereas heuristic algorithms such as Louvain and Leiden can scale to networks with millions of nodes and beyond~\cite{blondel2008fast}, mathematical-programming-based methods have generally been demonstrated on substantially smaller instances; for example, Aloise et al.~\cite{aloise2010column} considered networks with up to 512 nodes and 819 edges, Sato and Izunaga~\cite{sato2019enhanced} solved modularity-density instances with over 100 nodes, and Bayan targets exact or bounded modularity optimization for networks with up to about 3000 edges in the largest connected component~\cite{aref2024bayan}. Recent advances in column-generation and pricing strategies suggest that this computational gap can be reduced~\cite{maher2023integer}, but achieving both scalable computation and the direct treatment of constraints within a unified modularity-maximization framework remains challenging.

This gap motivates the present work. We focus on the constrained modularity maximization problem and develop a mathematical-programming-based optimization framework for the explicit treatment of constraints with optimality guarantees. We consider constraints on the number of communities, pairwise must-link and cannot-link relations, and general resource requirements that can represent community size, node attributes, composition, capacity, and pairwise cohesion. We formally formulate this problem as a \acrfull{milp} model, providing an explicit representation of both the modularity objective and the admissible community structures.

To address the resulting computational challenge, we develop a hierarchy of solution methods. First, \acrfull{clouvain} complements the optimization framework by providing high-quality initial solutions for subsequent optimization. Next, \acrfull{cggp} combines column generation with graph-partitioning-based heuristic pricing to improve the initial solution obtained by \acrshort{clouvain}. For optimality guarantees, \acrfull{bapgpl} and \acrfull{bapgpq} embed column generation within a branch-and-price framework and use exact pricing to establish global optimality. Together, these methods provide a flexible computational framework that combines fast constraint-aware heuristics with a graph-partitioning-based solution refinement and mathematical-programming-based exact optimization.

\begin{figure}[hbt!]
\centering
\includegraphics[page=1,width=0.95\linewidth]{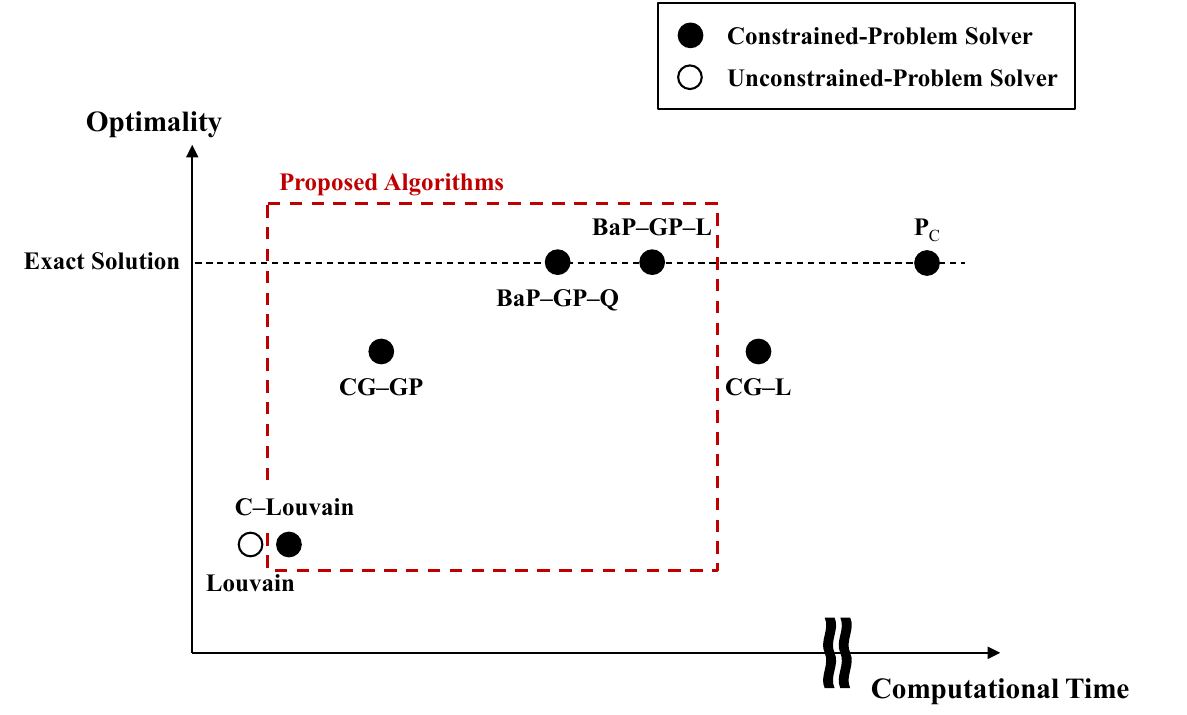}
\caption{Schematic positioning of community-detection methods in terms of computational time and optimality. $\prob{\mathrm{C}}$ and \acrshort{cgl} are included as naive baseline approaches for comparison in Sec.~\ref{sec:numerical-study}.}
\label{fig:algorithm_overview}
\end{figure}

Figure~\ref{fig:algorithm_overview} summarizes this spectrum of solution approaches. Conventional heuristic community-detection algorithms emphasize scalability but generally provide no guarantee of optimality, whereas exact mathematical-programming approaches can provide optimality guarantees at substantially greater computational cost. In the figure, Louvain is shown as a representative conventional heuristic approach, with other heuristics, such as the Girvan--Newman algorithm~\cite{girvan2002community}, expected to occupy nearby positions along the spectrum. In contrast, $\prob{\mathrm{C}}$ represents a direct mathematical-programming approach solved using commercial solvers, based on the MILP formulation introduced later in this paper, while \acrshort{cgl} represents a baseline column-generation approach without the proposed enhancements. The proposed methods occupy complementary positions along this spectrum. Rather than prescribing a single solution strategy, the framework allows constrained community structure to be identified using a method appropriate to the desired balance between computational efficiency and solution-quality guarantees.

The contributions of this work are twofold. First, we formulate constrained modularity maximization within a unified framework that accommodates multiple classes of constraints and develop a family of heuristic, column-generation, and branch-and-price algorithms for solving the resulting community-detection problem. Second, we systematically evaluate these methods on unconstrained and constrained benchmark networks to characterize their computational efficiency and solution quality. These experiments provide practical guidance on the trade-offs between computational efficiency and optimality across heuristic and exact methods for constrained modularity maximization.

The rest of this paper is organized as follows. Section~\ref{sec:problem-description} defines the constrained modularity-maximization problem and presents its \acrshort{milp} formulation. Section~\ref{sec:solution-methodology} develops the proposed optimization framework for solving and analyzing the problem. Section~\ref{sec:numerical-study} evaluates the proposed methods on benchmark networks under different classes of constraints. Finally, Sec.~\ref{sec:conclusion} summarizes the main findings and concludes the paper.

\section{Problem Description}
\label{sec:problem-description}

This section formulates the constrained modularity-maximization problem studied in this paper. We first present a baseline \acrshort{milp} representation of modularity maximization without constraints. We then introduce the constraints considered in this work and incorporate them into the extended formulation.

\subsection{Basic Formulation}
\label{subsec:basic-formulation}

Consider a directed graph $\graph=(\nodes,\edges)$, where $\nodes$ is the set of $\numnodes$ nodes and $\edges$ is the set of $\numedges$ directed edges.\footnote{We adopt the directed formulation as the general case; the minor modifications required for undirected networks are introduced where relevant.} Let $\adjM$ denote the adjacency matrix of $\graph$, where $\adjm{\nidx\niidx}=1$ if there is an arc from node $\nidx$ to node $\niidx$ and $\adjm{\nidx\niidx}=0$ otherwise. For each node $\nidx\in\nodes$, the in-degree and out-degree are defined as
\begin{equation}
\degi{\nidx}=\sum_{\niidx\in\nodes}\adjm{\niidx\nidx},\qquad\dego{\nidx}=\sum_{\niidx\in\nodes}\adjm{\nidx\niidx}.
\end{equation}

The directed modularity of a partition is defined as~\cite{leicht2008community}
\begin{equation}
\qVal=\frac{1}{\numedges}\sum_{\nidx,\niidx\in \nodes}\left(\adjm{\nidx\niidx}-\frac{\dego{\nidx}\degi{\niidx}}{\numedges}\right)\delta\left(\nidx,\niidx\right),
\label{eq:directed-modularity}
\end{equation}
where $\delta\left(\nidx,\niidx\right)$ equals $1$ if nodes $\nidx$ and $\niidx$ belong to the same community and $0$ otherwise. For an undirected network, the standard modularity formulation is recovered by replacing the in- and out-degrees with the node degree and replacing $\numedges$ with $2\numedges$ in both the normalization factor and the null-model term~\cite{newman2004finding}.

The diagonal terms and the positive multiplier $1/\numedges$ do not affect the optimal partition. Thus, modularity maximization can be written, up to an additive constant and positive scaling factor, using unordered-pair coefficients. For each unordered pair $\left\{\nidx,\niidx\right\}$, define
\begin{equation}\label{eq:qij-definition}
\qval{\nidx\niidx}=\adjm{\nidx\niidx}+\adjm{\niidx\nidx}-\frac{\dego{\nidx}\degi{\niidx}}{\numedges}-\frac{\dego{\niidx}\degi{\nidx}}{\numedges}.
\end{equation}
Then, maximizing modularity is equivalent to maximizing
\begin{equation}
\sum_{\nidx,\niidx\in\nodes,\ \nidx<\niidx}\qval{\nidx\niidx}\delta\left(\nidx,\niidx\right).
\end{equation}
Here, $\nidx<\niidx$ ensures that each unordered node pair is counted only once. For notational convenience, we henceforth assume that every unordered pair $\left\{\nidx,\niidx\right\}$ is indexed such that $\nidx<\niidx$.

Let $\communities$ denote the set of candidate community indices. Since the maximum possible number of communities equals the number of nodes, we set $\communities=\left\{1,2,\dots,\numnodes\right\}$. We introduce binary variables $\decx{\nidx\cidx}$, which equal $1$ if node $\nidx$ is assigned to community $\cidx \in \communities$ and $0$ otherwise. Because these variables explicitly represent node-to-community assignments, they provide a flexible basis for adding node-level, pairwise, and community-level constraints later. In addition, binary variables $\decv{\cidx}$ are introduced, indicating whether community $\cidx$ contains at least one node. The auxiliary variables $\decy{\nidx\niidx\cidx}$ indicate whether nodes $\nidx$ and $\niidx$ are both assigned to community $\cidx$, for each node pair $\left\{\nidx,\niidx\right\}$ and community $\cidx$. Equivalently, $\decy{\nidx\niidx\cidx}$ represents the product $\decx{\nidx\cidx}\decx{\niidx\cidx}$. The basic \acrshort{milp} formulation is given as follows:
\newline\newline\noindent
[$\prob{\mathrm{B}}$] Basic formulation for modularity maximization
\begin{subequations}
\begin{equation}
\max\ \sum_{\nidx,\niidx\in\nodes,\ \nidx<\niidx}\qval{\nidx\niidx}\sum_{\cidx\in\communities}\decy{\nidx\niidx\cidx}
\label{eq:basic-obj}
\end{equation}
subject to
\begin{equation}
\sum_{\cidx\in\communities}\decx{\nidx\cidx}=1\quad \forall \nidx\in \nodes,
\label{eq:basic-assignment}
\end{equation}
\begin{equation}
\decv{\cidx}\le \sum_{\nidx\in\nodes}\decx{\nidx\cidx}\quad \forall\cidx\in\communities,
\label{eq:basic-used-lower}
\end{equation}
\begin{equation}
\decv{\cidx} \ge \decx{\nidx\cidx}\quad \forall \nidx\in\nodes,\ \forall\cidx\in\communities,
\label{eq:basic-used-upper}
\end{equation}
\begin{equation}
\decv{\cidx+1}\le \decv{\cidx}\quad \forall\cidx\in\communities\setminus\{\numnodes\},
\label{eq:basic-symmetry}
\end{equation}
\begin{equation}
\decy{\nidx\niidx\cidx}\le\decx{\nidx\cidx}\quad \forall \nidx,\niidx\in\nodes,\ \nidx<\niidx,\ \forall \cidx\in\communities,
\label{eq:basic-linearization-1}
\end{equation}
\begin{equation}
\decy{\nidx\niidx\cidx}\le\decx{\niidx\cidx}\quad \forall\nidx,\niidx\in\nodes,\ \nidx<\niidx,\ \forall\cidx\in\communities,
\label{eq:basic-linearization-2}
\end{equation}
\begin{equation}
\decy{\nidx\niidx\cidx}\ge\decx{\nidx\cidx}+\decx{\niidx\cidx}-1\quad \forall\nidx,\niidx\in \nodes,\ \nidx<\niidx,\ \forall\cidx\in\communities,
\label{eq:basic-linearization-3}
\end{equation}
\begin{equation}
\decx{\nidx\cidx}\in\{0,1\}\quad \forall\nidx\in\nodes,\ \forall\cidx\in\communities,
\label{eq:basic-x-domain}
\end{equation}
\begin{equation}
\decv{\cidx}\in\{0,1\}\quad \forall\cidx\in\communities,
\label{eq:basic-v-domain}
\end{equation}
\begin{equation}
\decy{\nidx\niidx\cidx}\in\mathbb{R}_{\ge0}\quad \forall\nidx,\niidx\in\nodes,\ \nidx<\niidx,\ \forall\cidx\in\communities.
\label{eq:basic-y-domain}
\end{equation}
\end{subequations}
Here, the objective function in Eq.~\eqref{eq:basic-obj} maximizes modularity, replacing $\delta\left(\nidx,\niidx\right)$ with $\sum_{\cidx\in\communities}\decy{\nidx\niidx\cidx}$. Equation~\eqref{eq:basic-assignment} assigns each node to exactly one community, so overlapping communities are not allowed in this study. Equations~\eqref{eq:basic-used-lower} and~\eqref{eq:basic-used-upper} define whether each community index is used. Equation~\eqref{eq:basic-symmetry} imposes a simple symmetry-breaking rule: if community index $\cidx+1$ is used, then community index $\cidx$ must also be used. This rule does not remove all label symmetries among communities, but it eliminates solutions that differ only by gaps in the ordered community indices. Equations~\eqref{eq:basic-linearization-1}--\eqref{eq:basic-linearization-3} constitute the standard Fortet linearization enforcing $\decy{\nidx\niidx\cidx}=\decx{\nidx\cidx}\decx{\niidx\cidx}$~\cite{fortet1960applications,conforti2014integer}.

\subsection{Additional Constraints}
\label{subsec:additional-constraints}

In many applications, the community structure must satisfy constraints beyond modularity maximization itself. We introduce three common classes of such constraints, including the number of communities, pairwise relations, and resource requirements.

\subsubsection{Number of Communities}
\label{subsubsec:number-of-communities}

The number of communities can be constrained to lie within a prescribed range or fixed to an exact value~\cite{ganji2017declarative}. Given a maximum number of communities $\numcom_{\max}$ and a minimum number $\numcom_{\min}$, the candidate set is defined as
\begin{equation}
\communities=\left\{1,2,\dots,\numcom_{\max}\right\},
\end{equation}
and the following constraint is imposed:
\begin{equation}\label{eq:number-of-communities}
\decv{\numcom_{\min}}=1.
\end{equation}
Together with the sequential-use constraint in Eq.~\eqref{eq:basic-symmetry}, Eq.~\eqref{eq:number-of-communities} ensures that at least $\numcom_{\min}$ communities are used. If the exact number of communities is specified as $\numcom_{e}$, it can be enforced by setting $\numcom_{\min}=\numcom_{\max}=\numcom_{e}$.

\begin{figure}[hbt!]
\centering
\includegraphics[page=2,width=0.80\linewidth]{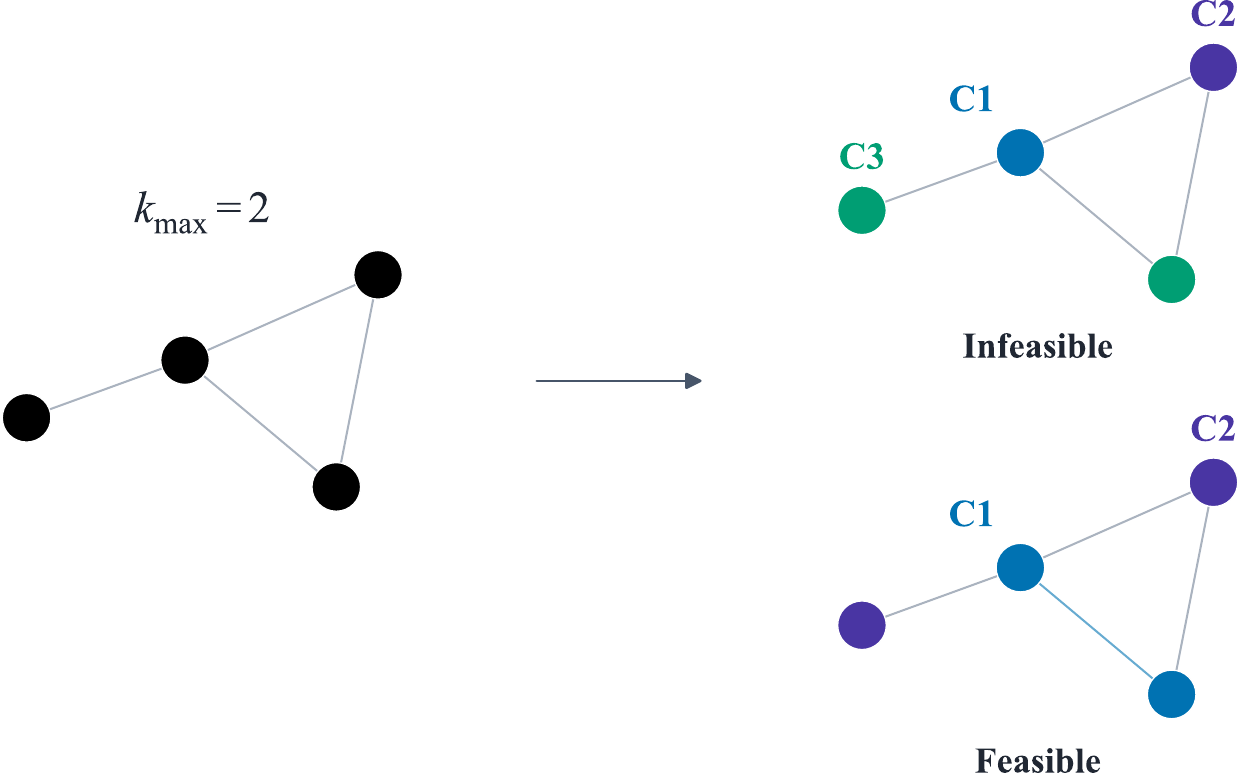}
\caption{Illustration of the maximum-community constraint with $\numcom_{\max}=2$.}
\label{fig:constraints-number_of_communities}
\end{figure}

Figure~\ref{fig:constraints-number_of_communities} illustrates the maximum-community constraint with $\numcom_{\max}=2$. Here, C1, C2, and C3 denote the communities in the resulting partitions, and nodes are color-coded according to their community membership. The upper partition is infeasible because it contains three communities, whereas the lower partition is feasible with two communities.

\subsubsection{Pairwise Constraints}
\label{subsubsec:pairwise-constraints}

Pairwise constraints consist of must-link and cannot-link constraints. Must-link constraints require specified pairs of nodes to belong to the same community, whereas cannot-link constraints require specified pairs to belong to different communities. Such pairwise constraints are commonly used to incorporate additional information into semi-supervised and constrained community detection~\cite{eaton2012spin,cheng2014active,yang2017exploring,ganji2018lagrangian}.

Let $\mlpairset$ and $\clpairset$ denote the sets of unordered node pairs subject to must-link and cannot-link constraints, respectively. Must-link constraints are represented as
\begin{equation}\label{eq:must-link-hard}
\sum_{\cidx\in\communities}\decy{\nidx\niidx\cidx}=1
\quad\forall \{\nidx,\niidx\}\in\mlpairset,
\end{equation}
while cannot-link constraints are represented as
\begin{equation}\label{eq:cannot-link-hard}
\decy{\nidx\niidx\cidx}=0
\quad\forall \{\nidx,\niidx\}\in\clpairset,\ \forall \cidx\in\communities.
\end{equation}
Because $\decy{\nidx\niidx\cidx}=1$ exactly when nodes $\nidx$ and $\niidx$ are both assigned to community $\cidx$, Eq.~\eqref{eq:must-link-hard} requires each must-link pair to share exactly one community. Equation~\eqref{eq:cannot-link-hard} prevents any cannot-link pair from being assigned to the same community.

\begin{figure}[hbt!]
\centering
\includegraphics[page=3,width=0.80\linewidth]{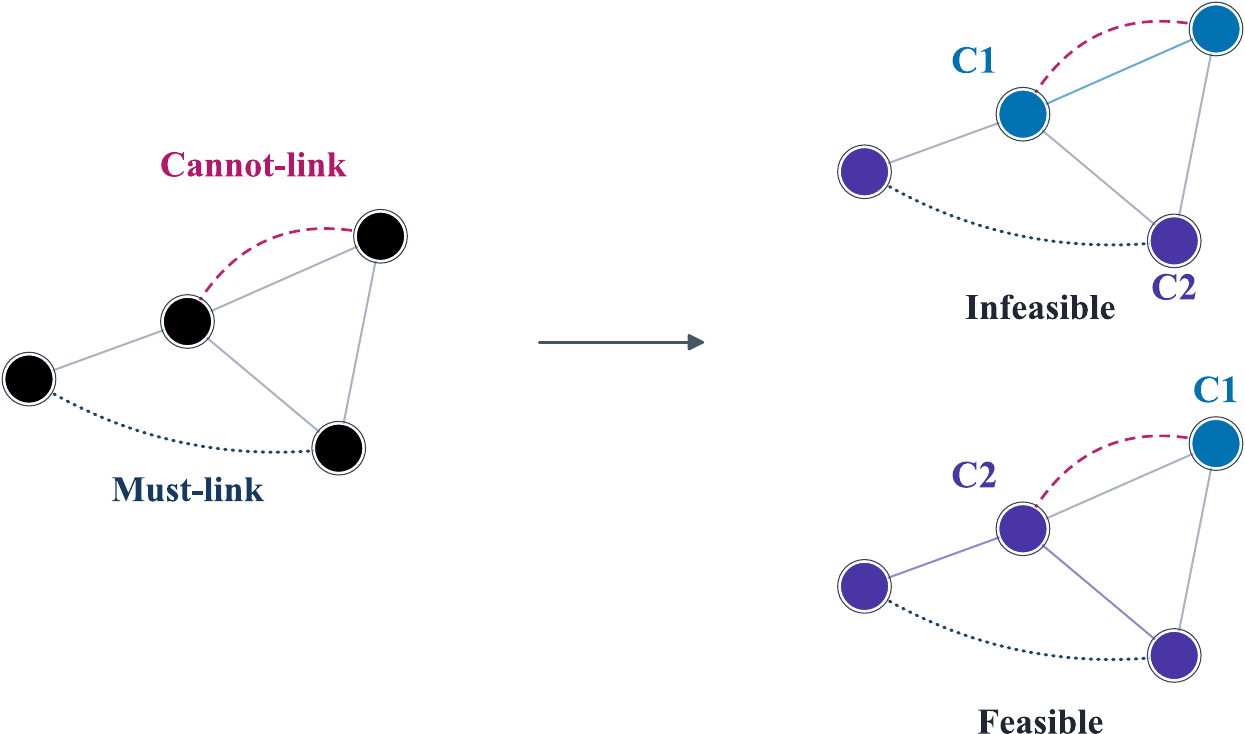}
\caption{Illustration of must-link and cannot-link constraints.}
\label{fig:constraints-pairwise}
\end{figure}

Figure~\ref{fig:constraints-pairwise} illustrates the pairwise constraints. The upper partition is infeasible because the cannot-link pair is assigned to the same community, whereas the lower partition is feasible because both the must-link and cannot-link constraints are satisfied.

\subsubsection{Resource Constraints}
\label{subsubsec:bin-packing-constraints}

Resource constraints regulate aggregate quantities within each community. These quantities may be based on node attributes, pairwise attributes, or a combination of both. Such constraints are useful for representing community-size bounds, attribute-balance requirements, capacity restrictions, or cohesion-related conditions. Community-size and balance constraints have been incorporated directly into mathematical-programming formulations for community detection~\cite{xu2007finding}, while broader design and resource restrictions have been considered in engineering design problems~\cite{sinha2020design,kim2025mathematical}. Cohesion-related restrictions are also closely related to structural definitions of communities~\cite{radicchi2004defining,hu2008comparative,cafieri2012finding} and have been incorporated explicitly as constraints in modularity-maximization models~\cite{cafieri2015adding}.

Let $\rval{\nidx}{\ridx}$ and $\rval{\nidx\niidx}{\ridx}$ denote the node-level and pairwise resource coefficients associated with node $\nidx$ and node pair $\left\{\nidx,\niidx\right\}$, respectively, for resource class $\ridx\in\setres=\{1,\dots,\numres\}$. Let $\maxres{\ridx}$ denote the corresponding upper bound. A general resource constraint can then be represented as
\begin{equation}\label{eq:bin-packing-general}
\sum_{\nidx\in\nodes}\rval{\nidx}{\ridx}\decx{\nidx\cidx}
+\sum_{\substack{\nidx,\niidx\in\nodes,\ \nidx<\niidx}}
\rval{\nidx\niidx}{\ridx}\decy{\nidx\niidx\cidx}
\le\maxres{\ridx}\decv{\cidx}
\quad\forall \cidx\in\communities,\ \forall \ridx\in\setres.
\end{equation}

Note that node-only and pairwise-only constraints are special cases of Eq.~\eqref{eq:bin-packing-general}: node-only constraints are obtained by setting $\rvec{\nidx\niidx}=\mathbf{0}$ for all $\nidx,\niidx\in\nodes$ with $\nidx<\niidx$, whereas pairwise-only constraints are obtained by setting $\rvec{\nidx}=\mathbf{0}$ for all $\nidx\in\nodes$. In addition, lower-bound requirements can be represented by multiplying the corresponding resource coefficients and bound by $-1$. For example, if each community must contain between $4$ and $6$ nodes, one may set $\rvec{\nidx}=\left[1,-1\right]^\top$ for all $\nidx\in\nodes$, $\rvec{\nidx\niidx}=\mathbf{0}$ for all $\nidx,\niidx\in\nodes$ with $\nidx<\niidx$, and $\maxVres=\left[6,-4\right]^\top$.

\begin{figure}[hbt!]
\centering
\includegraphics[page=4,width=0.80\linewidth]{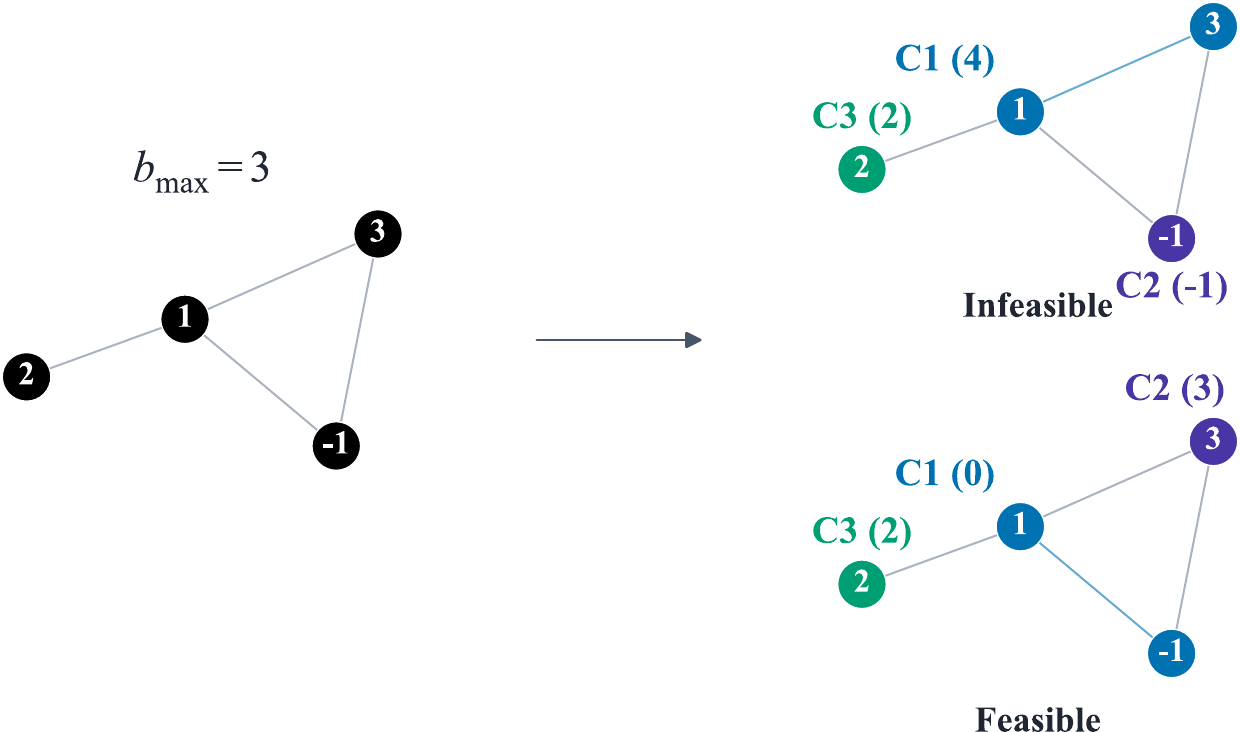}
\caption{Illustration of a node-level resource constraint with an upper bound of $b_{\max}=3$.}
\label{fig:constraints-resource}
\end{figure}

Figure~\ref{fig:constraints-resource} illustrates a node-level resource constraint with $b_{\max}=3$. The upper partition is infeasible because community C1 has a total resource value of $4$, whereas the lower partition is feasible because the resource value of each community does not exceed $3$.

\subsection{Extended Formulation}
\label{subsec:extended-formulation}

Combining the basic formulation with constraints introduced above yields the following extended formulation, with the candidate community set $\communities=\{1,2,\dots,\numcom_{\max}\}$:
\newline\newline\noindent
[$\prob{\mathrm{C}}$] Extended formulation for constrained modularity maximization
\begin{subequations}
\begin{equation}\label{eq:extended-obj}
\max\ \sum_{\nidx,\niidx\in\nodes,\ \nidx<\niidx}
\qval{\nidx\niidx}
\sum_{\cidx\in\communities}\decy{\nidx\niidx\cidx}
\end{equation}
subject to Eqs.~\eqref{eq:basic-assignment}--\eqref{eq:basic-y-domain} and
\begin{equation}\label{eq:extended-number}
\decv{\numcom_{\min}}=1,
\end{equation}
\begin{equation}\label{eq:extended-must-link}
\sum_{\cidx\in\communities}\decy{\nidx\niidx\cidx}=1
\quad\forall \{\nidx,\niidx\}\in\mlpairset,
\end{equation}
\begin{equation}\label{eq:extended-cannot-link}
\decy{\nidx\niidx\cidx}=0
\quad\forall \{\nidx,\niidx\}\in\clpairset,\ \forall \cidx\in\communities,
\end{equation}
\begin{equation}\label{eq:extended-resource}
\sum_{\nidx\in\nodes}
\rval{\nidx}{\ridx}\decx{\nidx\cidx}
+\sum_{\substack{\nidx,\niidx\in\nodes,\ \nidx<\niidx}}
\rval{\nidx\niidx}{\ridx}\decy{\nidx\niidx\cidx}
\le\maxres{\ridx}\decv{\cidx}
\quad\forall \cidx\in\communities,\ \forall \ridx\in\setres.
\end{equation}
\end{subequations}

Here, Eq.~\eqref{eq:extended-number} ensures that at least $\numcom_{\min}$ communities are used. Equations~\eqref{eq:extended-must-link} and~\eqref{eq:extended-cannot-link} represent the must-link and cannot-link constraints, respectively, while Eq.~\eqref{eq:extended-resource} enforces the resource constraints for each used community. In the absence of the constraints in Eqs.~\eqref{eq:extended-number}--\eqref{eq:extended-resource}, $\prob{\mathrm{C}}$ reduces to the basic formulation $\prob{\mathrm{B}}$.

\section{Solution Methodology}
\label{sec:solution-methodology}

\begin{figure}[hbt!]
\centering
\includegraphics[page=5,width=0.45\linewidth]{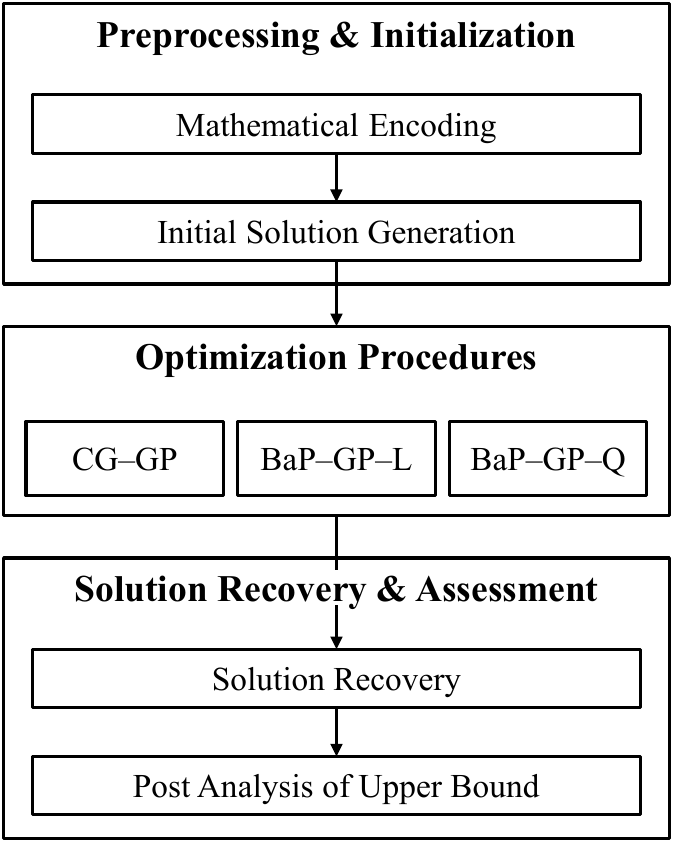}
\caption{Overview of the proposed optimization framework.}
\label{fig:framework_overview}
\end{figure}

The extended formulation $\prob{\mathrm{C}}$ provides an explicit \acrshort{milp} formulation of constrained modularity maximization. However, the problem size grows rapidly as the graph size increases. Consequently, direct solution using general-purpose \acrshort{milp} solvers such as Gurobi~\cite{gurobi} can become computationally intractable as the problem size increases. To address this limitation while retaining the modeling flexibility and solution-quality guarantees of mathematical programming, we develop an integrated optimization framework consisting of three modules: \textit{preprocessing and initialization}, \textit{optimization procedures}, and \textit{solution recovery and assessment}, as illustrated in Fig. \ref{fig:framework_overview}.

The preprocessing and initialization module prepares the problem for optimization. It first defines the model parameters, decision variables, objective function, and constraints. Must-link constraints are then handled through graph contraction during mathematical encoding to reduce the size of the graph. A feasible initial partition is then constructed using the proposed \acrshort{clouvain}. Other initialization methods can also be incorporated depending on the problem characteristics and computational requirements.

The optimization procedures improve the initial solution by decomposing the complex problem into two subproblems: generating promising communities and selecting a combination of these communities to form a partition. Specifically, $\prob{\mathrm{C}}$ is reformulated as a set-partitioning problem where each column represents a feasible community and the associated decision variable determines whether that community is selected. Because enumerating all feasible communities is generally impractical, \acrshort{cggp} and the proposed branch-and-price algorithms---\acrshort{bapgpl} and \acrshort{bapgpq}---dynamically generate promising columns through column generation.

The solution recovery and assessment module maps the optimized partition from the contracted graph back to the original graph. The recovered solution is then evaluated using available upper bounds, including the cutting-plane upper bound, to quantify the remaining optimality gap when global optimality is not established directly. The cutting-plane procedure starts from an \acrshort{lp} relaxation and iteratively adds violated triangle inequalities as cuts.

The remainder of this section provides details of the framework.

\subsection{Preprocessing and Initialization}

\subsubsection{Mathematical Encoding}

The problem of interest is formally defined by a graph $\graph=\left(\nodes,\edges\right)$ together with a set of constraints. These inputs are mathematically encoded to construct $\prob{\mathrm{C}}$. Table~\ref{tab:problem-encoding} summarizes the principal elements required for this encoding, including the modularity coefficients, bounds on the number of communities, pairwise relations, and resource-constraint parameters.

\begin{table}[hbt!]
\centering
\caption{Problem elements to be defined during mathematical encoding.}
\label{tab:problem-encoding}
\footnotesize
\begin{tabular}{lll}
\hline
Element & Notation & Description \\
\hline
Modularity coefficient
& $\qval{\nidx\niidx}$
& Coefficient in the modularity objective \\

Candidate communities
& $\communities$
& Set of available community indices \\

Community-count bounds
& $(\numcom_{\min},\numcom_{\max})$
& Minimum and maximum numbers of communities \\

Must-link pairs
& $\mlpairset$
& Node pairs required to belong to the same community \\

Cannot-link pairs
& $\clpairset$
& Node pairs required to belong to different communities \\

Resource classes
& $\setres$
& Set of resource-constraint classes \\

Node resource
& $\rval{\nidx}{\ridx}$
& Node contribution to resource $\ridx$ \\

Pairwise resource
& $\rval{\nidx\niidx}{\ridx}$
& Pairwise contribution to resource $\ridx$ \\

Resource bound
& $\maxres{\ridx}$
& Maximum allowable amount of resource $\ridx$ in each community \\
\hline
\end{tabular}
\end{table}

If must-link constraints are present, graph contraction is subsequently applied to reduce the size of the graph. Nodes connected directly or indirectly through must-link relations must belong to the same community. Accordingly, each connected component of the undirected graph $(\nodes,\mlpairset)$ is merged into a contracted node, while nodes not involved in any must-link relation remain as singleton components. For example, if $\mlpairset=\left\{\left\{1,2\right\},\left\{2,3\right\},\left\{4,5\right\}\right\}$, then nodes $1$, $2$, and $3$ form one contracted node, while nodes $4$ and $5$ form another. The graph and the corresponding problem elements listed in Table~\ref{tab:problem-encoding} are updated as follows.

Let $\bar{\graph}$ denote the contracted graph obtained after must-link contraction. The index set of the contracted nodes is
\begin{equation}
\bar{\nodes}=\{1,2,\ldots,\bar{\numnodes}\},
\end{equation}
where $\bar{\numnodes}$ is the number of contracted nodes. Throughout this construction, indices $\nidx,\niidx$ refer to contracted nodes, and $\nodes_{\nidx}$ denotes the set of original nodes contracted to node $\nidx$, whereas Greek indices $\onidx,\oniidx$ refer to nodes of the original graph. Let $\bar{\communities}$ denote the community index set for the contracted graph, with the maximum number of communities adjusted to $\bar{\numcom}_{\max}$:
\begin{equation}
\bar{\communities}=\{1,2,\ldots,\bar{\numcom}_{\max}\},
\qquad
\bar{\numcom}_{\max}=\min(\numcom_{\max},\bar{\numnodes}).
\end{equation}
The minimum number of communities remains unchanged. If $\numcom_{\min}>\bar{\numcom}_{\max}$, the problem is identified as infeasible, and the procedure terminates.

The adjacency coefficients of the contracted graph $\bar{\graph}$ are defined as
\begin{equation}\label{eq:contracted-adjacency}
\baradjm{\nidx\niidx}
=\sum_{\onidx\in\nodes_{\nidx}}
\sum_{\oniidx\in\nodes_{\niidx}}
\adjm{\onidx\oniidx}
\quad\forall \nidx,\niidx\in\bar{\nodes},\ \nidx\neq\niidx,
\end{equation}
and the corresponding contracted in- and out-degrees are given by
\begin{equation}\label{eq:contracted-degree}
\bardegi{\nidx}
=\sum_{\onidx\in\nodes_{\nidx}}\degi{\onidx},
\qquad
\bardego{\nidx}
=\sum_{\onidx\in\nodes_{\nidx}}\dego{\onidx}
\quad\forall \nidx\in\bar{\nodes}.
\end{equation}
These degree quantities retain the contributions of edges internal to each contracted node, and $\numedges$ retains its original-graph value throughout the subsequent optimization. The corresponding modularity coefficient is defined as
\begin{equation}\label{eq:aggregated-q}
\barqval{\nidx\niidx}
=\sum_{\onidx\in\nodes_{\nidx}}
\sum_{\oniidx\in\nodes_{\niidx}}
\qval{\min\left\{\onidx,\oniidx\right\}\max\left\{\onidx,\oniidx\right\}}
\quad\forall \nidx,\niidx\in\bar{\nodes},\ \nidx<\niidx.
\end{equation}
Equivalently, using the contracted adjacency and degree quantities defined above, the coefficient can be expressed as
\begin{equation}\label{eq:contracted-modularity-coefficient}
\barqval{\nidx\niidx}
=\baradjm{\nidx\niidx}+\baradjm{\niidx\nidx}
-\frac{
\bardego{\nidx}\bardegi{\niidx}
+\bardego{\niidx}\bardegi{\nidx}
}{\numedges}
\quad\forall \nidx,\niidx\in\bar{\nodes},\ \nidx<\niidx.
\end{equation}

To account for the contributions of original-node pairs within each contracted node, we define the internal modularity contribution as
\begin{equation}\label{eq:contracted-modularity-coefficient-internal}
\barqval{\nidx}
=\sum_{\substack{\onidx,\oniidx\in\nodes_{\nidx}\\\onidx<\oniidx}}
\qval{\onidx\oniidx}
\quad\forall \nidx\in\bar{\nodes}.
\end{equation}
Under the must-link constraints, all original nodes within $\nodes_{\nidx}$ belong to the same community, so the indicator $\delta$ introduced in Eq.~\eqref{eq:directed-modularity} satisfies $\delta(\onidx,\oniidx)=1$ for every pair $\onidx,\oniidx\in\nodes_{\nidx}$. For any partition satisfying these constraints, let $\bar{\delta}(\nidx,\niidx)$ indicate whether contracted nodes $\nidx$ and $\niidx$ belong to the same community in the corresponding partition of the contracted graph. The original objective in Eq.~\eqref{eq:extended-obj} can then be decomposed as
\begin{equation}\label{eq:contracted-objective-decomposition}
\begin{aligned}
&\sum_{\substack{\onidx,\oniidx\in\nodes\\\onidx<\oniidx}}
\qval{\onidx\oniidx}\delta(\onidx,\oniidx)=
\sum_{\nidx\in\bar{\nodes}}\barqval{\nidx}
+
\sum_{\substack{\nidx,\niidx\in\bar{\nodes}\\\nidx<\niidx}}
\barqval{\nidx\niidx}\bar{\delta}(\nidx,\niidx).
\end{aligned}
\end{equation}
The first term on the right-hand side is independent of how the contracted nodes are partitioned and is therefore constant. Consequently, the total internal modularity contribution $\sum_{\nidx\in\bar{\nodes}}\barqval{\nidx}$ can be omitted during optimization and added back when recovering the objective value in Eq.~\eqref{eq:extended-obj} for the original graph.

The cannot-link set is also mapped onto the contracted graph. If there exists a cannot-link pair $\left\{\onidx,\oniidx\right\}\in\clpairset$ such that $\onidx,\oniidx\in\nodes_{\nidx}$ for some $\nidx$, then the must-link and cannot-link constraints conflict. In this case, the problem is identified as infeasible, and the procedure terminates. Otherwise, the reduced cannot-link set is defined as
\begin{equation}\label{eq:reduced-cannot-link-set}
\begin{aligned}
\barclpairset=\bigl\{
\{\nidx,\niidx\}:
\nidx,\niidx\in\bar{\nodes},\ \nidx<\niidx,\
\exists\,\onidx\in\nodes_{\nidx},\
\exists\,\oniidx\in\nodes_{\niidx}
\text{ such that }
\{\onidx,\oniidx\}\in\clpairset
\bigr\}.
\end{aligned}
\end{equation}
The cannot-link constraints in Eq.~\eqref{eq:extended-cannot-link} are then applied to $\barclpairset$ on the contracted graph.

Resource coefficients are updated to preserve the resource usage of each community. Node-level resource coefficients are aggregated as
\begin{equation}\label{eq:aggregated-node-attribute}
\begin{aligned}
\barrval{\nidx}{\ridx}
=\sum_{\onidx\in\nodes_{\nidx}}\rval{\onidx}{\ridx}
+\sum_{\substack{\onidx,\oniidx\in\nodes_{\nidx}\\\onidx<\oniidx}}
\rval{\onidx\oniidx}{\ridx}
\quad\forall \nidx\in\bar{\nodes},\ \forall \ridx\in\setres,
\end{aligned}
\end{equation}
where the second term accounts for pairwise resource contributions internal to each contracted node. Pairwise resource coefficients between distinct contracted nodes are aggregated as
\begin{equation}\label{eq:aggregated-pair-attribute}
\begin{aligned}
\barrval{\nidx\niidx}{\ridx}
=\sum_{\onidx\in\nodes_{\nidx}}
\sum_{\oniidx\in\nodes_{\niidx}}
\rval{\min\{\onidx,\oniidx\}\max\{\onidx,\oniidx\}}{\ridx}
\quad\forall \nidx,\niidx\in\bar{\nodes},\ \nidx<\niidx,\
\forall \ridx\in\setres.
\end{aligned}
\end{equation}
The resource bounds $\maxres{\ridx}$ remain unchanged, and the aggregated coefficients are used in Eq.~\eqref{eq:extended-resource}.

These transformations preserve feasibility and the optimization objective up to the constant internal modularity contribution. Must-link constraints are satisfied by graph contraction, so the explicit constraints in Eq.~\eqref{eq:extended-must-link} are no longer required in the formulation on the contracted graph. The resulting contracted graph and associated model parameters define the working optimization problem used in the subsequent modules. 

For notational simplicity, we henceforth treat $\bar{\graph}$ as the working graph\footnote{If must-link constraints are absent, no contraction occurs, and the working graph is the original graph.} and suppress bars from all contracted-graph quantities. Thus, $\bar{\graph}$, $\bar{\nodes}$, $\bar{\communities}$, $\barclpairset$, $\bar{\numnodes}$, $\bar{\numcom}_{\max}$, $\barqval{}$, and $\bar{b}$ are written as $\graph$, $\nodes$, $\communities$, $\clpairset$, $\numnodes$, $\numcom_{\max}$, $\qval{}$, and $b$, respectively, and nodes of the working graph are indexed by $\nidx,\niidx$. This convention changes only the notation and does not alter the underlying optimization problem.

\subsubsection{Initial Solution Generation}

Once the working graph has been constructed, a feasible initial partition should be supplied as a starting point for the subsequent optimization process. Any procedure capable of producing a partition that satisfies all constraints can be adopted at this stage.

For this purpose, we propose \acrshort{clouvain} as an efficient initialization procedure that exploits the graph structure while explicitly accounting for the constraints. The algorithm mainly follows the conventional Louvain method~\cite{blondel2008fast}, thereby retaining its computational efficiency, while incorporating minimal repair logic to ensure that the resulting partition is feasible. Details of the algorithm are provided in Appendix~\ref{sec:constrained-louvain}.

\subsection{Optimization Procedures}

We propose three core algorithms: \acrshort{cggp}, \acrshort{bapgpl}, and \acrshort{bapgpq}. These algorithms share a common structure that decomposes the partitioning problem into two coupled tasks: identifying promising communities (\textit{pricing subproblem}) and selecting a compatible set of communities to form a partition (\textit{master problem}). Each feasible community is treated as a candidate building block, or a \emph{column}. Because the number of feasible columns grows exponentially with the number of nodes, this decomposition avoids explicitly enumerating them.

The set-partitioning formulation introduced later in this section serves as the master problem, with one decision variable for each feasible community indicating whether that community is selected to form a partition. Since constructing the full master problem is generally impractical, column generation considers only a restricted subset of columns at a time. Specifically, it solves the \acrshort{lp} relaxation over the currently available columns, referred to as the \textit{relaxed restricted master problem}, and uses the resulting dual information to solve the pricing subproblem. Improving columns identified by pricing are added to the restricted master problem, and this process is repeated.

This column-generation mechanism is common to all three algorithms. They differ in how pricing is performed and whether column generation is embedded within a branch-and-bound framework. \acrshort{cggp} relies on graph-partitioning-based heuristic pricing to generate promising columns efficiently, whereas \acrshort{bapgpl} and \acrshort{bapgpq} additionally employ exact pricing and embed column generation within branch-and-bound, yielding branch-and-price algorithms. The remainder of this section develops the common set-partitioning and column-generation framework before detailing the graph-partitioning-based heuristic pricing, exact pricing formulations, and branch-and-price algorithms.

\subsubsection{Set-Partitioning Reformulation}
\label{subsec:set-partitioning-reformulation}

Let $\columns$ denote the index set of all feasible communities, or columns, on the working graph. For simplicity, we refer to $\columns$ as the set of feasible columns hereafter. For each $\colidx\in\columns$, the corresponding community is represented by a binary column vector with elements $\decw{\nidx\colidx}\in\left\{0,1\right\}$, where $\decw{\nidx\colidx}=1$ if node $\nidx\in\nodes$ belongs to the community indexed by $\colidx$, and $\decw{\nidx\colidx}=0$ otherwise. A feasible column is nonempty and satisfies all community-level constraints of the problem, including pairwise and resource constraints. The modularity contribution of the column indexed by $\colidx$ is
\begin{equation}\label{eq:bnp-column-contribution}
    \qcol{\colidx}=\sum_{\nidx,\niidx\in\nodes,\ \nidx<\niidx}\qval{\nidx\niidx}\decw{\nidx\colidx}    \decw{\niidx\colidx}.
\end{equation}

Let $\decVz$ denote the vector of binary decision variables, where $\decz{\colidx}=1$ if column $\colidx$ is selected and $\decz{\colidx}=0$ otherwise. The resulting set-partitioning formulation is
\newline\newline\noindent
[$\prob{\mathrm{MP}}$] Set-partitioning formulation
\begin{subequations}
\begin{equation}\label{eq:bnp-mp-obj}
    \max\ \sum_{\colidx\in\columns}\qcol{\colidx}\decz{\colidx}
\end{equation}
subject to
\begin{equation}\label{eq:bnp-mp-cover}
    \sum_{\colidx\in\columns}\decw{\nidx\colidx}\decz{\colidx}=1\quad \forall \nidx\in\nodes,
\end{equation}
\begin{equation}\label{eq:bnp-mp-cmin}
    \sum_{\colidx\in\columns}\decz{\colidx}\ge \numcom_{\min},
\end{equation}
\begin{equation}\label{eq:bnp-mp-cmax}
    \sum_{\colidx\in\columns}\decz{\colidx}\le \numcom_{\max},
\end{equation}
\begin{equation}\label{eq:bnp-mp-domain}
    \decz{\colidx}\in\left\{0,1\right\}\quad \forall \colidx\in\columns.
\end{equation}
\end{subequations}
Here, the objective function in Eq.~\eqref{eq:bnp-mp-obj} maximizes the total modularity contribution of the selected columns. Equation~\eqref{eq:bnp-mp-cover} assigns each node to exactly one selected community, while Eqs.~\eqref{eq:bnp-mp-cmin}--\eqref{eq:bnp-mp-cmax} are for bounds on the number of communities.

\subsubsection{Column Generation at the Root Branch-and-Price Node}
\label{subsec:bnp-node-column-generation}

For simplicity, we first focus on column generation at the root branch-and-price node. The number of feasible columns can grow exponentially with the number of nodes, making it impractical to enumerate them all in advance. Therefore, column generation works with a restricted column set $\columns_\iteridx\subseteq\columns$ and repeatedly determines whether an omitted feasible column can improve the current \acrshort{lp} relaxation. At the root branch-and-price node, the initial column set $\columns_0$ corresponds to the communities forming the feasible initial partition.

Given the current column set $\columns_\iteridx$, column generation solves the \acrshort{lp} relaxation of the set-partitioning formulation $\prob{\mathrm{MP}}$ using only the columns in $\columns_\iteridx$. The resulting problem, referred to as the relaxed restricted master problem and denoted by $\prob{\mathrm{RMP}}(\columns_\iteridx)$, is as follows:
\newline\newline\noindent
[$\prob{\mathrm{RMP}}(\columns_\iteridx)$] Relaxed restricted master problem
\begin{subequations}
\begin{equation}\label{eq:bnp-rmp-obj}
    \max\ \sum_{\colidx\in\columns_\iteridx}\qcol{\colidx}\decz{\colidx}
\end{equation}
subject to
\begin{equation}\label{eq:bnp-rmp-cover}
    \sum_{\colidx\in\columns_\iteridx}\decw{\nidx\colidx}\decz{\colidx}=1\quad \forall \nidx\in\nodes,
\end{equation}
\begin{equation}\label{eq:bnp-rmp-cmin}
    \sum_{\colidx\in\columns_\iteridx}\decz{\colidx}\ge \numcom_{\min},
\end{equation}
\begin{equation}\label{eq:bnp-rmp-cmax}
    \sum_{\colidx\in\columns_\iteridx}\decz{\colidx}\le \numcom_{\max},
\end{equation}
\begin{equation}\label{eq:bnp-rmp-domain}
    \decz{\colidx}\in\left[0,1\right]\quad \forall \colidx\in\columns_\iteridx.
\end{equation}
\end{subequations}
Here, $\dual{\nidx}$ denotes the dual variable associated with Eq.~\eqref{eq:bnp-rmp-cover}, while $\ddual{1}$ and $\ddual{2}$ denote the dual variables associated with Eqs.~\eqref{eq:bnp-rmp-cmin} and \eqref{eq:bnp-rmp-cmax}, respectively. Solving $\prob{\mathrm{RMP}}(\columns_\iteridx)$ yields the optimal dual variables $\dual{\nidx}^*$ for $\nidx\in\nodes$, $\ddual{1}^*$, and $\ddual{2}^*$. Under the maximization convention used here, $\ddual{1}^*\le0$ and $\ddual{2}^*\ge0$. Using these optimal dual variables, the pricing subproblem searches for a feasible column with the maximum reduced cost\footnote{We use the standard term \emph{reduced cost}, although in this maximization setting it can be interpreted more intuitively as the marginal contribution of a column.}:
\newline\newline\noindent
[$\prob{\mathrm{PSP}}$] Pricing subproblem
\begin{subequations}
\begin{equation}\label{eq:bnp-psp-obj}
    \max\ \sum_{\nidx,\niidx\in\nodes,\ \nidx<\niidx}\qval{\nidx\niidx}\dece{\nidx}\dece{\niidx}-\sum_{\nidx\in\nodes}\dual{\nidx}^*\dece{\nidx}-\ddual{1}^*-\ddual{2}^*
\end{equation}
subject to
\begin{equation}\label{eq:bnp-psp-nonempty}
    \sum_{\nidx\in\nodes}\dece{\nidx}\ge1,
\end{equation}
\begin{equation}\label{eq:bnp-psp-cannotlink}
    \dece{\nidx}+\dece{\niidx}\le1\quad\forall\left\{\nidx,\niidx\right\}\in\clpairset,
\end{equation}
\begin{equation}\label{eq:bnp-psp-bin-packing}
    \sum_{\nidx\in\nodes}\rval{\nidx}{\ridx}\dece{\nidx}+\sum_{\nidx,\niidx\in\nodes,\ \nidx<\niidx}\rval{\nidx\niidx}{\ridx}\dece{\nidx}\dece{\niidx}\le\maxres{\ridx}\quad\forall\ridx\in\setres,
\end{equation}
\begin{equation}
    \dece{\nidx}\in\left\{0,1\right\}
    \quad
    \forall\nidx\in\nodes.
    \label{eq:bnp-psp-domain}
\end{equation}
\end{subequations}
Here, $\dece{\nidx}=1$ indicates that node $\nidx$ is included in the candidate column, and $\dece{\nidx}=0$ otherwise. The objective function in Eq.~\eqref{eq:bnp-psp-obj} gives the reduced cost of the candidate column. The first term is its modularity contribution, the second term accounts for the dual values associated with node assignment, and the last two terms account for the lower and upper bounds on the number of communities.

If the optimal pricing solution $\decVe^*$ has a positive objective value, the resulting column is assigned a new index $\hat{\colidx}$, with
\begin{equation}
    \decw{\nidx\hat{\colidx}}=\dece{\nidx}^*\quad\forall\nidx\in\nodes.
\end{equation}
The column set is then updated as
\begin{equation}
    \columns_{\iteridx+1}=\columns_\iteridx\cup\left\{\hat{\colidx}\right\},
\end{equation}
and $\prob{\mathrm{RMP}}(\columns_{\iteridx+1})$ is solved again. If the optimal objective value of $\prob{\mathrm{PSP}}$ is nonpositive, no omitted feasible column has a positive reduced cost. Therefore, the current solution of $\prob{\mathrm{RMP}}(\columns_\iteridx)$ is optimal for the \acrshort{lp} relaxation of the full master problem.

Until this termination condition is satisfied, column generation repeatedly solves $\prob{\mathrm{RMP}}$ and $\prob{\mathrm{PSP}}$. Among these two problems, $\prob{\mathrm{PSP}}$ can become the computational bottleneck, since $\prob{\mathrm{RMP}}$ is an \acrshort{lp}, for which polynomial-time solution algorithms are available, whereas $\prob{\mathrm{PSP}}$ is a binary quadratic optimization problem and is generally more difficult to solve exactly.\footnote{This difficulty has also motivated previous studies to employ heuristic methods for the pricing problem; for example, Aloise et al.~\cite{aloise2010column} used a heuristic pricing procedure before resorting to exact pricing.} Motivated by this observation, we use graph-partitioning-based heuristic pricing to identify improving columns efficiently before invoking exact pricing.

\subsubsection{Graph-Partitioning-Based Heuristic Pricing} 
\label{subsec:bnp-heuristic-pricing}

Graph partitioning seeks to divide the nodes of a graph into a prescribed number of parts while minimizing the total weight of edges cut between parts and maintaining balance in the node weights assigned to each part. Although its goal differs from that of modularity maximization, graph partitioning shares the general principle of keeping strongly related nodes together, and related partitioning approaches have been employed for modularity-based community detection~\cite{newman2006modularity}. A standard balanced graph-partitioning problem can be formulated as follows:
\newline\newline\noindent
[$\prob{\mathrm{GP}}$] Graph partitioning problem
\begin{subequations}
\begin{equation}\label{eq:gp-obj}
    \min\ \sum_{\left\{\nidx,\niidx\right\}\in\edges}\wedg{\nidx\niidx}\decy{\nidx\niidx}
\end{equation}
subject to
\begin{equation}\label{eq:gp-assignment}
    \sum_{\cidx\in\communities}\decx{\nidx\cidx}=1\quad \forall \nidx\in\nodes,
\end{equation}
\begin{equation}\label{eq:gp-cut-1}
    \decy{\nidx\niidx}\ge\decx{\nidx\cidx}-\decx{\niidx\cidx}\quad\forall \left\{\nidx,\niidx\right\}\in\edges,\ \forall \cidx\in\communities,
\end{equation}
\begin{equation}\label{eq:gp-cut-2}
    \decy{\nidx\niidx}\ge\decx{\niidx\cidx}-\decx{\nidx\cidx}\quad\forall \left\{\nidx,\niidx\right\}\in\edges,\ \forall \cidx\in\communities,
\end{equation}
\begin{equation}\label{eq:gp-nonempty}
    \sum_{\nidx\in\nodes}\decx{\nidx\cidx}\ge1\quad \forall \cidx\in\communities,
\end{equation}
\begin{equation}\label{eq:gp-balance}
    \sum_{\nidx\in\nodes}\wnod{\nidx}\decx{\nidx\cidx}\le\frac{1+\relimb}{\gpc}\sum_{\nidx\in\nodes}\wnod{\nidx}\quad\forall \cidx\in\communities,
\end{equation}
\begin{equation}\label{eq:gp-domain-x}
    0\le\decy{\nidx\niidx}\le1\quad\forall\left\{\nidx,\niidx\right\}\in\edges,
\end{equation}
\begin{equation}\label{eq:gp-domain-w}
    \decx{\nidx\cidx}\in\left\{0,1\right\}\quad\forall\nidx\in\nodes,\ \forall\cidx\in\communities.
\end{equation}
\end{subequations}
Here, $\graph=(\nodes,\edges)$ is a given undirected graph, $\gpc$ is the prescribed number of parts,\footnote{The parts in $\prob{\mathrm{GP}}$ are auxiliary groups generated for heuristic pricing and should not be confused with the communities selected by the master problem.} $\wedg{\nidx\niidx}\ge0$ is the edge weight, $\wnod{\nidx}\ge0$ is the node weight, and $\relimb\ge0$ is the allowable relative imbalance. The binary variable $\decx{\nidx\cidx}$ indicates whether node $\nidx$ is assigned to part $\cidx$, while $\decy{\nidx\niidx}$ indicates whether the endpoints of edge $\{\nidx,\niidx\}$ are assigned to different parts. The objective function in Eq.~\eqref{eq:gp-obj} minimizes the total weight of cut edges. Equation~\eqref{eq:gp-assignment} assigns each node to exactly one part, Eqs.~\eqref{eq:gp-cut-1}--\eqref{eq:gp-cut-2} identify cut edges, and Eq.~\eqref{eq:gp-nonempty} ensures that each prescribed part contains at least one node. Equation~\eqref{eq:gp-balance} limits the total node weight in each part to $(1+\relimb)$ times the average weight per part, thereby promoting balanced parts.\footnote{In this study, we set $\relimb=0.001$.}

In our framework, $\prob{\mathrm{GP}}$ serves as a surrogate for generating candidate columns for $\prob{\mathrm{PSP}}$. However, balanced graph partitioning is NP-hard in general~\cite{karypis1998multilevel}. We therefore employ efficient graph-partitioning heuristics to obtain high-quality partitions, rather than solving $\prob{\mathrm{GP}}$ to optimality.

Graph partitioning is particularly attractive as the primary heuristic pricing engine for two reasons. First, efficient multilevel algorithms such as METIS can solve large weighted graph-partitioning problems at relatively low computational cost~\cite{karypis1998fast}. Second, a single $\gpc$-way partitioning run produces $\gpc$ node groups, each of which can serve as a candidate column. Thus, one graph-partitioning solve can generate multiple structurally coherent candidate columns simultaneously, providing $\prob{\mathrm{RMP}}$ with several alternatives at relatively low computational cost. This contrasts with an exact solve of $\prob{\mathrm{PSP}}$, which typically returns a single optimal pricing column per solve. To exploit these advantages, we adapt $\prob{\mathrm{GP}}$ to the pricing problem using both the modularity coefficients and the current dual solution of $\prob{\mathrm{RMP}}$. The resulting heuristic consists of three steps: constructing modularity-based edge weights, constructing dual-guided node weights, and generating and validating candidate columns.

\textbf{Step 1: Modularity-based edge weights.}
At each branch-and-price node, we construct a sparse auxiliary graph $\graph_{\mathrm{GP}}=(\nodes,\edges_{\mathrm{GP}})$ with the same node set as the working graph and an edge set containing only node pairs with positive pairwise modularity contributions:
\begin{equation}\label{eq:gp-edge-set}
    \edges_{\mathrm{GP}}=\left\{\{\nidx,\niidx\}:\nidx,\niidx\in\nodes,\ \qval{\nidx\niidx}>0\right\}.
\end{equation}
For each $\{\nidx,\niidx\}\in\edges_{\mathrm{GP}}$, the edge weight is defined as
\begin{equation}
    \wedg{\nidx\niidx}=\qval{\nidx\niidx}.
    \label{eq:gp-edge-weight}
\end{equation}
Thus, node pairs with non-positive modularity contributions are omitted from $\edges_{\mathrm{GP}}$. Minimizing the total cut weight in $\prob{\mathrm{GP}}$ then encourages nodes with large positive pairwise modularity contributions to remain in the same candidate column.

\textbf{Step 2: Dual-guided node weights.}
The edge weights of $\graph_{\mathrm{GP}}$ remain fixed, whereas the optimal dual variables of $\prob{\mathrm{RMP}}(\columns_\iteridx)$ change across column-generation iterations. In the pricing objective, a positive $\dual{\nidx}^*$ penalizes the inclusion of node $\nidx$ in a candidate column, while
$\sum_{\niidx:\{\nidx,\niidx\}\in\edges_{\mathrm{GP}}}\wedg{\nidx\niidx}$
measures the total positive modularity support incident to that node. We combine these two quantities to define the node weight
\begin{equation}\label{eq:gp-node-weight}
    \wnod{\nidx}=\frac{\max\left\{\dual{\nidx}^*,0\right\}}{\scon{\mathrm{GP}}+\sum_{\niidx:\{\nidx,\niidx\}\in\edges_{\mathrm{GP}}}\wedg{\nidx\niidx}},
\end{equation}
where $\scon{\mathrm{GP}}>0$ is a small constant that prevents division by zero.\footnote{In this study, we set $\scon{\mathrm{GP}}=10^{-9}$.} The resulting node weights are normalized to lie between 1 and 10 before solving $\prob{\mathrm{GP}}$. Through Constraint~\eqref{eq:gp-balance}, these weights discourage a part from concentrating nodes with large dual penalties relative to their positive modularity support. The dual variables $\ddual{1}^*$ and $\ddual{2}^*$ are not included in the node weights because their contributions to the pricing objective are constant across all candidate columns within a given column-generation iteration.

\textbf{Step 3: Candidate generation and validation.}
For each admissible $\gpc$,---all integer values from $\numcom_{\min}$ to $\numcom_{\max}$ are considered in this research---a graph-partitioning solution produces $\gpc$ node groups, each of which is treated as a candidate column. The feasibility and reduced cost of the generated candidates are then evaluated because solutions of $\prob{\mathrm{GP}}$ do not necessarily satisfy the constraints of $\prob{\mathrm{PSP}}$ or yield a positive pricing objective. Only candidates that are feasible for $\prob{\mathrm{PSP}}$ and have a positive objective value in Eq.~\eqref{eq:bnp-psp-obj} are included in $\columns_{\mathrm{cand}}$, the set of candidate columns to be added to $\columns_{\iteridx}$ for solving $\prob{\mathrm{RMP}}$. If no such column is identified, graph partitioning is repeated up to a prescribed number of runs, $\gprun$\footnote{In this study, we set $\gprun=5$.}, to exploit variations in the partitioning solution. Algorithm~\ref{alg:heuristic-pricing} summarizes the graph-partitioning-based heuristic pricing procedure.

\begin{ruledalgorithm}{Graph-partitioning-based heuristic pricing}
\label{alg:heuristic-pricing}
\begin{algorithmic}[1]
\Require Modularity coefficients $\qval{\nidx\niidx}$, optimal dual variables $\dual{\nidx}^*$, admissible values of $\gpc$, maximum number of graph-partitioning runs $\gprun$.
\State Construct $\graph_{\mathrm{GP}}=(\nodes,\edges_{\mathrm{GP}})$ and edge weights $\wedg{\nidx\niidx}$ according to Eqs.~\eqref{eq:gp-edge-set}--\eqref{eq:gp-edge-weight}.
\State Compute node weights $\wnod{\nidx}$ from the current dual solution according to Eq.~\eqref{eq:gp-node-weight}.
\State Initialize the set of improving candidate columns $\columns_{\mathrm{cand}}\leftarrow\emptyset$ and $\gpridx\leftarrow1$.
\Repeat
    \For{each admissible $\gpc$}
        \State Run METIS on $\graph_{\mathrm{GP}}$ with edge weights $\{\wedg{\nidx\niidx}\}$, node weights $\{\wnod{\nidx}\}$, and target $\gpc$.
        \For{each resulting candidate column $\decVe$}
            \If{$\decVe$ is feasible for $\prob{\mathrm{PSP}}$ and has positive reduced cost}
                \State Add $\decVe$ to $\columns_{\mathrm{cand}}$.
            \EndIf
        \EndFor
    \EndFor
    \State $\gpridx\leftarrow\gpridx+1$.
\Until{$\columns_{\mathrm{cand}}\neq\emptyset$ or $\gpridx>\gprun$}
\State \Return $\columns_{\mathrm{cand}}$.
\end{algorithmic}
\end{ruledalgorithm}

\subsubsection{Enhanced Exact Pricing}
\label{subsec:bnp-exact-pricing}

Graph-partitioning-based heuristic pricing is used to identify improving columns quickly, but failure to find one does not imply that no such column exists. Therefore, when heuristic pricing fails, we solve $\prob{\mathrm{PSP}}$ exactly to either identify an improving column or certify that none remains.

Directly solving $\prob{\mathrm{PSP}}$ as a binary quadratic problem becomes difficult as the number of nodes increases. We therefore use two formulations that are equivalent to $\prob{\mathrm{PSP}}$ but more suitable for mathematical programming solvers. The first is a general \acrshort{milp} formulation obtained by linearizing all pairwise products. The second is a sparse \acrfull{miqp} formulation that exploits the modularity structure by retaining the dense null-model term in an aggregate quadratic form. Both formulations can be solved to global optimality using commercial mathematical programming solvers. Each exact-pricing solve returns a single optimal column $\decVe^*$, which is added to the candidate column set $\columns_{\mathrm{cand}}$ if its objective value is positive.

\textbf{Linearized Exact Pricing.}
Similar to $\prob{\mathrm{PSP}}$, the linearized formulation introduces a binary variable $\dece{\nidx}$ for each node and a continuous variable $\decy{\nidx\niidx}$ for each unordered node pair $\nidx<\niidx$. The variable $\decy{\nidx\niidx}$ represents the product $\dece{\nidx}\dece{\niidx}$. The resulting \acrshort{milp} formulation is
[$\prob{\text{PSP-MILP}}$] Linearized exact pricing
\begin{subequations}
\begin{equation}\label{eq:bnPSP-MILP-pricing-obj}
\max\ \sum_{\nidx,\niidx\in\nodes,\ \nidx<\niidx}\qval{\nidx\niidx}\decy{\nidx\niidx}-\sum_{\nidx\in\nodes}\dual{\nidx}^*\dece{\nidx}-\ddual{1}^*-\ddual{2}^*
\end{equation}
subject to Eqs. \eqref{eq:bnp-psp-nonempty}, \eqref{eq:bnp-psp-cannotlink}, and \eqref{eq:bnp-psp-domain},
\begin{equation}\label{eq:bnPSP-MILP-bin-packing}
\sum_{\nidx\in\nodes}\rval{\nidx}{\ridx}\dece{\nidx}+\sum_{\nidx,\niidx\in\nodes,\ \nidx<\niidx}\rval{\nidx\niidx}{\ridx}\decy{\nidx\niidx}\le\maxres{\ridx}\quad\forall\ridx\in\setres,
\end{equation}
\begin{equation}\label{eq:bnPSP-MILP-linearization-1}
\decy{\nidx\niidx}\le\dece{\nidx}\quad\forall\nidx,\niidx\in\nodes,\ \nidx<\niidx,
\end{equation}
\begin{equation}\label{eq:bnPSP-MILP-linearization-2}
\decy{\nidx\niidx}\le\dece{\niidx}\quad\forall\nidx,\niidx\in\nodes,\ \nidx<\niidx,
\end{equation}
\begin{equation}\label{eq:bnPSP-MILP-linearization-3}
\decy{\nidx\niidx}\ge\dece{\nidx}+\dece{\niidx}-1\quad\forall\nidx,\niidx\in\nodes,\ \nidx<\niidx,
\end{equation}
\begin{equation}\label{eq:bnPSP-MILP-y-domain}
0\le\decy{\nidx\niidx}\le1\quad
\forall\nidx,\niidx\in\nodes,\ \nidx<\niidx.
\end{equation}
\end{subequations}
The objective function in Eq.~\eqref{eq:bnPSP-MILP-pricing-obj} gives the reduced cost of the candidate column. Equations~\eqref{eq:bnp-psp-nonempty}, \eqref{eq:bnp-psp-cannotlink}, and \eqref{eq:bnp-psp-domain} carry over the nonempty-column, cannot-link, and binary constraints from $\prob{\mathrm{PSP}}$. Equation~\eqref{eq:bnPSP-MILP-bin-packing} is the linearized form of Eq.~\eqref{eq:bnp-psp-bin-packing}, obtained by replacing each product $\dece{\nidx}\dece{\niidx}$ with $\decy{\nidx\niidx}$. Equations~\eqref{eq:bnPSP-MILP-linearization-1}--\eqref{eq:bnPSP-MILP-linearization-3} represent the standard exact linearization of $\decy{\nidx\niidx}=\dece{\nidx}\dece{\niidx}$ for binary $\dece{\nidx}$ and $\dece{\niidx}$ \cite{fortet1960applications,conforti2014integer}. Hence, $\prob{\text{PSP-MILP}}$ is equivalent to $\prob{\mathrm{PSP}}$.

\textbf{Sparse Bilinear Exact Pricing.} The second exact formulation exploits the structure of the modularity coefficient in Eq.~\eqref{eq:qij-definition}. In $\prob{\text{PSP-MILP}}$, a product variable $\decy{\nidx\niidx}$ is introduced for every unordered node pair, even though the adjacency coefficient is nonzero only for pairs connected in the working graph. For sparse networks, this results in many unnecessary product variables. We therefore linearize only the pairs required by the adjacency and resource terms, while retaining the dense degree-product term in an aggregate quadratic form.

To identify the pairs requiring explicit product variables, define the adjacency-supported pair set as
\begin{equation}\label{eq:bnp-adjacency-pair-set}
\edges_{\mathrm{A}}=\left\{\{\nidx,\niidx\}:\nidx,\niidx\in\nodes,\ \adjm{\nidx\niidx}+\adjm{\niidx\nidx}>0\right\}.
\end{equation}
Similarly, define the resource-supported pair set as
\begin{equation}\label{eq:bnp-resource-pair-set}
\edges_{\mathrm{B}}=\left\{\{\nidx,\niidx\}:\nidx,\niidx\in\nodes,\ \exists\,\ridx\in\setres\text{ such that }\rval{\nidx\niidx}{\ridx}\neq0\right\}.
\end{equation}
Thus, explicit product variables $\decy{\nidx\niidx}$ are required only for pairs $\left\{\nidx,\niidx\right\}$ in
\begin{equation}\label{eq:bnp-sparse-pair-set}
\edges_{\mathrm{Q}}=\edges_{\mathrm{A}}\cup\edges_{\mathrm{B}}.
\end{equation}

Using this structure, the sparse bilinear exact-pricing formulation is
\par\medskip\noindent
[$\prob{\text{PSP-SMIQP}}$] Sparse bilinear exact pricing
\begin{subequations}
\begin{equation}\label{eq:bnp-sparse-pricing-obj}
\max\ \sum_{\{\nidx,\niidx\}\in\edges_{\mathrm{A}}}\left(\adjm{\nidx\niidx}+\adjm{\niidx\nidx}\right)\decy{\nidx\niidx}+\sum_{\nidx\in\nodes}\left(\frac{\dego{\nidx}\degi{\nidx}}{\numedges}-\dual{\nidx}^*\right)\dece{\nidx}-\frac{1}{\numedges}\left(\sum_{\nidx\in\nodes}\dego{\nidx}\dece{\nidx}\right)\left(\sum_{\nidx\in\nodes}\degi{\nidx}\dece{\nidx}\right)-\ddual{1}^*-\ddual{2}^*
\end{equation}
subject to Eqs.~\eqref{eq:bnp-psp-nonempty},
\eqref{eq:bnp-psp-cannotlink}, and
\eqref{eq:bnp-psp-domain}, and
Eqs.~\eqref{eq:bnPSP-MILP-linearization-1}--\eqref{eq:bnPSP-MILP-y-domain}
applied only to pairs
$\{\nidx,\niidx\}\in\edges_{\mathrm{Q}}$,
\begin{equation}\label{eq:bnp-sparse-bin-packing}
\sum_{\nidx\in\nodes}\rval{\nidx}{\ridx}\dece{\nidx}+\sum_{\{\nidx,\niidx\}\in\edges_{\mathrm{B}}}\rval{\nidx\niidx}{\ridx}\decy{\nidx\niidx}\le\maxres{\ridx}\quad\forall\ridx\in\setres.
\end{equation}
\end{subequations}
Here, the objective function in Eq.~\eqref{eq:bnp-sparse-pricing-obj} separates the sparse adjacency contribution from the dense degree-product contribution. Consequently, explicit product variables are required only for pairs in $\edges_{\mathrm{Q}}$, rather than for all unordered node pairs as in $\prob{\text{PSP-MILP}}$. The nonempty-column, cannot-link, and binary constraints are inherited from $\prob{\mathrm{PSP}}$, while the same exact product linearization used in $\prob{\text{PSP-MILP}}$ is applied only to pairs in $\edges_{\mathrm{Q}}$. Equation~\eqref{eq:bnp-sparse-bin-packing} represents the resource constraints using only the pair variables required by the corresponding resource terms.

The main advantage of $\prob{\text{PSP-SMIQP}}$ is that it substantially reduces the number of explicit product variables, which is particularly beneficial for sparse networks. This reduction comes at the cost of retaining a quadratic term in the objective. For an undirected network, however, the in-degree and out-degree of each node coincide, i.e., $\degi{\nidx}=\dego{\nidx}=\deg{\nidx}$, and the objective reduces to
\begin{equation}\label{eq:bnp-sparse-undirected-obj}
\max\ 2\sum_{\{\nidx,\niidx\}\in\edges_{\mathrm{A}}}\adjm{\nidx\niidx}\decy{\nidx\niidx}+\sum_{\nidx\in\nodes}\left(\frac{\deg{\nidx}^2}{\numedges}-\dual{\nidx}^*\right)\dece{\nidx}-\frac{1}{\numedges}\left(\sum_{\nidx\in\nodes}\deg{\nidx}\dece{\nidx}\right)^2-\ddual{1}^*-\ddual{2}^*,
\end{equation}
where the remaining quadratic term is the negative square of a linear function and is therefore concave in maximization form, allowing commercial \acrshort{miqp} solvers to exploit this favorable structure. For directed networks, the corresponding aggregate product of the in-degree and out-degree terms is generally nonconvex, so this additional computational advantage is lost, although the reduction in product variables is retained. Consequently, $\prob{\text{PSP-SMIQP}}$ is expected to be particularly effective for sparse undirected networks.

\subsubsection{Branch-and-Price Framework}
\label{subsec:bnp-framework}

Column generation solves the \acrshort{lp} relaxation of $\prob{\mathrm{MP}}$, so the resulting solution may remain fractional and therefore may not constitute a feasible solution to the original set partitioning problem. Possible heuristic approaches include solving the integer restricted master problem using only the columns generated during column generation or retaining the best integer solution obtained during the solution process. Such approaches can provide good feasible partitions, but they do not guarantee global optimality. To obtain an exact integer solution, we embed column generation within a branch-and-bound framework~\cite{barnhart1998branch,sato2019enhanced}.\footnote{In contrast, the column-generation algorithms of Aloise et al.~\cite{aloise2010column} do not embed the master problem within a branch-and-bound tree and therefore do not provide an integer optimality guarantee through branch-and-price.}

\begin{figure}[hbhtbp]
\centering
\includegraphics[page=6,width=0.99\linewidth]{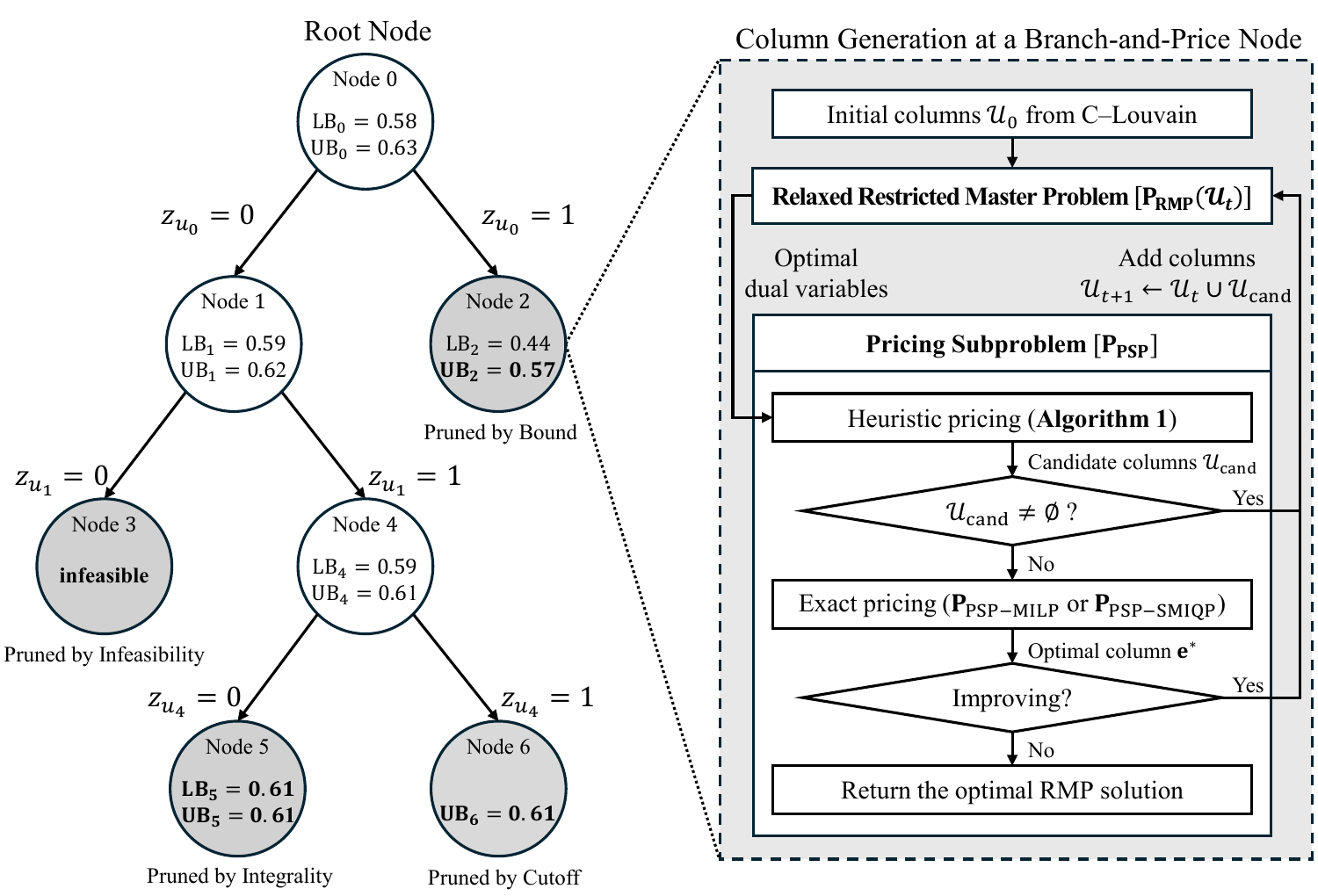}
\caption{Overview of the proposed branch-and-price framework and an illustrative example.}
\label{fig:bnp-framework-overview}
\end{figure}

Figure~\ref{fig:bnp-framework-overview} illustrates the proposed branch-and-price procedure. The left panel shows the branch-and-price tree, while the right panel summarizes the column-generation routine used to solve $\prob{\mathrm{RMP}}$ at each branch-and-price node. In the branch-and-price tree, $\mathrm{LB}_{\bnpidx}$ and $\mathrm{UB}_{\bnpidx}$ shown for a processed branch-and-price node $\bnpidx$ denote the lower and upper bounds available after column generation is completed at that node. Starting from the root branch-and-price node $\bnpidx=0$, column generation alternates between solving $\prob{\mathrm{RMP}}$ and generating improving columns through the heuristic and exact pricing procedures introduced previously. Once exact pricing certifies that no improving column remains, the resulting optimal solution for $\prob{\mathrm{RMP}}$ provides an upper bound $\mathrm{UB}_{\bnpidx}$ for that branch-and-price node.

If the resulting solution of $\prob{\mathrm{RMP}}$ is fractional, we apply column-based branching. Suppose that the solution at branch-and-price node $\bnpidx$ contains a fractional variable $\decz{\colidx_{\bnpidx}}$ for a column $\colidx_{\bnpidx}$. Two child branch-and-price nodes are created by imposing $\decz{\colidx_{\bnpidx}}=0$ and $\decz{\colidx_{\bnpidx}}=1$, respectively. In the first branch, column $\colidx_{\bnpidx}$ is excluded, and the pricing procedure is prevented from generating the same column again. In the second branch, column $\colidx_{\bnpidx}$ is fixed as part of the partition. Its modularity contribution is fixed in the objective, and the nodes covered by the column are removed from the remaining problem. Let $\bnpnodes{}{\bnpidx}$ denote the node set of the graph at
the parent branch-and-price node $\bnpidx$. The child
branch-and-price node with $\decz{\colidx_{\bnpidx}}=1$
uses the residual graph with node set
\begin{equation}\label{eq:bnp-include-residual-set}
\left\{
\nidx\in\bnpnodes{}{\bnpidx}
:
\decw{\nidx\colidx_{\bnpidx}}=0
\right\}.
\end{equation}
The minimum and maximum community counts for the residual problem are each reduced by one. In the illustrative example in Fig.~\ref{fig:bnp-framework-overview}, the fractional solution at branch-and-price node $0$ generates the two child branch-and-price nodes $1$ and $2$.

At a non-root branch-and-price node, the initial column set is inherited from the final column set of its parent branch-and-price node, excluding columns incompatible with the branching decisions. For the branch in which a column is fixed to one, \acrshort{clouvain} is additionally applied to the residual graph obtained by removing the nodes covered by that column. The communities in the resulting partition are added to the initial column set.

Let $\abnpnodes$ denote the set of active branch-and-price nodes, comprising generated branch-and-price nodes that have not been pruned or processed. A branch-and-price node is considered processed once its column-generation procedure terminates. Each child branch-and-price node inherits its parent's upper bound until column generation establishes its own bound. The global upper bound is
\begin{equation}\label{eq:bnp-global-upper-bound}
\mathrm{UB}=
\begin{cases}
\max_{\bnpidx\in\abnpnodes}\mathrm{UB}_{\bnpidx}, & \abnpnodes\neq\emptyset,\\
\mathrm{LB}, & \abnpnodes=\emptyset.
\end{cases}
\end{equation}

A processed branch-and-price node $\bnpidx$ is pruned by bound if $\mathrm{UB}_{\bnpidx}\le\mathrm{LB}$. If its optimal \acrshort{lp} solution is integral, combining it with the columns fixed by its parents branch-and-price node yields a feasible partition. We update $\mathrm{LB}$ if this partition improves the incumbent and prune the branch-and-price node by integrality. An unprocessed branch-and-price node is pruned by cutoff when its inherited upper bound does not exceed the updated $\mathrm{LB}$, avoiding unnecessary column generation at that branch-and-price node. Thus, pruning by bound and cutoff use the same bound comparison but occur after and before processing, respectively. A branch-and-price node is also pruned if infeasibility is established.

The procedure certifies optimality when $\mathrm{UB}=\mathrm{LB}$, including when no active branch-and-price nodes remain. If the time limit is reached earlier, the difference $\mathrm{UB}-\mathrm{LB}$ provides an absolute optimality gap for the incumbent solution. Algorithm~\ref{alg:bnp-final} summarizes the branch-and-price procedure.

We illustrate the branch-and-price procedure using the example in Fig.~\ref{fig:bnp-framework-overview}. Assume that branch-and-price nodes are processed in numerical order. The lower bound at the root branch-and-price node initializes $\mathrm{LB}=0.58$. Branch-and-price node $2$ is pruned by bound because $\mathrm{UB}_2=0.57<\mathrm{LB}$, while branch-and-price node $3$ is pruned by infeasibility. At branch-and-price node $5$, the optimal \acrshort{lp} solution is integral, updating $\mathrm{LB}$ to $0.61$ and allowing pruning by integrality. Branch-and-price node $6$, whose inherited upper bound is $\mathrm{UB}_6=0.61$, is then pruned by cutoff without performing column generation.

In practice, the branching and branch-and-price node-selection rules can affect computational time without changing the optimality guarantee when the search is completed. In our numerical experiments, we branch on the fractional column variable whose value in the optimal \acrshort{lp} solution is closest to one. We select the active branch-and-price node with the largest upper bound for processing.

\begin{ruledalgorithm}{Branch-and-price for constrained modularity maximization}
\label{alg:bnp-final}
\begin{algorithmic}[1]
\Require Root branch-and-price node $\bnpidx=0$, initial partition and its modularity $Q^{\mathrm{init}}$, time limit.
\State Initialize $\abnpnodes\leftarrow\{0\}$.
\State Set $\mathrm{LB}\leftarrow Q^{\mathrm{init}}$ and $\mathrm{UB}\leftarrow+\infty$.
\While{$\abnpnodes\neq\emptyset$, $\mathrm{UB}>\mathrm{LB}$, and the time limit is not reached}
    \State Select and remove a branch-and-price node $\bnpidx\in\abnpnodes$ with the largest upper bound.
    \State Solve column generation at branch-and-price node $\bnpidx$ to optimality.
    \If{$\prob{\mathrm{RMP}}$ at branch-and-price node $\bnpidx$ is infeasible}
        \State Prune branch-and-price node $\bnpidx$. \Comment{Pruned by Infeasibility}
    \Else
        \State Set $\mathrm{UB}_{\bnpidx}$ to the final \acrshort{lp} objective plus the contributions of fixed columns.
        \If{$\mathrm{UB}_{\bnpidx}\le\mathrm{LB}$} 
            \State Prune branch-and-price node $\bnpidx$. \Comment{Pruned by Bound}
        \ElsIf{the optimal \acrshort{lp} solution is integral}
            \State Update $\mathrm{LB}$ and the incumbent partition if improved.
            \State Prune branch-and-price node $\bnpidx$. \Comment{Pruned by Integrality}
        \Else
            \State Select a fractional column $\colidx_{\bnpidx}$.
            \State Create child branch-and-price nodes with $\decz{\colidx_{\bnpidx}}=0$ and $\decz{\colidx_{\bnpidx}}=1$.
            \State Add both child branch-and-price nodes to $\abnpnodes$ with inherited upper bound $\mathrm{UB}_{\bnpidx}$.
        \EndIf
    \EndIf
    \State Update the active node set $\abnpnodes\leftarrow\{\bnpidx\in\abnpnodes:\mathrm{UB}_{\bnpidx}>\mathrm{LB}\}$. \Comment{Pruned by Cutoff}
    \State Update $\mathrm{UB}$ according to Eq.~\eqref{eq:bnp-global-upper-bound}.
\EndWhile
\State \Return $\mathrm{LB}$, $\mathrm{UB}$, and the incumbent partition.
\end{algorithmic}
\end{ruledalgorithm}

In summary, Table~\ref{tab:proposed-solvers} compares the three proposed solvers and highlights the algorithmic components that distinguish them. All three solvers use the graph-partitioning-based heuristic pricing procedure in Algorithm~\ref{alg:heuristic-pricing}. \acrshort{cggp} performs column generation only at the root branch-and-price node and employs neither exact pricing nor branching. After column generation terminates, it solves the integer restricted master problem using only the columns generated during the procedure. Consequently, \acrshort{cggp} remains heuristic.

\begin{table}[hbt!]
\caption{Summary of the proposed solvers.}
\label{tab:proposed-solvers}
\footnotesize
\centering
\setlength{\tabcolsep}{5pt}
\begin{tabular*}{\textwidth}{@{\extracolsep{\fill}}lccccc@{}}
\hline\hline
Solver & Heuristic Pricing & Exact Pricing & Integer RMP & Branching & Optimality guarantee \\
\hline
\acrshort{cggp} & Algorithm~\ref{alg:heuristic-pricing} & No & Yes & No & No \\
\acrshort{bapgpl} & Algorithm~\ref{alg:heuristic-pricing} & $\prob{\text{PSP-MILP}}$ & No & Yes & Yes \\
\acrshort{bapgpq} & Algorithm~\ref{alg:heuristic-pricing} & $\prob{\text{PSP-SMIQP}}$ & No & Yes & Yes \\
\hline\hline
\end{tabular*}
\end{table}

In contrast, \acrshort{bapgpl} and \acrshort{bapgpq} embed column generation within the branch-and-price framework and invoke $\prob{\text{PSP-MILP}}$ and $\prob{\text{PSP-SMIQP}}$, respectively, as exact-pricing fallbacks. These exact-pricing procedures certify that no improving column remains at each branch-and-price node, while branching resolves fractional master solutions. Therefore, \acrshort{bapgpl} and \acrshort{bapgpq} provide a global optimality guarantee when the branch-and-price procedure is allowed to terminate normally. Additional acceleration techniques for heuristic and exact pricing, used in the proposed implementations and numerical experiments, are described in Appendix~\ref{sec:bnp-pricing-acceleration}.

\subsection{Solution Recovery and Assessment}

\subsubsection{Solution Recovery}

After the optimization procedure terminates, the resulting partition is defined on the working graph, where each node may represent either a single original node or a contracted component induced by must-link constraints. We therefore recover the corresponding partition on the original graph by expanding each working-graph node into the set of original nodes represented by that node. All original nodes belonging to the same contracted component are assigned to the community containing the corresponding working-graph node. Because must-link components were contracted before optimization, this expansion preserves all must-link relations by construction.

The resulting communities on the original graph form the final partition. The modularity value is then evaluated on the original graph. Equivalently, when the optimization objective is evaluated on the contracted working graph, the modularity contributions internal to the contracted must-link components, which are constant over all feasible partitions, are added back to obtain the objective value corresponding to the original graph. This recovery step does not alter the optimized community structure; it only maps the solution from the reduced working graph back to the original problem representation.

\subsubsection{Cutting-Plane Upper Bound}
\label{subsec:bnp-triangle-lp-bound}

For large instances, branch-and-price may reach the time limit before closing the optimality gap. If column generation at the root branch-and-price node is not completed, it provides no global upper bound. We therefore propose the cutting-plane upper bound for independent, a posteriori assessment of incumbent solution quality. When both bounds are available, we use the tighter one to evaluate the incumbent's optimality gap. In addition, for a solution obtained by a heuristic, the difference between this bound and its modularity provides an upper limit on further improvement.

We first consider a pairwise upper-bound formulation that relaxes the number of communities and resource constraints while retaining the partition structure and cannot-link constraints. Let $\decy{\nidx\niidx}=1$ if nodes $\nidx$ and $\niidx$ are assigned to the same community and $\decy{\nidx\niidx}=0$ otherwise. The resulting formulation is
\par\medskip\noindent
[$\prob{\text{UB}}$] Pairwise upper-bound formulation
\begin{subequations}
\begin{equation}\label{eq:ub-obj}
    \max\ \sum_{\nidx,\niidx\in\nodes,\ \nidx<\niidx}\qval{\nidx\niidx}\decy{\nidx\niidx}
\end{equation}
subject to
\begin{equation}\label{eq:ub-triangle-ineq1}
    \decy{\nidx\niidx}+\decy{\niidx\niiidx}-\decy{\nidx\niiidx}\le 1\quad\forall \nidx,\niidx,\niiidx\in\nodes,\ \nidx<\niidx<\niiidx,
\end{equation}
\begin{equation}\label{eq:ub-triangle-ineq2}
    \decy{\nidx\niidx}+\decy{\nidx\niiidx}-\decy{\niidx\niiidx}\le 1\quad\forall \nidx,\niidx,\niiidx\in\nodes,\ \nidx<\niidx<\niiidx,
\end{equation}
\begin{equation}\label{eq:ub-triangle-ineq3}
    \decy{\nidx\niiidx}+\decy{\niidx\niiidx}-\decy{\nidx\niidx}\le 1\quad\forall \nidx,\niidx,\niiidx\in\nodes,\ \nidx<\niidx<\niiidx,
\end{equation}
\begin{equation}\label{eq:ub-cannot-link}
    \decy{\nidx\niidx}=0\quad\forall \{\nidx,\niidx\}\in\clpairset,
\end{equation}
\begin{equation}\label{eq:ub-domain}
    \decy{\nidx\niidx}\in\left\{0,1\right\}\quad\forall\nidx,\niidx\in\nodes,\ \nidx<\niidx.
\end{equation}
\end{subequations}
Here, the objective function in Eq.~\eqref{eq:ub-obj} maximizes the modularity contribution represented by the same-community variables. Equations~\eqref{eq:ub-triangle-ineq1}--\eqref{eq:ub-triangle-ineq3}, referred to as the triangle inequalities, enforce transitivity of the pairwise same-community relation, while Eq.~\eqref{eq:ub-cannot-link} forces every cannot-link pair to belong to different communities. Together with the binary domain in Eq.~\eqref{eq:ub-domain}, these constraints define a valid partition satisfying all cannot-link restrictions. Because $\prob{\mathrm{UB}}$ removes the community-count and resource constraints from $\prob{\mathrm{C}}$, every feasible solution of $\prob{\mathrm{C}}$ remains feasible for $\prob{\mathrm{UB}}$. Hence, the optimal value of $\prob{\mathrm{UB}}$ provides a valid upper bound on the original constrained problem.

To obtain this bound more efficiently, we further relax the binary domain in Eq.~\eqref{eq:ub-domain} to
\begin{equation}
    0\le\decy{\nidx\niidx}\le1\quad\forall\nidx,\niidx\in\nodes,\ \nidx<\niidx,
\end{equation}
yielding an \acrshort{lp} relaxation and therefore another valid upper bound. Although the relaxed solution does not necessarily represent a valid partition, its objective value remains an upper bound because the feasible region is further enlarged.

The number of triangle inequalities in Eqs.~\eqref{eq:ub-triangle-ineq1}--\eqref{eq:ub-triangle-ineq3} grows cubically with the number of nodes, so we do not include all of them initially. Instead, we solve the \acrshort{lp} relaxation using a cutting-plane procedure. Starting from the \acrshort{lp} relaxation of $\prob{\mathrm{UB}}$ without triangle inequalities, we solve the current relaxation, identify violated triangle inequalities, add a selected subset of them, and repeat. Each intermediate \acrshort{lp} solution provides a valid upper bound, and the bound becomes tighter as violated triangle inequalities are added. Thus, the procedure can be used as an anytime algorithm, returning the best available upper bound when the prescribed time limit is reached. The final bound obtained by this procedure is the cutting-plane upper bound. Algorithm~\ref{alg:triangle-lp-bound} summarizes the procedure.

\begin{ruledalgorithm}{Cutting-plane upper bound}
\label{alg:triangle-lp-bound}
\begin{algorithmic}[1]
\State Initialize the LP relaxation of $\prob{\mathrm{UB}}$ without triangle inequalities.
\Repeat
    \State Solve the current \acrshort{lp} relaxation to obtain an optimal solution.
    \State Identify triangle inequalities violated by this solution.
    \If{violated inequalities are found}
        \State Add a selected subset of violated triangle inequalities to the LP.
    \EndIf
\Until{no violated triangle inequality is found or the time limit is reached}
\State Return the latest solved LP objective value as an upper bound.
\end{algorithmic}
\end{ruledalgorithm}

\section{Numerical Experiments}
\label{sec:numerical-study}

To demonstrate the solution quality and computational efficiency of the proposed algorithms, we conducted experiments on both unconstrained and constrained instances. We use two reference approaches for comparison. The direct baseline solves the extended formulation $\prob{\mathrm{C}}$ using Gurobi, while the heuristic baseline uses the standard Louvain algorithm~\cite{blondel2008fast} for unconstrained instances and the proposed \acrshort{clouvain} for constrained instances. The proposed optimization methods are \acrshort{cggp}, \acrshort{bapgpl}, and \acrshort{bapgpq}, with the cutting-plane method used to provide an upper bound. These column-generation methods are initialized with Louvain for unconstrained instances and with \acrshort{clouvain} for constrained instances. 

All algorithms were implemented in Python 3.12 and run under Linux Mint 22.3 on a workstation with an AMD Ryzen 9 9950X 16-core processor and 128 GB of RAM. Mathematical-programming models were solved using Gurobi 12.0.3 with up to 32 threads. Each run had a time limit of 3600 seconds. For the direct formulation $\prob{\mathrm{C}}$, model construction was subject to a separate 3600-second time limit, and its duration was excluded from the reported computational time.

\subsection{Benchmark Instances}
\label{subsec:benchmark-instances}

\subsubsection{Unconstrained Benchmark Networks}

We evaluated the algorithms on the 11 benchmark networks summarized in Table~\ref{tab:benchmark-instances}. The networks range from 27 to 1,005 nodes and from 31 to 24,929 edges. Eight are established social, ecological, collaboration, or communication networks: \emph{Karate}~\cite{zachary1977information}, \emph{Kapferer}~\cite{kapferer1972strategy}, \emph{Les Mis\'erables}~\cite{knuth1993stanford}, \emph{Football}~\cite{girvan2002community}, \emph{Food web}~\cite{martinez1991artifacts}, \emph{Jazz}~\cite{gleiser2003community}, \emph{CA--NetScience}~\cite{newman2006finding,rossi2015networkrepository}, and \emph{Email--Eu--core}~\cite{yin2017local,leskovec2007graph}. The remaining three, \emph{Front suspension}, \emph{ICE--HVAC cooling}, and \emph{Thermal energy}, are new networks from automotive manufacturing system architectures. The \emph{Front suspension} graph was introduced in our previous work~\cite{kim2025mathematical}, while the other two cases are described in Appendix~\ref{sec:automotive-graphs}. For notational simplicity, we assign an ID to each network. The benchmark set contains both undirected and directed networks.

\begin{table}[hbt!]
\caption{Summary of the unconstrained benchmark networks and their structural properties.}
\label{tab:benchmark-instances}
\centering
\scriptsize
\setlength{\tabcolsep}{4pt}
\begin{tabular}{@{}llrrc@{}}
\hline\hline
Network ID & Network & $|\nodes|$ & $|\edges|$ & Graph type \\
\hline
N01 & \emph{Front suspension} & 27 & 31 & Undirected \\
N02 & \emph{Karate} & 34 & 78 & Undirected \\
N03 & \emph{Kapferer} & 39 & 316 & Directed \\
N04 & \emph{ICE--HVAC cooling} & 40 & 41 & Undirected \\
N05 & \emph{Thermal energy} & 76 & 162 & Undirected \\
N06 & \emph{Les Mis\'erables} & 77 & 254 & Undirected \\
N07 & \emph{Football} & 115 & 613 & Undirected \\
N08 & \emph{Food web} & 183 & 2494 & Directed \\
N09 & \emph{Jazz} & 198 & 2742 & Undirected \\
N10 & \emph{CA--NetScience} & 379 & 914 & Undirected \\
N11 & \emph{Email--Eu--core} & 1005 & 24929 & Directed \\
\hline\hline
\end{tabular}
\end{table}

\subsubsection{Constrained Benchmark Instances}

Based on the benchmark networks listed in Table~\ref{tab:benchmark-instances}, we derived 30 constrained benchmark instances from three representative undirected networks: N06, N07, and N10. Each constrained instance pairs a network with a specific constraint case. We use \emph{constraint class} for a broad type of restriction and \emph{case} for a specific parameterization within that class. Each network yields 10 constrained instances: three community-count instances, three pairwise-constraint instances, three resource-constraint instances, and one instance combining all three classes. Table~\ref{tab:constrained-instances} summarizes these specifications.

\begin{table*}[hbt!]
\caption{Summary of the constraint specifications for the constrained benchmark instances.}
\label{tab:constrained-instances}
\centering
\scriptsize
\setlength{\tabcolsep}{4pt}
\begin{tabular*}{\textwidth}{@{\extracolsep{\fill}}lllll@{}}
\hline\hline
Constraint class & Case & N06 & N07 & N10 \\
\hline
Community count & Min. community count & $\numcom_{\min}=22$ & $\numcom_{\min}=24$ & $\numcom_{\min}=160$ \\
 & Max. community count & $\numcom_{\max}=5$ & $\numcom_{\max}=9$ & $\numcom_{\max}=18$ \\
 & Fixed community count & $\numcom_{\min}=\numcom_{\max}=7$ & $\numcom_{\min}=\numcom_{\max}=11$ & $\numcom_{\min}=\numcom_{\max}=20$ \\
\hline
Pairwise constraints & Cannot-link & 10 CL pairs & 20 CL pairs & 60 CL pairs \\
 & Must-link & 2 ML pairs & 10 ML pairs & 20 ML pairs \\
 & Mixed & 2 ML; 10 CL & 10 ML; 20 CL & 20 ML; 60 CL \\
\hline
Resource constraints & Node resource & $\maxVres=(15)$ & $\maxVres=(20)$ & $\maxVres=(25)$ \\
 & Pairwise resource & $\maxVres=(5)$ & $\maxVres=(20)$ & $\maxVres=(40)$ \\
 & Combined resources & $\maxVres=(50,20,60)$ & $\maxVres=(45,25,60)$ & $\maxVres=(95,75,160)$ \\
\hline
\noalign{\vskip 1.5pt}
All constraints & Combined & \begin{tabular}[c]{@{}l@{}}$\numcom=7$; 2 ML, 10 CL;\\$\maxVres=(50,20,60)$\end{tabular} & \begin{tabular}[c]{@{}l@{}}$\numcom=11$; 10 ML, 20 CL;\\$\maxVres=(45,25,60)$\end{tabular} & \begin{tabular}[c]{@{}l@{}}$\numcom=20$; 20 ML, 60 CL;\\$\maxVres=(95,75,160)$\end{tabular} \\[1.5pt]
\hline\hline
\end{tabular*}
\end{table*}

The community-count instances include a minimum, maximum, or fixed number of communities. For example, for N10, $\numcom_{\min}=160$ requires at least 160 communities, $\numcom_{\max}=18$ permits at most 18 communities, and $\numcom_{\min}=\numcom_{\max}=20$ requires exactly 20 communities. For comparison, the unconstrained optimal solutions of N06, N07, and N10 contain 6, 10, and 19 communities, respectively. Thus, the maximum-count bounds are one below these values, the fixed-count targets are one above, and the minimum-count bounds are more than twice the corresponding unconstrained counts.

The pairwise-constraint instances include cannot-link, must-link, or both types of relations. For example, the cannot-link instances contain 10, 20, and 60 cannot-link pairs for N06, N07, and N10, respectively. The mixed instances combine the must-link and cannot-link specifications for the corresponding network.

Unlike community-count and pairwise constraints, resource constraints use node and pairwise resource coefficients. The three resource cases differ in how these coefficient types are used. The node-resource case uses one resource constraint with node coefficients $\rval{\nidx}{\ridx}$ only, whereas the pairwise-resource case uses one constraint with pairwise coefficients $\rval{\nidx\niidx}{\ridx}$ only. The combined-resource case uses three resource constraints: one defined by node coefficients, one by pairwise coefficients, and one by both. Accordingly, the three components of $\maxVres$ reported for each combined-resource instance in Table~\ref{tab:constrained-instances} correspond to these classes in the same order. All resource coefficients are signed integers in $[-5,5]$. Finally, the all-constraints case combines the fixed-count, mixed-pairwise, and combined-resource specifications to define one constrained instance for each network.

\subsection{Results on Unconstrained Networks}
\label{subsec:unconstrained-results}

Table~\ref{tab:unconstrained-results} reports modularity $\qVal$ and computational (CPU) time in seconds for the unconstrained networks. A value of 3600 indicates that the time limit was reached; OOM denotes termination because available memory was exhausted during model construction or optimization; T/O denotes a timeout during model construction before optimization began; and -- denotes an unavailable result. A superscript $*$ indicates that global optimality was proved by exact methods (i.e., $\prob{\mathrm{C}}$, \acrshort{bapgpl}, or \acrshort{bapgpq}); the best feasible modularity in each row is shown in bold. The Cutting-plane columns report upper bounds rather than feasible modularity values. Average rows use the available results, with time-limited runs contributing 3600 seconds.

\begin{table*}[hbt!]
\caption{Modularity ($\qVal$) and CPU time (in seconds) for the unconstrained benchmark networks.}
\label{tab:unconstrained-results}
\centering
\scriptsize
\setlength{\tabcolsep}{1.8pt}
\begin{tabular*}{\textwidth}{@{\extracolsep{\fill}}l*{6}{lr}@{}}
\hline\hline
& \multicolumn{2}{c}{$\prob{\mathrm{C}}$ (Naive)} & \multicolumn{2}{c}{Louvain} & \multicolumn{2}{c}{CG--GP} & \multicolumn{2}{c}{BaP--GP--L} & \multicolumn{2}{c}{BaP--GP--Q} & \multicolumn{2}{c}{Cutting-plane} \\
Network & \multicolumn{1}{c}{$\qVal$} & \multicolumn{1}{c}{CPU} & \multicolumn{1}{c}{$\qVal$} & \multicolumn{1}{c}{CPU} & \multicolumn{1}{c}{$\qVal$} & \multicolumn{1}{c}{CPU} & \multicolumn{1}{c}{$\qVal$} & \multicolumn{1}{c}{CPU} & \multicolumn{1}{c}{$\qVal$} & \multicolumn{1}{c}{CPU} & \multicolumn{1}{c}{UB} & \multicolumn{1}{c}{CPU} \\
\cmidrule(l{1pt}r{1pt}){2-3}\cmidrule(l{1pt}r{1pt}){4-5}\cmidrule(l{1pt}r{1pt}){6-7}\cmidrule(l{1pt}r{1pt}){8-9}\cmidrule(l{1pt}r{1pt}){10-11}\cmidrule(l{1pt}r{1pt}){12-13}
N01 & \textbf{0.5947}$^{*}$ & 5.2 & \textbf{0.5947} & 0.00 & \textbf{0.5947} & 0.2 & \textbf{0.5947}$^{*}$ & 0.3 & \textbf{0.5947}$^{*}$ & 0.3 & 0.5947 & 0.01 \\
N02 & \textbf{0.4198}$^{*}$ & 56.2 & 0.3949 & 0.00 & 0.4188 & 0.1 & \textbf{0.4198}$^{*}$ & 0.5 & \textbf{0.4198}$^{*}$ & 0.7 & 0.4198 & 0.04 \\
N03 & \textbf{0.3215}$^{*}$ & 902.9 & \textbf{0.3215} & 0.00 & \textbf{0.3215} & 0.2 & \textbf{0.3215}$^{*}$ & 1.9 & \textbf{0.3215}$^{*}$ & 2.9 & 0.3215 & 0.07 \\
N04 & \textbf{0.6713}$^{*}$ & 16.2 & \textbf{0.6713} & 0.00 & \textbf{0.6713} & 0.2 & \textbf{0.6713}$^{*}$ & 1.0 & \textbf{0.6713}$^{*}$ & 1.1 & 0.6713 & 0.01 \\
N05 & 0.5861 & 3600 & 0.6233 & 0.00 & \textbf{0.6258} & 0.8 & \textbf{0.6258}$^{*}$ & 36 & \textbf{0.6258}$^{*}$ & 52 & 0.6264 & 0.1 \\
N06 & 0.3741 & 3600 & 0.5583 & 0.00 & 0.5597 & 0.6 & \textbf{0.5600}$^{*}$ & 9.0 & \textbf{0.5600}$^{*}$ & 15 & 0.5609 & 0.2 \\
N07 & 0.0000 & 3600 & 0.6032 & 0.00 & \textbf{0.6046} & 3.6 & \textbf{0.6046}$^{*}$ & 102 & \textbf{0.6046}$^{*}$ & 106 & 0.6056 & 2.1 \\
N08 & 0.2916 & 3600 & 0.3626 & 0.01 & \textbf{0.3689} & 15 & 0.3626 & 3600 & 0.3626 & 3600 & 0.3737 & 6.6 \\
N09 & OOM & -- & \textbf{0.4410} & 0.01 & \textbf{0.4410} & 17 & \textbf{0.4410} & 3600 & \textbf{0.4410} & 3600 & 0.4455 & 54 \\
N10 & T/O & -- & 0.8433 & 0.01 & 0.8485 & 113 & 0.8433 & 3600 & \textbf{0.8486}$^{*}$ & 597 & 0.8498 & 16 \\
N11 & OOM & -- & \textbf{0.4018} & 0.07 & \textbf{0.4018} & 3600 & \textbf{0.4018} & 3600 & \textbf{0.4018} & 3600 & 0.4986 & 3600 \\
\hline
\emph{Average} & 0.4074 & 1925.1 & 0.5287 & 0.01 & \textbf{0.5324} & 341 & 0.5315 & 1323 & 0.5320 & 1052 & 0.5425 & 334 \\
\hline\hline
\end{tabular*}
\end{table*}

Although the direct formulation $\prob{\mathrm{C}}$ can guarantee global optimality, it does so only for the four relatively small networks and reaches the time limit for the remaining instances. Its scalability limitations become more evident as the network size increases, with runs terminating with weak or no feasible incumbents, memory exhaustion, or model-construction timeouts.

In contrast, both branch-and-price methods solve instances across a broader range of network sizes to optimality. \acrshort{bapgpq} also solves the relatively large, sparse N10 network in 597 s, whereas \acrshort{bapgpl} reaches the time limit with an incumbent of $\qVal=0.8433$. Across the seven networks solved to optimality by both methods, the mean CPU times are 21.5 s for \acrshort{bapgpl} and 25.4 s for \acrshort{bapgpq}; hence, \acrshort{bapgpl} requires 15.3\% less CPU time on average. At this scale, the benefit of solving the \acrshort{milp} model rather than the \acrshort{miqp} model outweighs the computational overhead of the additional variables introduced by linearization. As the network size increases, however, the \acrshort{miqp} formulation can become more effective by preserving sparsity, as observed for N10.

Louvain produces high-quality solutions in negligible CPU time, but leaves room for improvement on some networks. \acrshort{cggp} exploits this potential while remaining computationally efficient. It terminates within 5 s for N01 to N07 and improves upon Louvain on four of the seven networks. In particular, it attains the optimal modularity for N05 and N07. Most notably, for the relatively large and dense N08 network, it obtains $\qVal=0.3689$ in 15 s, which exceeds the $Q=0.3626$ incumbent returned by each exact branch-and-price method upon reaching the time limit. By bypassing the exact-pricing bottleneck and solving the final integer restricted master problem over the columns available at the end of heuristic pricing, \acrshort{cggp} rapidly produces strong feasible solutions. Across the eight networks for which \acrshort{bapgpq} certifies optimality, its mean absolute difference from the optimum is $1.71\times10^{-4}$, with a maximum difference of $10^{-3}$. These results support \acrshort{cggp} as a scalable primal heuristic, while exact pricing remains necessary to certify optimality.

\subsection{Results on Constrained Networks}
\label{subsec:constrained-results}

Table~\ref{tab:constrained-count-results} reports the results for the nine community-count instances. Although $\prob{\mathrm{C}}$ reaches the time limit on the unconstrained N06 network, it finds an optimal solution in 8.3 and 22 s for the maximum- and fixed-count cases, respectively. For N07, it obtains a substantially stronger incumbent than in the unconstrained case, reaching $\qVal=0.6044$ compared with $\qVal=0.4800$. However, $\prob{\mathrm{C}}$ still reaches the time limit in six instances.

\begin{table*}[hbt!]
\caption{Modularity ($\qVal$) and CPU time (in seconds) for the community-count constraint instances.}
\label{tab:constrained-count-results}
\centering
\scriptsize
\setlength{\tabcolsep}{1.8pt}
\begin{tabular*}{\textwidth}{@{\extracolsep{\fill}}ll*{6}{lr}@{}}
\hline\hline
& & \multicolumn{2}{c}{$\prob{\mathrm{C}}$ (Naive)} & \multicolumn{2}{c}{C--Louvain} & \multicolumn{2}{c}{CG--GP} & \multicolumn{2}{c}{BaP--GP--L} & \multicolumn{2}{c}{BaP--GP--Q} & \multicolumn{2}{c}{Cutting-plane} \\
Network ID & Case & \multicolumn{1}{c}{$\qVal$} & \multicolumn{1}{c}{CPU} & \multicolumn{1}{c}{$\qVal$} & \multicolumn{1}{c}{CPU} & \multicolumn{1}{c}{$\qVal$} & \multicolumn{1}{c}{CPU} & \multicolumn{1}{c}{$\qVal$} & \multicolumn{1}{c}{CPU} & \multicolumn{1}{c}{$\qVal$} & \multicolumn{1}{c}{CPU} & \multicolumn{1}{c}{UB} & \multicolumn{1}{c}{CPU} \\
\cmidrule(l{1pt}r{1pt}){3-4}\cmidrule(l{1pt}r{1pt}){5-6}\cmidrule(l{1pt}r{1pt}){7-8}\cmidrule(l{1pt}r{1pt}){9-10}\cmidrule(l{1pt}r{1pt}){11-12}\cmidrule(l{1pt}r{1pt}){13-14}
N06 & Min. count & 0.4931 & 3600 & 0.4471 & 0.02 & 0.5130 & 0.6 & \textbf{0.5134}$^{*}$ & 2.2 & \textbf{0.5134}$^{*}$ & 6.0 & 0.5609 & 0.2 \\
 & Max. count & \textbf{0.5562}$^{*}$ & 9.2 & 0.5561 & 0.09 & \textbf{0.5562} & 1.0 & \textbf{0.5562}$^{*}$ & 22 & \textbf{0.5562}$^{*}$ & 35 & 0.5609 & 0.2 \\
 & Fixed count & \textbf{0.5596}$^{*}$ & 21.6 & \textbf{0.5596} & 0.09 & \textbf{0.5596} & 0.9 & \textbf{0.5596}$^{*}$ & 4.3 & \textbf{0.5596}$^{*}$ & 7.4 & 0.5609 & 0.2 \\
\hline
N07 & Min. count & 0.3784 & 3600 & 0.4782 & 0.05 & \textbf{0.5518} & 1.0 & \textbf{0.5518}$^{*}$ & 12 & \textbf{0.5518}$^{*}$ & 10 & 0.6056 & 2.1 \\
 & Max. count & \textbf{0.6044} & 3600 & 0.6042 & 0.14 & \textbf{0.6044} & 2.1 & \textbf{0.6044}$^{*}$ & 126 & \textbf{0.6044}$^{*}$ & 114 & 0.6056 & 2.0 \\
 & Fixed count & \textbf{0.6032}$^{*}$ & 2811 & 0.6030 & 0.14 & 0.6030 & 1.3 & \textbf{0.6032}$^{*}$ & 66 & \textbf{0.6032}$^{*}$ & 63 & 0.6056 & 2.0 \\
\hline
N10 & Min. count & T/O & -- & 0.4633 & 0.18 & 0.6319 & 64 & \textbf{0.6351}$^{*}$ & 342 & \textbf{0.6351}$^{*}$ & 184 & 0.8498 & 16 \\
 & Max. count & 0.0773 & 3600 & 0.8472 & 3.85 & 0.8483 & 79 & 0.8472 & 3600 & \textbf{0.8484}$^{*}$ & 381 & 0.8498 & 16 \\
 & Fixed count & 0.4037 & 3600 & 0.8474 & 3.84 & \textbf{0.8485} & 85 & 0.8474 & 3600 & \textbf{0.8485}$^{*}$ & 414 & 0.8498 & 16 \\
\hline\hline
\end{tabular*}
\end{table*}

\acrshort{clouvain} generally provides high-quality solutions at negligible computational cost, but its performance varies across constraint cases. In particular, for the minimum-count instances, its relative modularity gaps from the optimal solutions are 12.9\%, 13.3\%, and 27.1\% for N06, N07, and N10, respectively, with an average gap of 17.8\%. \acrshort{cggp} substantially closes these gaps by improving the \acrshort{clouvain} solutions through column generation. For example, in the N07 minimum-count instance, \acrshort{cggp} improves the modularity from 0.4782 to 0.5518 and reaches the optimal solution. These results show that the effectiveness of \acrshort{clouvain} can depend strongly on the community-count constraint case, while column generation provides an effective mechanism for improving its initial solution.

Both branch-and-price methods solve all N06 and N07 instances to optimality within approximately two minutes. As the network size increases, however, their computational performance diverges. For N10, \acrshort{bapgpl} reaches the time limit for the maximum- and fixed-count instances, whereas \acrshort{bapgpq} solves all three community-count instances to optimality within seven minutes. This result demonstrates the computational advantage of the sparse \acrshort{miqp} formulation on relatively large, sparse networks.

Across the community-count cases, the maximum- and fixed-count cases reduce the optimal modularity by at most 0.0038 relative to the corresponding unconstrained optima, whereas the minimum-count cases result in substantially larger reductions of 0.0466, 0.0527, and 0.2135 for N06, N07, and N10, respectively. These results suggest that, in these instances, optimal modularity is relatively insensitive when the community-count requirement remains close to the unconstrained optimum, but can deteriorate substantially when the feasible community count is forced far from it.

Table~\ref{tab:constrained-pairwise-results} reports the results for the nine pairwise-constraint instances. The direct formulation $\prob{\mathrm{C}}$ reaches the time limit in all nine instances. In contrast, \acrshort{clouvain} handles the pairwise constraints effectively, producing solutions with an average relative modularity gap of only 0.87\% from the optimal solutions across the nine instances. As in the community-count instances, the computational performance of \acrshort{bapgpl} deteriorates on the larger N10 network, reaching the time limit in all three cases. \acrshort{bapgpq}, however, solves all nine instances to optimality within 453 s and remains computationally effective on N10.

\begin{table*}[hbt!]
\caption{Modularity ($\qVal$) and CPU time (in seconds) for the pairwise-constraint instances.}
\label{tab:constrained-pairwise-results}
\centering
\scriptsize
\setlength{\tabcolsep}{1.8pt}
\begin{tabular*}{\textwidth}{@{\extracolsep{\fill}}ll*{6}{lr}@{}}
\hline\hline
& & \multicolumn{2}{c}{$\prob{\mathrm{C}}$ (Naive)} & \multicolumn{2}{c}{C--Louvain} & \multicolumn{2}{c}{CG--GP} & \multicolumn{2}{c}{BaP--GP--L} & \multicolumn{2}{c}{BaP--GP--Q} & \multicolumn{2}{c}{Cutting-plane} \\
Network ID & Case & \multicolumn{1}{c}{$\qVal$} & \multicolumn{1}{c}{CPU} & \multicolumn{1}{c}{$\qVal$} & \multicolumn{1}{r}{CPU} & \multicolumn{1}{c}{$\qVal$} & \multicolumn{1}{c}{CPU} & \multicolumn{1}{c}{$\qVal$} & \multicolumn{1}{r}{CPU} & \multicolumn{1}{c}{$\qVal$} & \multicolumn{1}{c}{CPU} & \multicolumn{1}{c}{UB} & \multicolumn{1}{c}{CPU} \\
\cmidrule(l{1pt}r{1pt}){3-4}\cmidrule(l{1pt}r{1pt}){5-6}\cmidrule(l{1pt}r{1pt}){7-8}\cmidrule(l{1pt}r{1pt}){9-10}\cmidrule(l{1pt}r{1pt}){11-12}\cmidrule(l{1pt}r{1pt}){13-14}
N06 & Cannot-link & 0.5283 & 3600 & 0.5553 & 0.02 & \textbf{0.5572} & 0.5 & \textbf{0.5572}$^{*}$ & 5.6 & \textbf{0.5572}$^{*}$ & 9.4 & 0.5581 & 0.2 \\
 & Must-link & 0.4972 & 3600 & 0.4948 & 0.02 & 0.4955 & 1.2 & \textbf{0.4979}$^{*}$ & 7.2 & \textbf{0.4979}$^{*}$ & 13 & 0.5028 & 0.4 \\
 & Mixed & 0.4974 & 3600 & 0.4821 & 0.02 & 0.4964 & 0.6 & \textbf{0.4979}$^{*}$ & 6.1 & \textbf{0.4979}$^{*}$ & 10 & 0.5026 & 0.4 \\
\hline
N07 & Cannot-link & 0.0370 & 3600 & 0.5929 & 0.04 & 0.5929 & 1.0 & \textbf{0.5930}$^{*}$ & 51 & \textbf{0.5930}$^{*}$ & 57 & 0.5943 & 2.1 \\
 & Must-link & 0.0779 & 3600 & \textbf{0.5229} & 0.04 & \textbf{0.5229} & 1.7 & \textbf{0.5229}$^{*}$ & 90 & \textbf{0.5229}$^{*}$ & 67 & 0.5244 & 3.2 \\
 & Mixed & 0.1591 & 3600 & 0.5026 & 0.04 & 0.5053 & 1.1 & \textbf{0.5059}$^{*}$ & 99 & \textbf{0.5059}$^{*}$ & 92 & 0.5118 & 2.2 \\
\hline
N10 & Cannot-link & T/O & -- & 0.8182 & 0.14 & 0.8227 & 30 & 0.8182 & 3600 & \textbf{0.8235}$^{*}$ & 401 & 0.8255 & 19 \\
 & Must-link & T/O & -- & 0.7721 & 0.15 & 0.7766 & 84 & 0.7721 & 3600 & \textbf{0.7774}$^{*}$ & 453 & 0.7823 & 53 \\
 & Mixed & T/O & -- & 0.7509 & 0.15 & 0.7577 & 53 & 0.7509 & 3600 & \textbf{0.7636}$^{*}$ & 327 & 0.7692 & 89 \\
\hline\hline
\end{tabular*}
\end{table*}

Across the pairwise-constraint cases, the cannot-link cases reduce optimal modularity by at most 0.0251 relative to the corresponding unconstrained optima, whereas the must-link and mixed cases result in larger reductions of 0.0621--0.0987. Thus, under the tested settings, requiring specified node pairs to belong to the same community has a greater effect on optimal modularity than requiring them to belong to different communities.

Table~\ref{tab:constrained-resource-results} reports the results for the nine resource-constraint instances. The direct formulation $\prob{\mathrm{C}}$ returns weak incumbents for N07 and finds no feasible solution for N10 within the time limit. In contrast, \acrshort{bapgpq} solves all instances to optimality except the N07 node-resource case. For this instance, \acrshort{cggp} obtains $\qVal=0.4957$ in 5.3 s, outperforming the incumbent of $\qVal=0.4341$ returned by both branch-and-price methods at the time limit. By solving an integer restricted master problem over all generated columns after column generation, \acrshort{cggp} obtains a stronger feasible partition. This result highlights its ability to provide high-quality solutions rapidly when exact optimization is computationally challenging.

\begin{table*}[hbt!]
\caption{Modularity ($\qVal$) and CPU time (in seconds) for the resource-constraint instances.}
\label{tab:constrained-resource-results}
\centering
\scriptsize
\setlength{\tabcolsep}{1.8pt}
\begin{tabular*}{\textwidth}{@{\extracolsep{\fill}}ll*{6}{lr}@{}}
\hline\hline
& & \multicolumn{2}{c}{$\prob{\mathrm{C}}$ (Naive)} & \multicolumn{2}{c}{C--Louvain} & \multicolumn{2}{c}{CG--GP} & \multicolumn{2}{c}{BaP--GP--L} & \multicolumn{2}{c}{BaP--GP--Q} & \multicolumn{2}{c}{Cutting-plane} \\
Network ID & Case & \multicolumn{1}{c}{$\qVal$} & \multicolumn{1}{c}{CPU} & \multicolumn{1}{c}{$\qVal$} & \multicolumn{1}{c}{CPU} & \multicolumn{1}{c}{$\qVal$} & \multicolumn{1}{c}{CPU} & \multicolumn{1}{c}{$\qVal$} & \multicolumn{1}{c}{CPU} & \multicolumn{1}{c}{$\qVal$} & \multicolumn{1}{c}{CPU} & \multicolumn{1}{c}{UB} & \multicolumn{1}{c}{CPU} \\
\cmidrule(l{1pt}r{1pt}){3-4}\cmidrule(l{1pt}r{1pt}){5-6}\cmidrule(l{1pt}r{1pt}){7-8}\cmidrule(l{1pt}r{1pt}){9-10}\cmidrule(l{1pt}r{1pt}){11-12}\cmidrule(l{1pt}r{1pt}){13-14}
N06 & Node resource & 0.3591 & 3600 & 0.4388 & 0.03 & 0.4393 & 0.6 & \textbf{0.4913}$^{*}$ & 35 & \textbf{0.4913}$^{*}$ & 50 & 0.5609 & 0.2 \\
 & Pair resource & 0.0794 & 3600 & 0.5276 & 0.05 & 0.5468 & 0.6 & \textbf{0.5525}$^{*}$ & 35 & \textbf{0.5525}$^{*}$ & 47 & 0.5609 & 0.2 \\
 & Mixed resources & \textbf{0.5600} & 3600 & \textbf{0.5600} & 0.05 & \textbf{0.5600} & 0.6 & \textbf{0.5600}$^{*}$ & 6.3 & \textbf{0.5600}$^{*}$ & 9.3 & 0.5609 & 0.2 \\
\hline
N07 & Node resource & 0.0295 & 3600 & 0.4341 & 0.06 & \textbf{0.4957} & 5.3 & 0.4341 & 3600 & 0.4341 & 3600 & 0.6056 & 2.0 \\
 & Pair resource & 0.3929 & 3600 & 0.5908 & 0.15 & 0.5980 & 1.4 & \textbf{0.5981}$^{*}$ & 3426 & \textbf{0.5981}$^{*}$ & 1559 & 0.6056 & 2.0 \\
 & Mixed resources & 0.0100 & 3600 & 0.5904 & 0.10 & \textbf{0.6032} & 1.3 & \textbf{0.6032}$^{*}$ & 121 & \textbf{0.6032}$^{*}$ & 87 & 0.6056 & 2.0 \\
\hline
N10 & Node resource & T/O & -- & 0.7789 & 0.19 & 0.8081 & 139 & 0.7789 & 3600 & \textbf{0.8205}$^{*}$ & 2160 & 0.8498 & 16 \\
 & Pair resource & T/O & -- & 0.8471 & 0.47 & 0.8485 & 76 & 0.8471 & 3600 & \textbf{0.8486}$^{*}$ & 354 & 0.8498 & 16 \\
 & Mixed resources & T/O & -- & 0.8471 & 0.41 & 0.8485 & 71 & \textbf{0.8486}$^{*}$ & 2938 & \textbf{0.8486}$^{*}$ & 346 & 0.8498 & 16 \\
\hline\hline
\end{tabular*}
\end{table*}

For N10, the pairwise-resource and mixed-resource cases attain the same optimal modularity, $\qVal=0.8486$, as the unconstrained network. However, \acrshort{bapgpq} certifies optimality in 354 and 346 s, respectively, compared with 597 s for the unconstrained network. These results suggest that prior knowledge consistent with an optimal community structure can be encoded as optimality-preserving constraints to restrict the search space and accelerate optimality certification.

Table~\ref{tab:constrained-combined-results} reports the results for the three instances that combine all constraint classes. $\prob{\mathrm{C}}$ solves only the N06 instance to optimality. \acrshort{cggp} returns a high-quality solution within 50 s for every instance, with modularity at most 0.0064 below the optimum. On the larger N10 network, \acrshort{bapgpl} reaches the time limit with a weaker incumbent, whereas \acrshort{bapgpq} solves the instance to optimality in 268 s.

\begin{table*}[hbt!]
\caption{Modularity ($\qVal$) and CPU time (in seconds) for the instances combining all constraint classes.}
\label{tab:constrained-combined-results}
\centering
\scriptsize
\setlength{\tabcolsep}{1.8pt}
\begin{tabular*}{\textwidth}{@{\extracolsep{\fill}}ll*{6}{lr}@{}}
\hline\hline
& & \multicolumn{2}{c}{$\prob{\mathrm{C}}$ (Naive)} & \multicolumn{2}{c}{C--Louvain} & \multicolumn{2}{c}{CG--GP} & \multicolumn{2}{c}{BaP--GP--L} & \multicolumn{2}{c}{BaP--GP--Q} & \multicolumn{2}{c}{Cutting-plane} \\
Network ID & Case & \multicolumn{1}{c}{$\qVal$} & \multicolumn{1}{c}{CPU} & \multicolumn{1}{c}{$\qVal$} & \multicolumn{1}{c}{CPU} & \multicolumn{1}{c}{$\qVal$} & \multicolumn{1}{c}{CPU} & \multicolumn{1}{c}{$\qVal$} & \multicolumn{1}{c}{CPU} & \multicolumn{1}{c}{$\qVal$} & \multicolumn{1}{c}{CPU} & \multicolumn{1}{c}{UB} & \multicolumn{1}{c}{CPU} \\
\cmidrule(l{1pt}r{1pt}){3-4}\cmidrule(l{1pt}r{1pt}){5-6}\cmidrule(l{1pt}r{1pt}){7-8}\cmidrule(l{1pt}r{1pt}){9-10}\cmidrule(l{1pt}r{1pt}){11-12}\cmidrule(l{1pt}r{1pt}){13-14}
N06 & All constraints & \textbf{0.4974}$^{*}$ & 34 & 0.4815 & 0.21 & 0.4910 & 0.7 & \textbf{0.4974}$^{*}$ & 8.5 & \textbf{0.4974}$^{*}$ & 12 & 0.5026 & 0.4 \\
\hline
N07 & All constraints & 0.4963 & 3600 & 0.5038 & 0.44 & 0.5057 & 1.1 & \textbf{0.5058}$^{*}$ & 49 & \textbf{0.5058}$^{*}$ & 58 & 0.5118 & 2.2 \\
\hline
N10 & All constraints & 0.3184 & 3600 & 0.7528 & 10.97 & 0.7578 & 50 & 0.7528 & 3600 & \textbf{0.7636}$^{*}$ & 268 & 0.7692 & 89 \\
\hline\hline
\end{tabular*}
\end{table*}

\begin{figure}[hbt!]
\centering
\includegraphics[page=7,width=0.99\linewidth]{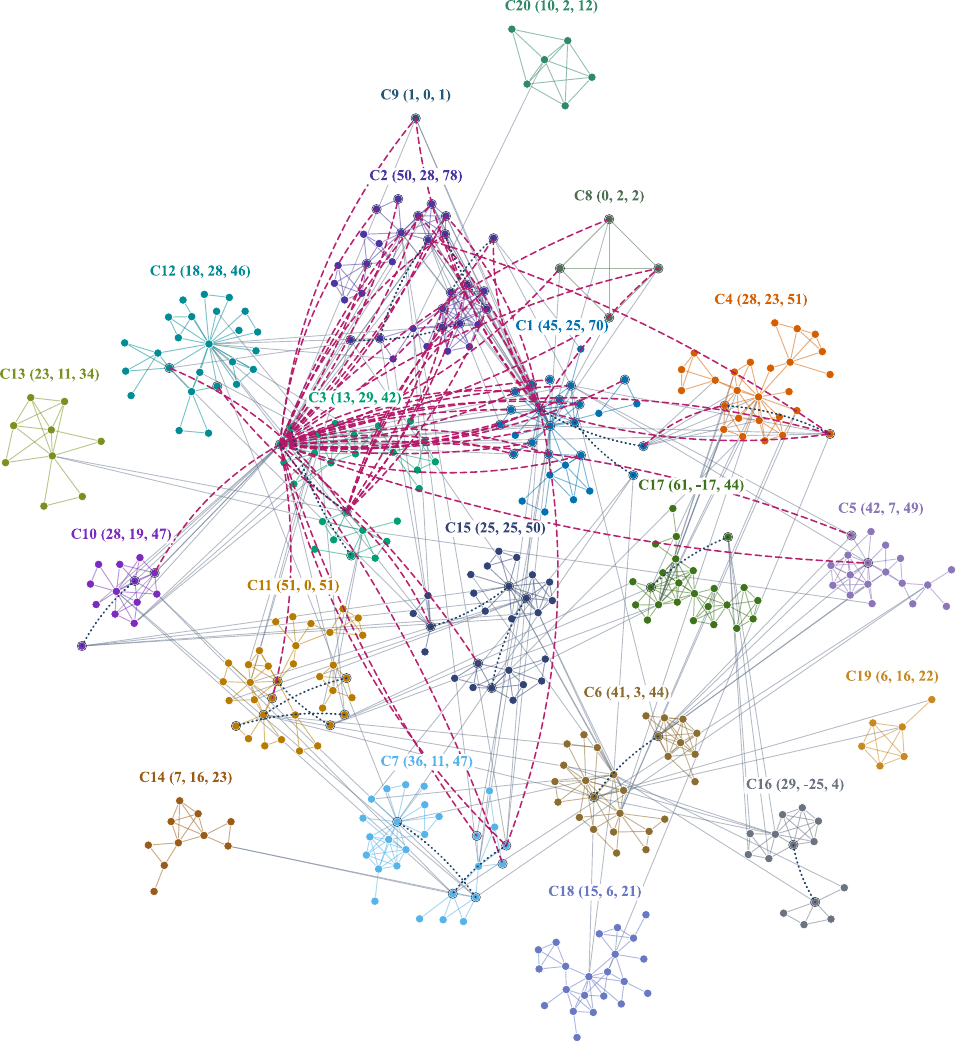}
\caption{Optimal community structure obtained by \acrshort{bapgpq} for the all-constraints instance of N10.}
\label{fig:n10-all-constraints-visual}
\end{figure}

Figure~\ref{fig:n10-all-constraints-visual} visualizes the optimal solution obtained by \acrshort{bapgpq} for the all-constraints instance of N10. Node colors indicate community assignments, while dark navy dotted lines denote must-link pairs and magenta dashed lines denote cannot-link pairs. The resource vector shown next to each community label reports its three resource values. The resulting partition consists of 20 communities and satisfies all constraints, as detailed in Table~\ref{tab:n10-constraint-satisfaction}. Specifically, all 20 must-link and 60 cannot-link constraints are satisfied, and the resource values of every community remain within the prescribed upper bounds of $(95,75,160)$.

\begin{table*}[hbt!]
\caption{Constraint specifications and satisfaction of the optimal \acrshort{bapgpq} solution for the all-constraints instance of N10.}
\label{tab:n10-constraint-satisfaction}
\centering
\scriptsize
\setlength{\tabcolsep}{1.8pt}
\renewcommand{\arraystretch}{1.12}
\begin{tabular*}{\textwidth}{@{\extracolsep{\fill}}lp{0.6\textwidth}r@{}}
\hline\hline
Constraint & Specifications & Satisfied \\
\hline
Community count & $\numcom_{\min}=\numcom_{\max}=20$ & Yes \\
\hline
Pairwise: Must-link\footnote{For both pairwise constraint types, each unordered pair identifies two nodes of the original graph by their node indices.}
&
$\{3,252\}$, $\{5,60\}$, $\{6,74\}$, $\{7,221\}$, $\{9,108\}$, $\{10,191\}$, $\{13,76\}$, $\{15,121\}$, $\{16,254\}$, $\{43,144\}$, $\{46,109\}$, $\{48,64\}$, $\{55,95\}$, $\{77,120\}$, $\{96,237\}$, $\{97,234\}$, $\{98,251\}$, $\{124,143\}$, $\{168,189\}$, $\{175,236\}$
& Yes (20/20) \\
\hline
Pairwise: Cannot-link &
$\{2,5\}$, $\{3,5\}$, $\{3,6\}$, $\{4,5\}$, $\{5,6\}$, $\{5,7\}$, $\{5,8\}$, $\{5,11\}$, $\{5,13\}$, $\{5,14\}$, $\{5,15\}$, $\{5,16\}$, $\{5,17\}$, $\{5,18\}$, $\{5,19\}$, $\{5,23\}$, $\{5,24\}$, $\{5,25\}$, $\{5,26\}$, $\{5,27\}$, $\{5,28\}$, $\{5,29\}$, $\{5,30\}$, $\{5,31\}$, $\{5,32\}$, $\{5,33\}$, $\{5,34\}$, $\{5,35\}$, $\{5,36\}$, $\{5,37\}$, $\{5,39\}$, $\{5,40\}$, $\{5,41\}$, $\{5,79\}$, $\{5,80\}$, $\{5,81\}$, $\{5,82\}$, $\{5,93\}$, $\{5,327\}$, $\{5,328\}$, $\{6,7\}$, $\{6,13\}$, $\{6,15\}$, $\{6,16\}$, $\{6,17\}$, $\{6,18\}$, $\{6,30\}$, $\{6,31\}$, $\{7,15\}$, $\{60,124\}$, $\{60,125\}$, $\{60,126\}$, $\{60,248\}$, $\{60,249\}$, $\{60,250\}$, $\{60,251\}$, $\{60,252\}$, $\{60,253\}$, $\{60,254\}$, $\{74,221\}$
& Yes (60/60) \\
\hline
Resource constraints\footnote{Each community's resource vector reports the left-hand side of Eq.~\eqref{eq:bin-packing-general} for resource classes $\ridx=1,2,3$.}
&
Upper bounds: $(95,75,160)$; resource-class order: $\ridx=1,2,3$.\newline
\makebox[0.14\textwidth][l]{C1 $(45,25,70)$}
\makebox[0.14\textwidth][l]{C2 $(50,28,78)$}
\makebox[0.14\textwidth][l]{C3 $(13,29,42)$}
\makebox[0.14\textwidth][l]{C4 $(28,23,51)$}\newline
\makebox[0.14\textwidth][l]{C5 $(42,7,49)$}
\makebox[0.14\textwidth][l]{C6 $(41,3,44)$}
\makebox[0.14\textwidth][l]{C7 $(36,11,47)$}
\makebox[0.14\textwidth][l]{C8 $(0,2,2)$}\newline
\makebox[0.14\textwidth][l]{C9 $(1,0,1)$}
\makebox[0.14\textwidth][l]{C10 $(28,19,47)$}
\makebox[0.14\textwidth][l]{C11 $(51,0,51)$}
\makebox[0.14\textwidth][l]{C12 $(18,28,46)$}\newline
\makebox[0.14\textwidth][l]{C13 $(23,11,34)$}
\makebox[0.14\textwidth][l]{C14 $(7,16,23)$}
\makebox[0.14\textwidth][l]{C15 $(25,25,50)$}
\makebox[0.14\textwidth][l]{C16 $(29,-25,4)$}\newline
\makebox[0.14\textwidth][l]{C17 $(61,-17,44)$}
\makebox[0.14\textwidth][l]{C18 $(15,6,21)$}
\makebox[0.14\textwidth][l]{C19 $(6,16,22)$}
\makebox[0.14\textwidth][l]{C20 $(10,2,12)$}
& \parbox[t]{0.12\textwidth}{\raggedleft Yes}\\
\hline\hline
\end{tabular*}
\end{table*}

\begin{figure}[hbt!]
\centering
\includegraphics[page=8,width=0.95\linewidth]{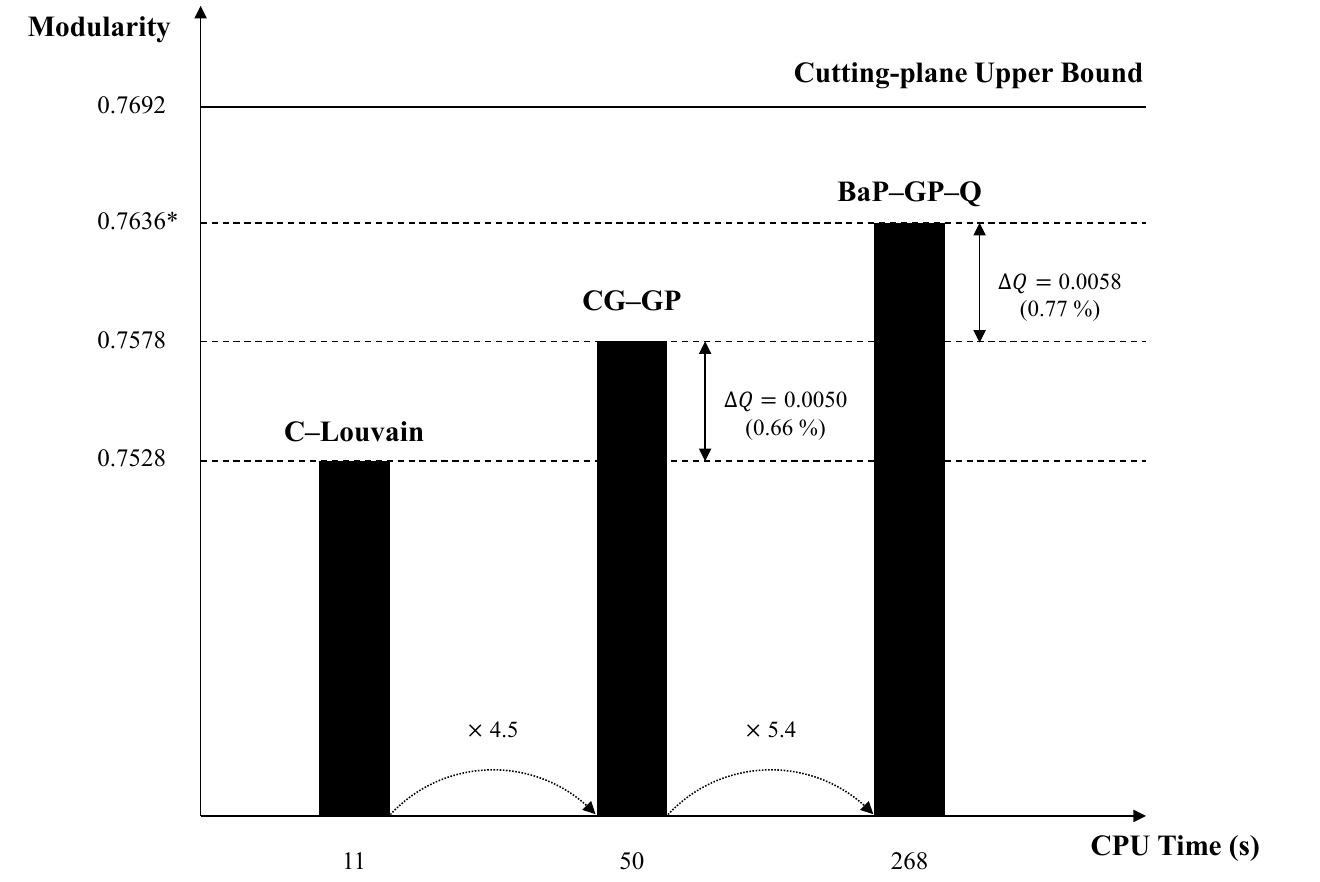}
\caption{Solution quality and CPU time for the all-constraints instance of N10. CPU time is plotted on a logarithmic scale.}
\label{fig:n10-all-constraints-tradeoff}
\end{figure}

Figure~\ref{fig:n10-all-constraints-tradeoff} illustrates the trade-off between solution quality and CPU time for the all-constraints instance of N10. Each bar reports an algorithm's final modularity and total CPU time in seconds, with time shown on a logarithmic scale. The solid horizontal line shows the cutting-plane upper bound, and the asterisk denotes proven optimality. \acrshort{clouvain} obtains $Q=0.7528$ in 11 s, and \acrshort{cggp} improves it by 0.0050 in 50 s. \acrshort{bapgpq} achieves a further gain of 0.0058 and guarantees optimality in 268 s, while \acrshort{cggp} requires 81.3\% less CPU time. Due to the high computational cost of exact pricing, \acrshort{bapgpl} reached the one-hour time limit without improving upon the initial \acrshort{clouvain} solution or proving optimality. It is therefore omitted from the figure, with its result reported in Table~\ref{tab:constrained-combined-results}.

Table~\ref{tab:constrained-quality-summary} extends this comparison to all 29 certified instances, summarizing the improvements achieved by \acrshort{cggp} and \acrshort{bapgpq} over the initial \acrshort{clouvain} solutions. The two gain columns report, for each network, the mean of the instance-level percentage improvements over \acrshort{clouvain}. Across these instances, \acrshort{cggp} and \acrshort{bapgpq} improve upon \acrshort{clouvain} by an average of 3.00\% and 3.67\%, respectively. Overall, \acrshort{bapgpq} proves optimality in 29 of the 30 constrained instances. Its effectiveness is particularly evident for N10. These results favor \acrshort{clouvain} for obtaining an initial solution at low computational cost, \acrshort{cggp} for improving solution quality under a limited computational budget, and \acrshort{bapgpq} when global optimality certification is required. Beyond its numerical gains, \acrshort{bapgpq} certifies that no feasible solution can attain higher modularity, providing a guarantee unavailable from heuristic methods.

\begin{table*}[hbt!]
\caption{Solution-quality summary for constrained instances certified by BaP--GP--Q.}
\label{tab:constrained-quality-summary}
\centering
\scriptsize
\begin{tabular*}{\textwidth}{@{\extracolsep{\fill}}lrrrrrr@{}}
\hline\hline
Network ID & Instances & C--Louvain & CG--GP & BaP--GP--Q & CG--GP gain (\%) & Exact gain (\%) \\
\hline
N06 & 10 & 0.5103 & 0.5215 & \textbf{0.5283} & 2.39 & 3.91 \\
N07 & 9 & 0.5543 & 0.5652 & \textbf{0.5654} & 2.19 & 2.22 \\
N10 & 10 & 0.7725 & 0.7949 & \textbf{0.7978} & 4.34 & 4.75 \\
\hline
\emph{Average} & 29 & 0.6144 & 0.6293 & \textbf{0.6327} & 3.00 & 3.67 \\
\hline\hline
\end{tabular*}
\end{table*}

\subsection{Computational Analysis of Pricing Strategies}
\label{subsec:pricing-analysis}

We compare the pricing strategies from two perspectives: (1) heuristic versus exact pricing to assess the efficiency of the proposed graph-partitioning-based heuristic during column generation, and (2) \acrshort{milp}-based versus sparse-\acrshort{miqp}-based exact pricing within the branch-and-price framework.

\subsubsection{Heuristic Pricing Performance}

The computational efficiency of branch-and-price depends strongly on the time spent on column generation at each branch-and-price node. Because exact pricing can be time-consuming and generates at most one improving column per iteration, we developed a graph-partitioning-based heuristic that generates multiple feasible, high-quality columns efficiently. For comparison, we use \acrfull{cgl}, which performs column generation at the root branch-and-price node without heuristic pricing. As a naive baseline, \acrshort{cgl} does not solve an integer restricted master problem over the generated columns after column generation terminates. This subsection compares the proposed heuristic pricing with \acrshort{cgl} to assess how effectively it handles the different constraint structures considered in this study.

Table~\ref{tab:heuristic-pricing-comparison} compares two column-generation methods applied at the root branch-and-price node for the 30 constrained instances. We report results by constraint class. Instances gives the group size, and all remaining entries are arithmetic means over the instances in that group. Columns denotes the total number of columns added to the RMP and Iterations denotes the number of RMP solves during column generation. Pricing denotes the total time spent in pricing calls, while CPU is the total computational time, including initialization and any final integer RMP solve. Both times are reported in seconds, with time-limited runs counted as 3600 s when computing mean CPU time.

\begin{table*}[hbt!]
\caption{Comparison of exact and graph-partitioning-based heuristic pricing on constrained instances.}
\label{tab:heuristic-pricing-comparison}
\centering
\scriptsize
\setlength{\tabcolsep}{1.8pt}
\begin{tabular*}{\textwidth}{@{\extracolsep{\fill}}l*{11}{r}@{}}
\hline\hline
& & \multicolumn{5}{c}{\acrshort{cgl}} & \multicolumn{5}{c}{\acrshort{cggp}} \\
\cmidrule(l{1pt}r{1pt}){3-7}\cmidrule(l{1pt}r{1pt}){8-12}
Constraint class & Instances & \multicolumn{1}{c}{$\qVal$} & Columns & Iterations & \shortstack{Pricing} & \shortstack{CPU (s)} & \multicolumn{1}{c}{$\qVal$} & Columns & Iterations & \shortstack{Pricing} & \shortstack{CPU (s)} \\
\hline
Community count & 9 & 0.6081 & 370 & 371 & 2337 & 2343 & 0.6352 & 1,557 & 15 & 24 & 26 \\
Pairwise & 9 & 0.6108 & 213 & 214 & 2426 & 2428 & 0.6141 & 1,534 & 16 & 18 & 19 \\
Resource & 9 & 0.6335 & 370 & 371 & 2486 & 2491 & 0.6387 & 1,623 & 22 & 31 & 33 \\
All constraints & 3 & 0.5853 & 304 & 305 & 1803 & 1810 & 0.5848 & 1,510 & 11 & 12 & 17 \\
\hline\hline
\end{tabular*}
\end{table*}

Across all four constraint classes, \acrshort{cggp} terminates within 33 s on average, whereas \acrshort{cgl} requires substantially more CPU time and often reaches the time limit before column generation converges. While \acrshort{cgl} generates one improving column per iteration, \acrshort{cggp} generates approximately 1500 improving columns in fewer than 22 iterations on average. This larger pool of columns provides more alternatives for the final integer RMP and can improve solution quality. For both methods, the time spent on pricing subproblems account for most of the total CPU time, highlighting the importance of efficient pricing to overall computational performance. These results show that the proposed graph-partitioning-based heuristic pricing can efficiently generate feasible, competitive columns across the tested constraint classes without directly solving the mathematical-programming-based pricing model.

\subsubsection{Comparison Among Exact Pricing Strategies}

Motivated by the sparsity commonly observed in complex networks for community detection, we develop the sparse \acrshort{miqp} formulation $\prob{\text{PSP-SMIQP}}$ to exploit this structure in exact pricing. Although an \acrshort{milp} formulation can often benefit from mature optimization solvers, the results in Tables~\ref{tab:unconstrained-results}--\ref{tab:constrained-combined-results} show that \acrshort{bapgpq} becomes increasingly effective relative to \acrshort{bapgpl} as the network size increases. This subsection examines the source of this computational advantage by comparing the sizes and structures of the two exact-pricing formulations, with particular emphasis on how the sparse \acrshort{miqp} formulation limits the growth of the pricing problem.

Table~\ref{tab:constrained-pricing-summary} compares the overall performance of the two branch-and-price methods and the sizes of their exact-pricing formulations at the root branch-and-price node. At a given branch-and-price node, both methods invoke exact pricing when heuristic pricing fails to identify an improving column, and the exact-pricing formulation retains the same variable set throughout column generation. For each network, the table reports the mean number of pair variables (Pair vars.) and the mean total number of variables (Total vars.) in the exact-pricing formulations at the root branch-and-price node over the 10 instances. The pair variables represent the product variables required by the modularity objective and pairwise-resource constraints, while the total variables additionally include the contracted-node selection variables. Density is the number of edges divided by the number of edges in a complete graph with the same number of nodes, expressed as a percentage. Optimal is the number of instances solved to optimality, and CPU is the mean total CPU time in seconds, with time-limited runs counted as 3600 s.

\begin{table*}[hbt!]
\caption{Performance and exact-pricing formulation size at the root branch-and-price node for constrained instances.}
\label{tab:constrained-pricing-summary}
\centering
\scriptsize
\setlength{\tabcolsep}{3pt}
\resizebox{\textwidth}{!}{%
\begin{tabular}{lrrrrrrrrr}
\hline\hline
& & \multicolumn{4}{c}{BaP--GP--L} & \multicolumn{4}{c}{BaP--GP--Q} \\
\cmidrule(l{1pt}r{1pt}){3-6}\cmidrule(l{1pt}r{1pt}){7-10}
Network ID & Density (\%) & Optimal & CPU (s) & Pair vars. & Total vars. & Optimal & CPU (s) & Pair vars. & Total vars. \\
\hline
N06 & 8.68 & 10/10 & 13 & 2,794 & 2,869 & 10/10 & 20 & 327 & 402 \\
N07 & 9.35 & 9/10 & 764 & 6,227 & 6,339 & 9/10 & 571 & 789 & 901 \\
N10 & 1.28 & 2/10 & 3208 & 69,420 & 69,793 & 10/10 & 529 & 1,183 & 1,556 \\
\hline\hline
\end{tabular}%
}
\end{table*}

\acrshort{bapgpl} introduces a product variable for every unordered pair of contracted nodes. In contrast, \acrshort{bapgpq} introduces such variables only for node pairs that correspond to graph edges or appear in pairwise-resource constraints, while retaining the degree-product term in quadratic form. This reduces the mean pair-variable count by 88.3\% for N06, 87.3\% for N07, and 98.3\% for N10. Because all three networks are undirected, the retained quadratic term is concave and can be exploited by \acrshort{miqp} solvers. Although the \acrshort{milp} formulation is more efficient on N06, the sparse \acrshort{miqp} formulation solves the N07 and N10 instances more efficiently by substantially reducing the number of pair variables.

These results indicate that the main advantage of the sparse \acrshort{miqp} formulation is the reduction in exact-pricing formulation size as the network size increases. By avoiding unnecessary auxiliary pair variables while retaining a tractable concave quadratic structure, \acrshort{bapgpq} yields a substantially more compact pricing problem than full linearization, particularly for the large, sparse N10 network. This reduction in formulation size lowers the computational time required for exact pricing, which accounts for most of the overall branch-and-price solution time.

\subsection{Effect of Branch-and-Price Search}
\label{subsec:branching-impact}

When the final solution at the root branch-and-price node is fractional, column generation itself is insufficient to certify global optimality and branching is required. We therefore examine the extent of the subsequent branch-and-price search through the numbers of branched and pruned nodes, together with the columns and computational time required per processed node. The analysis focuses on \acrshort{bapgpq}, which solves 29 of the 30 constrained instances to optimality and thus provides completed search trees for assessing the additional effort required beyond the root branch-and-price node.

Table~\ref{tab:branching-impact} reports the branch-and-price search statistics of \acrshort{bapgpq} for the 10 constrained N10 instances. Root RMP indicates whether the RMP solution obtained after column generation at the root branch-and-price node is integer or fractional. Created is the number of generated branch-and-price nodes, including the root. A branch-and-price node is considered processed when column generation terminates. Branched counts processed branch-and-price nodes whose RMP solutions are fractional and therefore generate child branch-and-price nodes. Under Pruned by, Integrality counts processed nodes with integer RMP solutions, Bound counts processed nodes whose upper bounds cannot improve the incumbent, Infeasibility counts processed nodes found to be infeasible, and Cutoff counts unprocessed open nodes removed after an incumbent update renders their upper bounds noncompetitive. Under Per processed node, Columns is the mean number of columns available at the end of a processed node, while RMP solves and CPU are the total number of RMP solves and total CPU time divided by the number of processed nodes, respectively. The Average row reports the mean of each statistic over the 10 instances.

\begin{table*}[hbt!]
\caption{BaP--GP--Q search statistics for the constrained N10 instances.}
\label{tab:branching-impact}
\centering
\scriptsize
\setlength{\tabcolsep}{3pt}
\begin{tabular*}{\textwidth}{@{\extracolsep{\fill}}llrrrrrrrrr@{}}
\hline\hline
& & & & \multicolumn{4}{c}{Pruned by} & \multicolumn{3}{c}{Per processed node} \\
\cmidrule(l{1pt}r{1pt}){5-8}\cmidrule(l{1pt}r{1pt}){9-11}
Case & Root RMP & Created & Branched & Integrality & Bound & Infeasibility & Cutoff & Columns & RMP solves & CPU (s) \\
\hline
Min. count & Integer & 1 & 0 & 1 & 0 & 0 & 0 & 3,787 & 24 & 184 \\
Max. count & Integer & 1 & 0 & 1 & 0 & 0 & 0 & 5,626 & 61 & 381 \\
Fixed count & Fractional & 3 & 1 & 2 & 0 & 0 & 0 & 5,142 & 21 & 138 \\
Cannot-link & Fractional & 7 & 3 & 1 & 0 & 0 & 3 & 6,351 & 23 & 100 \\
Must-link & Integer & 1 & 0 & 1 & 0 & 0 & 0 & 6,211 & 66 & 453 \\
Mixed & Integer & 1 & 0 & 1 & 0 & 0 & 0 & 8,014 & 80 & 327 \\
Node resource & Fractional & 9 & 4 & 1 & 0 & 0 & 4 & 6,276 & 50 & 432 \\
Pair resource & Integer & 1 & 0 & 1 & 0 & 0 & 0 & 6,334 & 64 & 354 \\
Mixed resources & Integer & 1 & 0 & 1 & 0 & 0 & 0 & 5,823 & 57 & 346 \\
All constraints & Integer & 1 & 0 & 1 & 0 & 0 & 0 & 5,479 & 62 & 268 \\
\hline
\emph{Average} & -- & 2.6 & 0.8 & 1.1 & 0.0 & 0.0 & 0.7 & 5,904 & 51 & 298 \\
\hline\hline
\end{tabular*}
\end{table*}

The root RMP solution is integer in seven instances, allowing \acrshort{bapgpq} to solve them to optimality without branching. Only the fixed-count, cannot-link, and node-resource instances require branching. Nearly half of the generated non-root branch-and-price nodes are pruned by cutoff before column generation, thereby avoiding substantial additional computation. These results indicate that the root-node relaxation is generally strong and that only a small branch-and-price tree is required to solve these instances to optimality.

\subsection{Upper-Bound Performance}
\label{subsec:upper-bound-results}

Despite their strong performance, the proposed branch-and-price methods may fail to establish global optimality on large or dense networks. If the time limit is reached during column generation at the root branch-and-price node, branching cannot begin and a valid global upper bound is unavailable. The proposed cutting-plane approach is designed to provide a tight upper bound for such difficult instances. This subsection evaluates its tightness by comparing the resulting upper bounds with those obtained from the other methods. For the constrained instances, \acrshort{bapgpq} already solves nearly all cases to optimality within the time limit, making an additional upper-bound analysis less informative. We therefore restrict the upper-bound analysis to the unconstrained networks.

Table~\ref{tab:bound-comparison} reports the best feasible modularity and the available upper bounds. Best $\qVal$ is the highest modularity obtained by any algorithm, while the remaining columns report upper bounds, for which a smaller value is tighter. The $\prob{\mathrm{C}}$ column gives the upper bound reported by the solver. A branch-and-price method provides an upper bound only after column generation at the root branch-and-price node is completed using exact pricing. A dash indicates that no finite upper bound was obtained because of memory exhaustion or a time limit. An asterisk marks a proven optimum, and bold indicates the smallest available upper bound.

\begin{table*}[hbt!]
\caption{Best modularity and upper bounds for the unconstrained networks.}
\label{tab:bound-comparison}
\centering
\scriptsize
\setlength{\tabcolsep}{4pt}
\begin{tabular*}{\textwidth}{@{\extracolsep{\fill}}l*{5}{l}@{}}
\hline\hline
& & \multicolumn{4}{c}{Upper bound} \\
\cmidrule(l{1pt}r{1pt}){3-6}
Network ID & \multicolumn{1}{c}{Best $\qVal$} & \multicolumn{1}{c}{$\prob{\mathrm{C}}$ (Naive)} & \multicolumn{1}{c}{BaP--GP--L} & \multicolumn{1}{c}{BaP--GP--Q} & \multicolumn{1}{c}{Cutting-plane} \\
\hline
N01 & 0.5947$^{*}$ & \textbf{0.5947} & \textbf{0.5947} & \textbf{0.5947} & \textbf{0.5947} \\
N02 & 0.4198$^{*}$ & \textbf{0.4198} & \textbf{0.4198} & \textbf{0.4198} & \textbf{0.4198} \\
N03 & 0.3215$^{*}$ & \textbf{0.3215} & \textbf{0.3215} & \textbf{0.3215} & \textbf{0.3215} \\
N04 & 0.6713$^{*}$ & \textbf{0.6713} & \textbf{0.6713} & \textbf{0.6713} & \textbf{0.6713} \\
N05 & 0.6258$^{*}$ & 0.6280 & \textbf{0.6258} & \textbf{0.6258} & 0.6264 \\
N06 & 0.5600$^{*}$ & 0.5778 & \textbf{0.5600} & \textbf{0.5600} & 0.5609 \\
N07 & 0.6046$^{*}$ & 0.8773 & \textbf{0.6046} & \textbf{0.6046} & 0.6056 \\
N08 & 0.3689 & 0.6095 & -- & -- & \textbf{0.3737} \\
N09 & 0.4410 & -- & -- & -- & \textbf{0.4455} \\
N10 & 0.8486$^{*}$ & -- & -- & \textbf{0.8486} & 0.8498 \\
N11 & 0.4018 & -- & -- & -- & \textbf{0.4986} \\
\hline\hline
\end{tabular*}
\end{table*}

The cutting-plane upper bound matches Best $\qVal$ on the four small networks. For every network except N11, it is no more than approximately 1.3\% above Best $\qVal$ and is obtained within 54 s, as reported in Table~\ref{tab:unconstrained-results}. For N11, the cutting-plane upper bound remains looser but finite at the time limit, whereas the other exact methods provide no bound. Thus, under a limited computational budget, Louvain (C--Louvain) or \acrshort{cggp} can be used to obtain a high-quality feasible solution and the cutting-plane upper bound can be used to assess its optimality gap independently. Moreover, because each cutting-plane iteration yields a valid upper bound that tightens as triangle inequalities are added, the procedure can be used for anytime bounding and return the best available upper bound at a prescribed time limit.

\subsection{Discussion and Solution Guidelines}
\label{subsec:discussion-and-solution-guidelines}

The numerical results demonstrate a clear trade-off between computational efficiency and solution quality. Solving the direct formulation $\prob{\mathrm{C}}$ with a general-purpose solver becomes computationally intractable for many benchmark instances as the network size and constraint complexity increase. In contrast, \acrshort{clouvain} provides high-quality feasible solutions within negligible computational time, making it an effective heuristic for rapid solution generation. Its solution quality, however, can deteriorate for challenging constraint cases. \acrshort{cggp} provides a more stable heuristic alternative by improving the initial \acrshort{clouvain} solution through graph-partitioning-based column generation, consistently producing high-quality solutions within a moderate computational budget.

For exact algorithms, the relative performance of the two branch-and-price methods becomes more distinct as the network size increases. On the relatively large and sparse network, \acrshort{bapgpq} consistently outperforms \acrshort{bapgpl} across the tested constraint classes. The computational analysis shows that this advantage stems from the sparse \acrshort{miqp} formulation, which avoids a large number of auxiliary pair variables and thereby limits the growth of the exact-pricing problem. Although \acrshort{bapgpl} can be faster on smaller networks, both methods require relatively short computational times in these cases. Considering its substantially better scalability on the larger sparse network, \acrshort{bapgpq} therefore provides a more robust choice as an exact algorithm for constrained modularity maximization.

Together with the analyses of pricing strategies and branch-and-price search, these results suggest a practical solution strategy according to the desired balance between computational time and optimality. When a solution is required almost immediately, \acrshort{clouvain} provides a strong starting point. When more consistent solution quality is desired under diverse constraints, \acrshort{cggp} can be used to improve this solution while retaining relatively low computational cost. When an optimality certificate is required, \acrshort{bapgpq} is the preferred exact method, particularly for large, sparse, undirected networks. When the available computational budget is not known in advance, a further option is to first combine \acrshort{clouvain} with the cutting-plane upper bound to obtain a feasible solution and an upper bound to estimate the optimality gap. If this gap is acceptable, the procedure can terminate; otherwise, \acrshort{cggp} and subsequently \acrshort{bapgpq} can be applied to progressively improve the solution and, if needed, solve the problem to optimality.

\section{Conclusion}
\label{sec:conclusion}

This study developed a unified optimization framework for constrained modularity maximization. The framework accommodates restrictions on the number of communities, must-link and cannot-link relations, and general resource constraints. It consists of three main components for formulating and preprocessing the constrained problem, efficiently searching for high-quality solutions, and assessing the quality of the resulting solutions. To support these components, we developed four algorithms: \acrfull{clouvain}, \acrfull{cggp}, \acrfull{bapgpl}, and \acrfull{bapgpq}. \acrshort{clouvain} provides fast constraint-feasible solutions, \acrshort{cggp} improves them through graph-partitioning-based heuristic pricing, and \acrshort{bapgpl} and \acrshort{bapgpq} incorporate exact pricing within branch-and-price to establish global optimality. We further proposed a cutting-plane upper bound for independently assessing solution quality when exact optimization cannot be completed within the available computational budget. Together, these methods provide a computational spectrum ranging from fast heuristic search to exact optimization and independent optimality assessment.

The numerical experiments demonstrate the effectiveness of the proposed framework across different network sizes and combinations of constraints. \acrshort{clouvain} efficiently generates feasible initial solutions, while \acrshort{cggp} rapidly produces high-quality solutions through heuristic column generation. The branch-and-price algorithms provide substantially stronger optimality guarantees, with \acrshort{bapgpq} proving global optimality for 29 of the 30 constrained instances within the one-hour time limit. The comparison between the two exact-pricing formulations further shows that \acrshort{bapgpq} becomes increasingly effective for larger sparse undirected networks by reducing the number of auxiliary product variables while retaining the quadratic structure of the pricing problem. The proposed acceleration techniques also reduce unnecessary column-generation iterations and exact-pricing effort without compromising exactness. For instances that remain computationally challenging, the cutting-plane upper bound provides an additional means of quantifying the quality of the obtained solutions. Overall, the proposed unified optimization framework provides a set of complementary algorithms that can be selected according to the desired trade-off between computational time and solution quality. In practice, \acrshort{clouvain} can be combined with the cutting-plane upper bound to obtain a feasible solution and estimate its optimality gap. If further improvement is required, \acrshort{cggp} and subsequently \acrshort{bapgpq} can be applied to improve solution quality and, when necessary, solve the problem to optimality.

Future research should focus on improving exact pricing for dense networks, where sparsity provides a smaller reduction in formulation size, and for directed networks, where the retained quadratic term is generally nonconvex. Promising directions include decomposition and parallelization strategies, adaptive selection between linearized and sparse exact-pricing formulations, and stronger constraint-aware upper bounds. Application to larger real-world networks with domain-specific constraints would further demonstrate the practical applicability of the proposed framework.

\begin{acknowledgments}
This work was supported by Hyundai Motor Company and Kia by the Industry-Academy Research Cooperation Project, the InnoCORE program of the Ministry of Science and ICT (\#N10260008), and the National Research Foundation of Korea (NRF) grant funded by the Korean government (MSIT) (No. RS-2026-25498353).
\end{acknowledgments}

\appendix

\section{Constrained Louvain Algorithm}
\label{sec:constrained-louvain}

We propose a constraint-aware variant of the Louvain algorithm~\cite{blondel2008fast} as an efficient feasible-solution generator. The algorithm retains the two main operations of standard Louvain---local node moving and graph aggregation---while restricting moves and merges to satisfy the constraints, supplemented by repair steps when necessary.

The algorithm proceeds over levels $\loulidx=0,1,\ldots$.\footnote{In this study, we set the maximum number of levels to $100$.} The level-zero working graph is the original graph when no must-link constraints are present; otherwise, it is obtained by contracting the must-link pairs. At each level, the algorithm performs constrained local moving, repairs any remaining feasibility violations when necessary, evaluates the resulting partition, and aggregates the accepted communities to construct the next-level graph.

\textbf{Level-Zero Initialization.}
Let $\graph^0=(\nodes^0,\edges^0)$ denote the level-zero working graph. The initial partition contains one community for each level-zero node:
\begin{equation}\label{eq:cl-initial-partition}
\partition^0=\left\{\{\nidx\}:\nidx\in\nodes^0\right\}.
\end{equation}
Thus, each original node or contracted must-link component initially forms a separate community.

The initialization first checks basic feasibility conditions. In particular, a level-zero node must not contain a cannot-link conflict, and the number of level-zero nodes must satisfy
\begin{equation}
|\partition^0|\ge\numcom_{\min}
\end{equation}
Violation of either condition is reported as infeasible.

A singleton community may nevertheless violate a resource constraint when signed resource coefficients are present. Such cases are handled by the initialization-repair procedure described below. If a resource violation cannot be reduced through any admissible merge, the initialization procedure reports failure.

\textbf{Level-Wise Score and Feasibility.}
Let $\graph^\loulidx=(\nodes^\loulidx,\edges^\loulidx)$ denote the working graph at level $\loulidx$. For each node $\nidx\in\nodes^\loulidx$, let $\qval{\nidx}^{\loulidx}$ denote its internal modularity contribution, and let $\qval{\nidx\niidx}^{\loulidx}$ denote the modularity contribution between distinct nodes $\nidx$ and $\niidx$, defined similarly to Eqs. \eqref{eq:contracted-modularity-coefficient-internal} and \eqref{eq:contracted-modularity-coefficient}, respectively. For a community $\community\subseteq\nodes^\loulidx$, define its score as
\begin{equation}\label{eq:cl-community-score}
\louscore^\loulidx(\community)=\sum_{\nidx\in\community}\qval{\nidx}^{\loulidx}+\sum_{\nidx,\niidx\in\community,\ \nidx<\niidx}\qval{\nidx\niidx}^{\loulidx}.
\end{equation}
The score of a partition $\partition^\loulidx$ is
\begin{equation}\label{eq:cl-partition-score}
\louscore^\loulidx(\partition^\loulidx)=\sum_{\community\in\partition^\loulidx}\louscore^\loulidx(\community).
\end{equation}

For each resource class $\ridx\in\setres$, let $\rval{\nidx}{\ridx,\loulidx}$ denote the resource contribution associated with node $\nidx$ at level $\loulidx$, and let $\rval{\nidx\niidx}{\ridx,\loulidx}$ denote the pairwise resource contribution associated with nodes $\nidx$ and $\niidx$, defined analogously to Eqs.~\eqref{eq:aggregated-node-attribute} and \eqref{eq:aggregated-pair-attribute}, respectively. Accordingly, the resource usage of community $\community\subseteq\nodes^\loulidx$ is
\begin{equation}\label{eq:cl-resource-usage}
\loures_{\ridx}^{\loulidx}(\community)=\sum_{\nidx\in\community}\rval{\nidx}{\ridx,\loulidx}+\sum_{\nidx,\niidx\in\community,\ \nidx<\niidx}\rval{\nidx\niidx}{\ridx,\loulidx}.
\end{equation}
A community is resource-feasible if
\begin{equation}\label{eq:cl-resource-feasibility}
\loures_{\ridx}^{\loulidx}(\community)\le\maxres{\ridx}\quad\forall\ridx\in\setres.
\end{equation}

Unless stated otherwise, every community resulting from a move or merge must satisfy the cannot-link and resource constraints. The bounds $\numcom_{\min}$ and $\numcom_{\max}$ are enforced at the partition level through the local-moving and repair procedures described below.

\textbf{Constrained Local Moving.}
Consider moving node $\nidx$ from its current source community $\community^{\mathrm{source}}$ to an existing target community $\community^{\mathrm{target}}$. The corresponding score gain is
\begin{equation}\label{eq:cl-move-gain}
\lougain_{\text{move}}^{\loulidx}\left(\nidx:\community^{\mathrm{source}}\rightarrow\community^{\mathrm{target}}\right)=\sum_{\niidx\in\community^{\mathrm{target}}}\qval{\nidx\niidx}^{\loulidx}-\sum_{\niidx\in\community^{\mathrm{source}}\setminus\{\nidx\}}\qval{\nidx\niidx}^{\loulidx}.
\end{equation}
Candidate targets are normally restricted to communities adjacent to $\nidx$ according to the edge structure of the current working graph. When an upper bound $\numcom_{\max}$ is enforced and the current number of communities exceeds it, all existing communities—including nonadjacent ones—are considered so that local moving may reduce the community count. During local moving, no new communities are created; each accepted move assigns a node to an existing target community.

A move is admissible only if the resulting nonempty communities satisfy the constraints. In particular, the target community must remain cannot-link- and resource-feasible. Because resource coefficients may be signed, the source community must also remain resource-feasible after removing $\nidx$. A move that empties the source community is allowed only if the resulting number of communities remains at least $\numcom_{\min}$.

Nodes are visited in randomized order. A \emph{pass} denotes one complete traversal of all nodes, during which each node is considered once for a local move. Let $\loumidx$ denote the number of accepted moves during a pass. Each node is moved to the admissible target with the largest gain whenever the gain exceeds the tolerance $\tol{\mathrm{Louvain}}$.\footnote{In this study, we set $\tol{\text{Louvain}}=10^{-9}$.} The local-moving phase is repeated until a complete pass yields $\loumidx=0$ or the maximum number of passes is reached.\footnote{In this study, we set the maximum number of local-moving passes at each level to $50$.} Algorithm~\ref{alg:constrained-louvain-local} summarizes the procedure.

\begin{ruledalgorithm}{Constrained local-moving phase}
\label{alg:constrained-louvain-local}
\begin{algorithmic}[1]
\Require Level-$\loulidx$ graph $\graph^\loulidx$, partition $\partition^\loulidx$, tolerance $\tol{\mathrm{Louvain}}$.
\Repeat
    \State Randomly order the nodes of $\graph^\loulidx$ and set $\loumidx\leftarrow0$.
    \For{each node $\nidx$ in the randomized order}
        \State Let $\community^{\mathrm{source}}$ be the community containing $\nidx$ in $\partition^\loulidx$.
        \State Construct the candidate target set $\tcommunity_{\nidx}$ from communities adjacent to $\nidx$.
        \If{$|\partition^\loulidx|>\numcom_{\max}$}
            \State Add all other existing communities to $\tcommunity_{\nidx}$.
        \EndIf
        \State Remove candidate communities that violate constraints.
        \State Select $\community^{\mathrm{target}}\in\tcommunity_{\nidx}$ with the largest
        $\lougain_{\text{move}}^\loulidx\!\left(\nidx:\community^{\mathrm{source}}\rightarrow\community^{\mathrm{target}}\right)$.
        \If{$\tcommunity_{\nidx} \neq \emptyset$ and
        $\lougain_{\text{move}}^\loulidx\!\left(\nidx:\community^{\mathrm{source}}\rightarrow\community^{\mathrm{target}}\right)
        >\tol{\text{Louvain}}$}
            \State Move $\nidx$ from $\community^{\mathrm{source}}$ to $\community^{\mathrm{target}}$ and update $\partition^\loulidx$.
            \State Set $\loumidx\leftarrow\loumidx+1$.
        \EndIf
    \EndFor
\Until{$\loumidx=0$ or the maximum number of passes is reached}
\State \Return $\partition^\loulidx$.
\end{algorithmic}
\end{ruledalgorithm}

\textbf{Repair Operations.}
The algorithm employs two repair operations targeting different types of constraint violations: a \emph{resource-feasibility repair} for resource-infeasible communities and a \emph{community-count repair} for partitions exceeding the maximum number of communities. The resource-feasibility repair is applied only at level zero before constrained local moving, whereas the community-count repair is applied after constrained local moving whenever $|\partition^\loulidx|>\numcom_{\max}$.

\emph{Resource-Feasibility Repair.}
At level zero, a singleton community may violate a resource constraint when signed resource coefficients are present. To restore resource feasibility, define the normalized resource excess of a community $\community$ as
\begin{equation}\label{eq:cl-resource-excess}
\louexcess(\community)=\sum_{\ridx\in\setres}\frac{\max\left\{\loures_{\ridx}^{0}(\community)-\maxres{\ridx},0\right\}}{\max\left\{|\maxres{\ridx}|,1\right\}}.
\end{equation}
While a resource-infeasible community remains, the algorithm considers admissible merges between two communities $\community_1$ and $\community_2$, with at least one of them being resource-infeasible. Among these candidate merges, it selects the pair that maximizes the reduction in normalized resource excess,
\begin{equation}\label{eq:cl-resource-excess-reduction}
\louexcess(\community_1)+\louexcess(\community_2)-\louexcess(\community_1\cup\community_2).
\end{equation}
Only merges with a positive excess reduction that preserve the cannot-link constraints and the minimum community-count bound $\numcom_{\min}$ are considered. The resource-feasibility repair fails if no such merge exists.

\emph{Community-Count Repair.}
After constrained local moving, the resulting partition may still contain more than $\numcom_{\max}$ communities. In this case, the algorithm repairs the partition by successively merging admissible communities. For two communities $\community_1$ and $\community_2$, define the merge gain as
\begin{equation}\label{eq:cl-merge-gain}
\lougain_{\text{merge}}^\loulidx(\community_1,\community_2)=\sum_{\nidx\in\community_1}\sum_{\niidx\in\community_2}\qval{\nidx\niidx}^{\loulidx}.
\end{equation}
While $|\partition^\loulidx|>\numcom_{\max}$, the algorithm repeatedly selects an admissible community pair with the largest merge gain. Each merge must preserve the cannot-link and resource constraints and must not reduce the number of communities below $\numcom_{\min}$. The selected merge may have negative gain because feasibility takes precedence over objective improvement. If no admissible merge can satisfy the required community-count bound, the current partition is rejected.

\textbf{Aggregation and Higher-Level Initialization.}
Once a feasible partition $\partition^\loulidx$ is accepted, each community $\community_\nidx\in\partition^\loulidx$ is contracted into a node $\nidx$ of the next-level graph. The modularity coefficients are updated as
\begin{equation}\label{eq:cl-internal-score-aggregation}
\qval{\nidx}^{\loulidx+1}=\louscore^\loulidx(\community_\nidx),
\end{equation}
and
\begin{equation}\label{eq:cl-score-aggregation}
\qval{\nidx\niidx}^{\loulidx+1}=\sum_{\onidx\in\community_\nidx}\sum_{\oniidx\in\community_\niidx}\qval{\onidx\oniidx}^{\loulidx}.
\end{equation}

For each resource class $\ridx\in\setres$, the corresponding resource coefficients are aggregated as
\begin{equation}\label{eq:cl-node-resource-aggregation}
\rval{\nidx}{\ridx,\loulidx+1}=\loures_{\ridx}^{\loulidx}(\community_\nidx),
\end{equation}
and
\begin{equation}\label{eq:cl-pair-resource-aggregation}
\rval{\nidx\niidx}{\ridx,\loulidx+1}=\sum_{\onidx\in\community_\nidx}\sum_{\oniidx\in\community_\niidx}\rval{\onidx\oniidx}{\ridx,\loulidx}.
\end{equation}
These aggregation rules preserve both the partition score and resource usage across levels. A cannot-link relation is assigned between two aggregated nodes whenever at least one cannot-link pair crosses their corresponding communities. Original-node memberships are also retained so that any higher-level partition can be mapped back to the original graph.

The next level is initialized with one community per aggregated node:
\begin{equation}\label{eq:cl-higher-level-initialization}
\partition^{\loulidx+1}=\left\{\{\nidx\}:\nidx\in\nodes^{\loulidx+1}\right\}.
\end{equation}
Because each aggregated node represents a feasible community accepted at the preceding level, the higher-level initialization preserves the cannot-link and resource feasibility of the preceding partition.

\textbf{Overall Procedure.}
Algorithm~\ref{alg:constrained-louvain} summarizes the complete procedure. At level zero, resource-infeasible communities are first repaired, if necessary. At each level, constrained local moving is followed by community-count repair when the resulting partition contains more than $\numcom_{\max}$ communities. The resulting partition is then mapped back to the original graph and checked for feasibility. A feasible partition replaces the incumbent only if its score improves by more than $\tol{\text{Louvain}}$. If no such improvement is obtained or the prescribed time limit is reached, the algorithm terminates and returns the incumbent.

\begin{ruledalgorithm}{Constrained Louvain heuristic}
\label{alg:constrained-louvain}
\begin{algorithmic}[1]
\Require Original graph $\graph$, tolerance $\tol{\text{Louvain}}$, time limit.
\State Construct the level-zero graph $\graph^{0}$ by contracting must-link pairs.
\State Initialize $\partition^{0}$ according to Eq.~\eqref{eq:cl-initial-partition}.
\State Repair resource-infeasible communities using Eq.~\eqref{eq:cl-resource-excess-reduction}. \Comment{Resource-feasibility repair}
\State Set $\loulidx\leftarrow0$ and initialize the incumbent score to $-\infty$.
\Loop
\State Apply Algorithm~\ref{alg:constrained-louvain-local} to $\partition^\loulidx$. \Comment{Constrained local moving}
\State Merge admissible communities until $|\partition^\loulidx|\le\numcom_{\max}$. \Comment{Community-count repair}
\State Map $\partition^\loulidx$ to the original graph and check feasibility.
\If{the partition is infeasible}
\State \Return the incumbent partition, or report failure if no incumbent partition exists.
\EndIf
\If{$\louscore^\loulidx(\partition^\loulidx)$ improves the incumbent score by more than $\tol{\text{Louvain}}$}
\State Update the incumbent partition and score.
\Else
\State \textbf{break}.
\EndIf
\If{the time limit is reached or $|\partition^\loulidx|\le\numcom_{\min}$}
\State \textbf{break}.
\EndIf
\State Aggregate $\partition^\loulidx$ using Eqs.~\eqref{eq:cl-internal-score-aggregation}--\eqref{eq:cl-pair-resource-aggregation}.
\State Initialize $\partition^{\loulidx+1}$ according to Eq.~\eqref{eq:cl-higher-level-initialization}.
\State Set $\loulidx\leftarrow\loulidx+1$.
\EndLoop
\State \Return the incumbent partition.
\end{algorithmic}
\end{ruledalgorithm}

\section{Pricing Acceleration Techniques}
\label{sec:bnp-pricing-acceleration}

This section introduces techniques that reduce unnecessary pricing iterations and accelerate exact pricing while preserving the exactness of the branch-and-price algorithm.

\textbf{Redundancy Control.}
A column already present in $\prob{\mathrm{RMP}}$ does not enlarge the column pool and therefore provides no further improvement to the master problem. In heuristic pricing, duplicate columns are simply removed from the generated candidate set. In exact pricing, if the optimal solution coincides with an existing or explicitly forbidden column $\bar{\decVe}$, we exclude that column by adding the redundancy cut
\begin{equation}\label{eq:bnp-redundancy-cut}
    \sum_{\nidx:\bardece{\nidx}=1}\left(1-\dece{\nidx}\right)+\sum_{\nidx:\bardece{\nidx}=0}\dece{\nidx}\ge1.
\end{equation}
This constraint requires the next pricing solution to differ from $\bar{\decVe}$ in at least one node assignment. The pricing problem is then reoptimized, and additional redundancy cuts are introduced only when needed. This avoids repeatedly generating the same columns without imposing a large set of exclusion constraints in advance.

\textbf{Heuristic Pricing Acceptance Threshold.}
Let $\columns_{\mathrm{cand}}$ denote the set of improving candidate columns obtained by graph-partitioning-based heuristic pricing at a given column-generation iteration, and let $\objpri(\decVe)$ denote the pricing objective value in Eq.~\eqref{eq:bnp-psp-obj} for $\decVe\in\columns_{\mathrm{cand}}$ under the current dual solution. For a nonempty $\columns_{\mathrm{cand}}$, heuristic pricing is considered to have failed to identify a sufficiently strong improving column if
\begin{equation}\label{eq:bnp-heuristic-threshold}
    \max_{\decVe\in\columns_{\mathrm{cand}}}\objpri(\decVe)\le\tol{\mathrm{GP}},
\end{equation}
where $\tol{\mathrm{GP}}=0.5$ in the implementation. When Eq.~\eqref{eq:bnp-heuristic-threshold} holds, the generated candidates are discarded even if some have small positive pricing objective values, and exact pricing is triggered. This rule avoids repeated column-generation iterations driven by weak columns and invokes exact pricing when heuristic pricing provides insufficient improvement.

\textbf{Exact Pricing with Warm Start.}
When heuristic pricing fails under the acceptance rule above, exact pricing is invoked. If heuristic pricing has produced at least one feasible candidate, the candidate with the largest pricing objective value is used as a warm start for the exact-pricing model.

\textbf{Variable Fixing with a Dominance Rule.}
Before solving exact pricing, selected pricing variables can be fixed to zero using a dominance rule. For node $\nidx$, the marginal improvement in the pricing objective from setting $\dece{\nidx}=1$ is bounded above by
\begin{equation}\label{eq:bnp-pricing-fixing-bound}
\sum_{\niidx\in\nodes\setminus\left\{\nidx\right\}}\max\left\{\qval{\nidx\niidx},0\right\}-\dual{\nidx}^*.
\end{equation}
If this upper bound is nonpositive, including node $\nidx$ cannot improve the pricing objective even under the most favorable contributions from the other nodes. After separately checking the singleton column and the branching and feasibility restrictions at the current branch-and-price node, the variable can therefore be fixed as
\begin{equation}\label{eq:bnp-pricing-fixing-cut}
\decw{\nidx}=0.
\end{equation}
This fixing reduces the size of the exact-pricing problem without affecting its optimality certificate.

\textbf{Post-Pricing Neighborhood Search.}
If exact pricing returns a column $\decVe^*$ with a positive objective value, the algorithm performs an additional one-flip neighborhood search. For each node $\nidx$, a neighboring candidate is generated by replacing $\dece{\nidx}^*$ with $1-\dece{\nidx}^*$ while leaving all other components unchanged. A neighboring column is added only if it is feasible, nonredundant, and has a positive pricing objective value. This step increases column diversity after exact pricing without affecting the optimality certificate provided by the exact-pricing solution.

\section{Automotive Subsystem Architecture Graphs}
\label{sec:automotive-graphs}

\begin{figure}[hbt!]
\centering
\subfloat[\textit{ICE--HVAC cooling}\label{fig:ice-hvac-cooling}]{
    \includegraphics[page=9,width=0.31\linewidth]{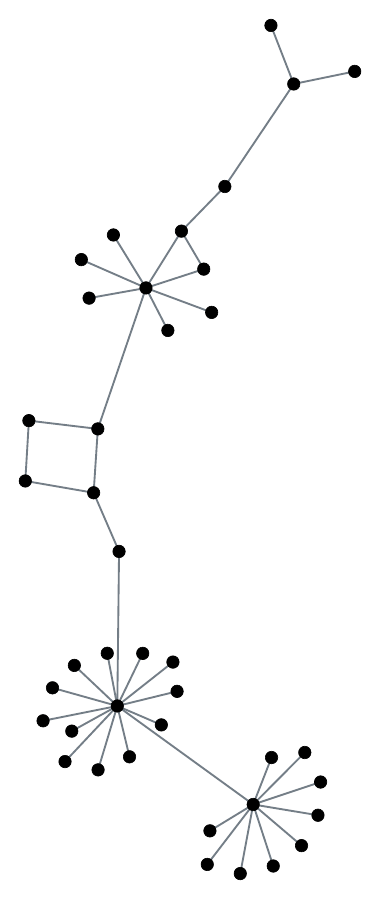}
}
\hfill
\subfloat[\textit{Thermal energy}\label{fig:thermal-energy}]{
    \includegraphics[page=10,width=0.64\linewidth]{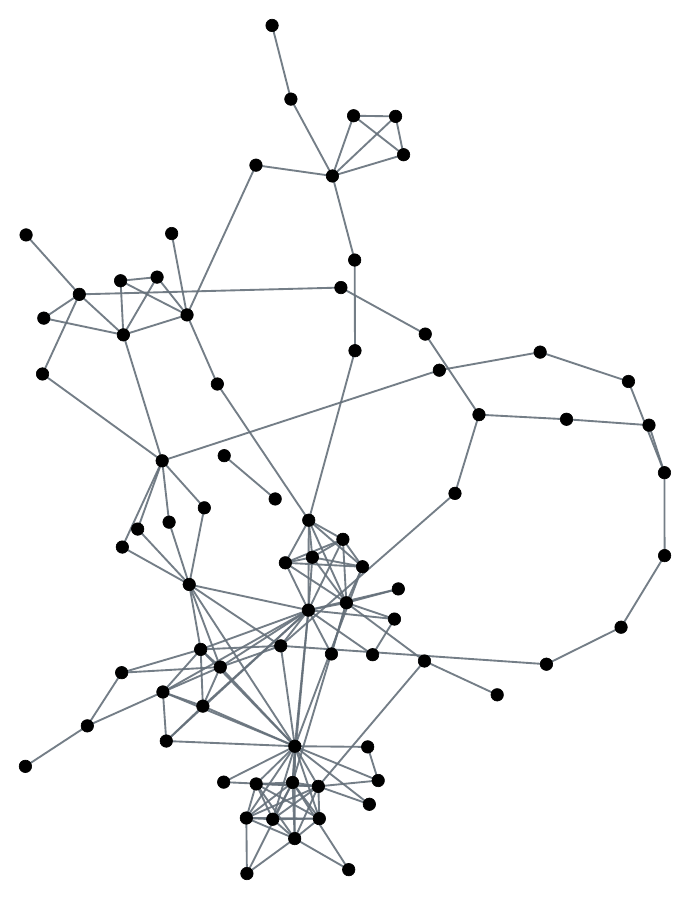}
}
\caption{Graph representations of the automotive subsystem architectures newly introduced in this work.}
\label{fig:automotive-graphs}
\end{figure}

The automotive graphs considered in the numerical experiments originate from modular architecture design problems for subsystems of \acrfull{hmc}. In this application, nodes represent system components and edges represent functional connections between them. The design problem seeks a modular decomposition where strongly interconnected components are grouped together. Modularity maximization provides a graph-based objective for this purpose, while additional requirements imposed by system designers and other decision makers can be incorporated as constraints. Automotive modular architecture design provides a practical application of the constrained modularity-maximization problem considered in this work.

Three subsystem graphs are considered: \textit{Front suspension}, \textit{ICE--HVAC cooling}, and \textit{Thermal energy}. The \textit{Front suspension} graph was previously introduced in Ref.~\cite{kim2025mathematical}, whereas the latter two are introduced in this work and illustrated in Fig.~\ref{fig:automotive-graphs}.

The three automotive subsystem graphs represent different functional domains of a vehicle. The \textit{Front suspension} system supports and controls the interaction between the vehicle body and the front wheels, contributing to ride comfort, tire--road contact, and handling stability. The \textit{ICE--HVAC cooling} system combines thermal management of the internal-combustion engine with cabin heating, ventilation, and air-conditioning, coordinating coolant and refrigerant circuits for heat rejection and cabin temperature control. The \textit{Thermal energy} system represents a broader thermal-management architecture for an electrified vehicle, coordinating heat transfer among the battery, electric powertrain and power electronics, and cabin HVAC to maintain suitable operating temperatures and improve energy efficiency.

\bibliography{references}

\end{document}